\documentclass[11pt]{elsarticle}
\usepackage{verbatim,amsmath,amssymb}
\usepackage{epsfig,float,color}
\usepackage{geometry}
\usepackage{setspace}
\usepackage{natbib}
\usepackage{amsthm,mathrsfs,amsfonts,dsfont}
\usepackage{hyperref}
\usepackage[yyyymmdd,hhmmss]{datetime}
\usepackage{cleveref}
\usepackage{caption}
\usepackage{subcaption}
\usepackage[section]{placeins}
\usepackage{fontenc}
\usepackage{mathbbol}
\usepackage{mmap}
\usepackage{footnote}
\definecolor{darkblue}{rgb}{0,0,1}
\hypersetup{pdftex=true, colorlinks=true, breaklinks=true, linkcolor=darkblue, menucolor=darkblue, pagecolor=darkblue, citecolor=darkblue, urlcolor=darkblue}
\newcommand{\bitm}{\begin{itemize}}
\newcommand{\eitm}{\end{itemize}}
\newcommand{\bnumr}{\begin{enumerate}}
\newcommand{\enumr}{\end{enumerate}}
\newcommand{\bolds}[1]{\boldsymbol{#1}}

\newcommand {\sigab}{\sigma^{\alpha\beta}}

\newcommand {\aab}{a^{\alpha\beta}}
\newcommand {\auab}{a_{\alpha\beta}}
\newcommand {\agd}{a^{\gamma\delta}}

\newcommand {\aag}{a^{\alpha\gamma}}

\newcommand {\augd}{a_{\gamma\delta}}

\newcommand {\adb}{a^{\delta\beta}}

\newcommand {\Aab}{A^{\alpha\beta}}
\newcommand {\Auab}{A_{\alpha\beta}}

\newcommand {\Agd}{A^{\gamma\delta}}

\newcommand {\Mab}{M^{\alpha\beta}}

\newcommand {\bab}{b^{\alpha\beta}}

\newcommand {\buab}{b_{\alpha\beta}}

\newcommand {\bgd}{b^{\gamma\delta}}

\newcommand {\bugd}{b_{\gamma\delta}}

\newcommand {\tauab}{\tau^{\alpha\beta}}

\newcommand {\eqb}[1]{\begin{equation}\begin{array}{#1}}
\newcommand {\eqe}{\end{array}\end{equation}}

\newcommand {\esb}[1]{\begin{equation*}\begin{array}{#1}}
\newcommand {\ese}{\end{array}\end{equation*}}
\newcommand {\ds}{\displaystyle}

\newcommand {\pa}[2]{\frac{\partial{#1}}{\partial{#2}}}

\newcommand {\paqq}[3]{\frac{\partial^2{#1}}{\partial{#2}\,\partial{#3}}}
\newcommand {\back}{\! \! \!}
\newcommand {\is}{\back &=& \back}
\newcommand {\dis}{\back &:=& \back}

\newcommand {\plus}{\back &+& \back}
\newcommand {\mi}{\back &-& \back}

\newcommand {\tr}{\mathrm{tr}\,}

\newcommand {\dif}{\mathrm{d}}

\newcommand {\II}{{I\kern-.3em I}}
\newcommand {\III}{{I\kern-.3em I\kern-.3em I}}

\newcommand {\mf}{\mathbf{f}}

\newcommand {\mx}{\mathbf{x}}

\newcommand {\ba}{\boldsymbol{a}}

\newcommand {\bi}{\boldsymbol{i}}

\newcommand {\bn}{\boldsymbol{n}}

\newcommand {\bx}{\boldsymbol{x}}
\newcommand {\by}{\boldsymbol{y}}

\newcommand {\mN}{\mathbf{N}}

\newcommand {\bA}{\boldsymbol{A}}
\newcommand {\bB}{\boldsymbol{B}}
\newcommand {\bC}{\boldsymbol{C}}

\newcommand {\bE}{\boldsymbol{E}}
\newcommand {\bF}{\boldsymbol{F}}

\newcommand {\bI}{\boldsymbol{I}}

\newcommand {\bM}{\boldsymbol{M}}
\newcommand {\bN}{\boldsymbol{N}}

\newcommand {\bS}{\boldsymbol{S}}

\newcommand {\bU}{\boldsymbol{U}}

\newcommand {\bX}{\boldsymbol{X}}
\newcommand {\bY}{\boldsymbol{Y}}
\newcommand {\bZ}{\boldsymbol{Z}}

\newcommand {\bsig}{\mbox{\boldmath$\sigma$}}

\newcommand {\btau}{\mbox{\boldmath$\tau$}}

\newcommand {\bbH}{\mathbb{H}}

\newcommand {\bbR}{\mathbb{R}}

\newcommand {\IR}{{\rm\kern.24em
   \vrule width.02em height1.53ex depth-.05ex
   \kern-.3em R}}
\newcommand {\ic}{{\rm\kern.20em
   \vrule width.02em height1.0ex depth-.05ex
   \kern-.22em c}}
\newcommand {\ia}{{\rm\kern.20em
   \vrule width.02em height1.05ex depth-.0ex
   \kern-.25em a}}
\newcommand {\IC}{{\rm\kern.24em
   \vrule width.02em height1.4ex depth-.05ex
   \kern-.26em C}}
\newcommand {\ID}{{\rm\kern.34em
   \vrule width.02em height1.5ex depth-.05ex
   \kern-.36em D}}
\newcommand {\IS}{{\rm\kern.24em
   \vrule width.02em height1.6ex depth.05ex
   \kern-.26em S}}
\newcommand {\IT}{{\rm\kern.50em
   \vrule width.02em height1.55ex depth-.05ex
   \kern-.52em T}}

\newcommand {\IE}{{\rm\kern.24em
   \vrule width.02em height1.55ex depth-.05ex
   \kern-.33em E}}
\newcommand {\IEa}{{\rm\kern.24em
   \vrule width.02em height1.55ex depth-.05ex
   \kern-.33em E}^{1}_{ijkl}}
\newcommand {\IEb}{{\rm\kern.24em
   \vrule width.02em height1.55ex depth-.05ex
   \kern-.33em E}^{2}_{ijkl}}

\newcommand {\sJ}{\mathcal{J}}

\newcommand {\vaua}{\ba_\alpha}
\newcommand {\vaa}{\ba^\alpha}

\newcommand {\vaub}{\ba_\beta}
\newcommand {\vab}{\ba^\beta}

\newcommand {\vag}{\ba^\gamma}

\newcommand {\vad}{\ba^\delta}

\newcommand {\Ass}[2]{\kern 0.9ex \vrule width0.45em height0.2ex depth0ex \kern -2.1ex \bigwedge_{#1}^{#2}}
\newcommand {\ASS}[2]{\kern 1.45ex \vrule width0.5em height0.2ex depth0ex \kern -2.65ex \bigwedge_{#1}^{#2}}

\newcommand {\aabgd}{{a}^{\alpha\beta\gamma\delta}}

\newcommand {\cabgd}{{c}^{\alpha\beta\gamma\delta}}
\newcommand {\dabgd}{{d}^{\alpha\beta\gamma\delta}}
\newcommand {\eabgd}{{e}^{\alpha\beta\gamma\delta}}
\newcommand {\fabgd}{{f}^{\alpha\beta\gamma\delta}}

\newcommand {\bkappa}{\boldsymbol{\kappa}}

\graphicspath{ {/} }

\definecolor{cgn}{rgb}{0,0,0}
\definecolor{cgm}{rgb}{0.0,.0,0}

\begin{document}

\begin{center}
\Large{\bf{A new efficient hyperelastic finite element model for graphene and its application to carbon nanotubes and nanocones}}\\

\end{center}

\begin{center}

\large{Reza Ghaffari\footnote{email: ghaffari@aices.rwth-aachen.de} and Roger A. Sauer\footnote{Corresponding author, email: sauer@aices.rwth-aachen.de}}\\
\vspace{4mm}

\small{\textit{
Aachen Institute for Advanced Study in Computational Engineering Science (AICES), \\
RWTH Aachen University, Templergraben 55, 52056 Aachen, Germany \\[1.1mm]}}

%

\vspace{4mm}

Published\footnote{This pdf is the personal version of an article whose final publication is available at \href{https://doi.org/10.1016/j.finel.2018.04.001}{http:/\!/sciencedirect.com}} in \textit{
Finite Elements in Analysis and Design}, \href{https://doi.org/10.1016/j.finel.2018.04.001}{DOI: 10.1016/j.finel.2018.04.001}\\
Submitted on 13.~January 2018, Revised on 2.~April 2018, Accepted on 3.~April 2018

\end{center}

\vspace{3mm}


\rule{\linewidth}{.15mm}
{\bf Abstract}
A new hyperelastic material model is proposed for graphene-based structures, such as graphene, carbon nanotubes (CNTs) and carbon nanocones (CNC). The proposed model is based on a set of invariants obtained from the right surface Cauchy-Green strain tensor and a structural tensor. The model is fully nonlinear and can simulate buckling and postbuckling behavior. It is calibrated from existing quantum data. It is implemented within a rotation-free isogeometric shell formulation. The speedup of the model is 1.5 relative to the finite element model of \citet{Ghaffari2017_01}, which is based on the logarithmic strain formulation of \citet{Kumar2014_01}. The material behavior is verified by testing uniaxial tension and pure shear. The performance of the material model is illustrated by several numerical examples. The examples include bending, twisting, and wall contact of CNTs and CNCs. The wall contact is modeled with a coarse grained contact model based on the Lennard-Jones potential. The buckling and post-buckling behavior is captured in the examples. The results are compared with reference results from the literature and there is good agreement.

{{\bf Keywords}: Anisotropic hyperelastic material models; buckling and post-buckling; carbon na\-no\-tube and nanocones; isogeometric finite elements; Kirchhoff-Love shell theory.}

\vspace{-4mm}
\rule{\linewidth}{.15mm}
\section{Introduction}
Graphene and graphene-based structures such as carbon nanotubes (CNT) and carbon nanocones (CNC) \citep{Merkulov2001,Geim2007,Yudasaka2008,Naess2009,ovidko2012,alwarappan2013} have unique mechanical \citep{Mokashi2007_01,Javvaji2016,Khoei2016_01}, thermal \citep{Balandin2011,Pop2012,Kim2016,Fan2017} and electrical \citep{Neto2009,Ulloa2013,Qian2013_01,Baringhaus2014} properties. They can be used in sensors \citep{natsuki2015}, energy storage devices \citep{Liu2014}, healthcare \citep{Kostarelos2014} and as a coating against corrosion \citep{Bohm2014}. They are used to improve mechanical, thermal and electrical properties of composites \citep{Marinho2012,Galindo2014,SPANOS2015,Atif2016,Xia2017}. CNTs and CNCs can be obtained by rolling of a graphene sheet \citep{Lee2012,Zhong2012_01}. Robust and efficient analysis methods should be developed in order to reduce the time and cost of design and production.\\
There are several different approaches in the literature for modeling graphene. One is based on the Cauchy-Born rule applied to intermolecular potentials. \citet{Arroyo2004_01} propose an exponential Cauchy-Born rule to simulate the mechanical behavior of CNTs. \citet{Guo2006} and \citet{Wang2006} use a higher order Cauchy-Born rule to model CNTs. \citet{Yan2012} use a higher order gradient continuum theory\footnote{This method is similar to \textcolor{cgm}{the} Cauchy-Born rule.} and the Tersoff-–Brenner potential to obtain the properties of single-walled CNCs.
A second approach is based on the quasi-continuum method \citep{Shenoy1999_01}. \citet{Yan_2013_01} apply the quasi-continuum to simulate buckling and post-buckling of CNCs. A temperature-related quasi-continuum model is proposed by \citet{Wang2016} to model the behavior of CNCs under axial compression.
A third approach is based on classical continuum material models. Those are popular for graphene in the context of isotropic linear material models. \citet{FIROUZABADI2011} obtain the natural frequencies of nanocones by using a nonlocal continuum theory and linear elasticity assumptions. Their work is extended to the stability analysis under external pressure and axial loads by \citet{FIROUZABADI2012} and the stability analysis of CNCs conveying fluid by \citet{Gandomani2016_01}. \citet{Lee2012} use the finite element (FE) method to obtain the natural frequencies of CNTs and CNCs. The interaction of carbon atoms is modeled as continuum frame elements. Graphene has an anisotropic behavior under large strains. There are several continuum material models for anisotropic behavior of graphene. \citet{Sfyris2014_01} and \citet{Sfyris2014_02} use Taylor expansion and a set of invariants to propose strain energy functionals for graphene based on its lattice structure. \citet{Delfani2013_01} and \citet{Delfani2015_01,Delfani2016_01} use a similar Taylor expansion for the strain energy and apply symmetry operators to the elasticity tensors in order to reduce the number of independent variables. Nonlinear membrane material models are proposed by \citet{Xu2012} and \citet{Kumar2014_01}. They use ab-initio results to calibrate their models. The model of \citet{Kumar2014_01} is based on the logarithmic strain and the symmetry group of the graphene lattice \citep{Zheng_1993_01,Zheng_1994_01,Zheng_1994_02}. This symmetry group reduces the number of parameters in the model of \citet{Xu2012} by a half. The membrane model of \citet{Kumar2014_01} is extended by \citet{Ghaffari2017_01} to a \textcolor{cgm}{FE} shell model by adding a bending energy term.
\textcolor{cgn}{Such FE models tend to be much more efficient than all-atom models: \citet{Ghaffari2017_01} study the indentation of a square sheet with length 550 nm and found that the FE model requires 122,412 nodes, while the corresponding atomistic system has about 12 million atoms, i.e. about 100 times more.} \citet{Ghaffari2018_01} conduct a modal analysis of graphene sheets and CNTs under various nonlinearities. A finite thickness for graphene is considered in the most of the mentioned works. Thus, an integration through the thickness needs be conducted to obtain the bending stiffness. The finite thickness assumption can be avoided by writing the strain energy density per unit area of the surface as in \citet{Xu2012,Kumar2014_01,Ghaffari2017_01} and \citet{Ghaffari2018_01}.\\
The material model of \citet{Ghaffari2017_01} and the proposed new material model in the current paper are implemented in the rotation-free isogeometric finite shell element formulation of \citet{Sauer2014_01}, \citet{Sauer2017_01} and \citet{Duong2016_01}. This formulation is based on displacement degree of freedoms (DOFs) and avoids rotation DOFs through the use of Kirchhoff-Love kinematics and NURBS discretization \citep{Hughes2005_01}. The avoidance of rotational DOFs increases efficiency and simplifies the formulation \citep{Simo1989_01}. A material model based on continuum mechanics is necessary for the development of a shell formulation. The model of \citet{Ghaffari2017_01} is quite complicated and computationally expensive. It is based on a logarithmic strain formulations, which requires using chain rule and summation over fourth and sixth order tensors (see Sec.~\ref{s:finite_element} for more details). This high computational cost is avoided in the new proposed material model.\\
In summary, the novelties of the current work are:
\begin{itemize}
  \item It can be used both in curvilinear and Cartesian shell formulations.
  \item It is simpler to implement and thus 1.5 faster\footnote{\textcolor{cgn}{In computing the stiffness matrix.}} than the model of \citet{Ghaffari2017_01}.
  \item It is fully nonlinear and can capture buckling and post-buckling behavior.
  \item It is suitable to simulate and study carbon nanocones under large deformations.
  \item It is applied to simulate contact of CNTs and CNCs with a Lennard-Jones wall.
  \item The latter example demonstrates that CNCs are ideal candidates for AFM tips.
\end{itemize}
The remainder of this paper is organized as follows: In Sec.~\ref{s:finite_element} the finite element formulation is summarized and the development of a new material model is motivated. In Sec.~\ref{s:metric_mat_model}, a new hyperelastic shell material model for graphene-based structures is proposed. In Sec.~\ref{s:elementary_behav}, the model is verified and compared with the model of \citet{Ghaffari2017_01} considering various test cases. Sec.~\ref{s:metric_num_result} presents several numerical examples involving buckling and contact of CNTs and CNCs. The behavior is compared with molecular dynamics and quasi-continuum results from the literature. The paper is concluded in Sec.~\ref{s:conclusion}.

\section{Finite element formulation for Kirchhoff-Love shells}\label{s:finite_element}
It this section, the discretized weak form is summarized and the development of a new material model is motivated. The Cauchy stress tensor of Kirchhoff-Love shell theory can be written as\footnote{\textcolor{cgm}{Subscript} KL is added here to distinguish the total Cauchy stress $\bsig_{\text{KL}}$ from its \textcolor{cgm}{membrane} contribution $\bsig := \sigab \vaua \otimes \vaub$.} \citep{Sauer2017_01}
\eqb{lll}
\bsig_{\text{KL}} \is N^{\alpha\beta}\,\ba_{\alpha}\otimes \ba_{\beta} + S^{\alpha}\,\ba_{\alpha}\otimes\bn~,
\label{e:shell_stress}
\eqe
where
\eqb{lll}
N^{\alpha\beta} \is \ds \sigma^{\alpha\beta}+b_{\gamma}^{\beta}\,M^{\gamma\alpha}~
\label{e:shell_mem_stress}
\eqe
and
\eqb{lll}
S^{\alpha} \is \ds -M_{;\beta}^{\beta\alpha}
\eqe
are the components of the membrane stress and out-of-plane shear. Here, ``;'' denotes the co-variant derivative, and $\ba_{\alpha}$ and $\bn$ are the tangent and normal vectors of the shell surface in the current configuration, see \ref{s:curvilinear_desc}.
For hyperelastic materials, $\sigma^{\alpha\beta}$ and $M^{\alpha\beta}$ are given by
\eqb{l}
\sigma^{\alpha\beta} = \tauab/J~,\quad \tauab = \ds \pa{W}{\auab}~,
\label{e:surf_Cauchy_compo}
\eqe
\eqb{l}
M^{\alpha\beta} = M_{0}^{\alpha\beta}/J~,\quad M_0^{\alpha\beta} = \ds \pa{W}{\buab}~,
\label{e:surf_Moment_compo}
\eqe
where $W$ is the strain energy density per unit area of the initial configuration, and $\auab$ and $\buab$ are the covariant components of the metric and curvature tensor \citep{Sauer2017_01}. $b_{\alpha}^{\beta}$ in Eq.~\eqref{e:shell_mem_stress} are the mixed components of the curvature tensor (see \ref{s:curvilinear_desc}).  The discretized weak form for Kirchhoff-Love shells can be written as \citep{Duong2016_01}
\eqb{l}
\ds \sum\limits_{e=1}^{n_{\text{el}}} { (G_{\text{int}}^{e}+G_{\text{c}}^{e}-G_{\text{ext}}^{e})}=0~, \forall~\delta \mx_{e} \in \mathcal{V}~,
\eqe
where $\delta\mx_{e}$ is the variation of the element nodes, $n_{\text{el}}$ is the number of elements and $\mathcal{V}$ is the space of admissible variations. $G_{\mathrm{c}}^{e}$ and $G_{\text{ext}}^{e}$ are related to contact and external forces \citep{Ghaffari2017_01}. $G_{\text{int}}^{e}$ is the internal virtual work of element $\Omega^e_0$ defined as
\eqb{llll}
G_{\text{int}}^{e} := \delta\mx_{e}^{\text{T}}\,(\mf_{\text{int}\tau}^{e}+\mf_{\text{int}M}^{e})~,
\eqe
with
\eqb{lll}
\ds \mf_{\text{int}\tau}^{e} \is \ds \int\limits_{\Omega^e_0 }{\tau^{\alpha\beta}\,\mN_{,\alpha}^{\text{T}}\,\ba_{\beta}~\dif A}~,\\[3mm]
\ds \mf_{\text{int}M}^{e} \is \ds \ds \int\limits_{\Omega^e_0 }{M_0^{\alpha\beta}\,\tilde{\mN}_{;\alpha\beta}^{\text{T}}\,\bn~\dif A}~,
\eqe
where $\mN$~, $\mN_{,\alpha}$ and $\tilde{\mN}_{;\alpha\beta}$ are the shape function arrays of the element that are defined as
\eqb{lll}
\mN \dis [N_1\,\boldsymbol{1}, N_2\,\boldsymbol{1}, \ldots, N_{n_e}\,\boldsymbol{1}]~,\\[1mm]
\mN_{,\alpha} \dis [N_{1,\alpha}\,\boldsymbol{1}, N_{2,\alpha}\,\boldsymbol{1}, \ldots, N_{n_e,\alpha}\,\boldsymbol{1}]~, \\[1mm]
\tilde{\mN}_{;\alpha\beta} \dis \mN_{,\alpha\beta}-\Gamma_{\alpha\beta}^{\gamma}\,\mN_{,\gamma}~,\\[1mm]
\mN_{,\alpha\beta} \dis [N_{1,\alpha\beta}\,\boldsymbol{1}, N_{2,\alpha\beta}\,\boldsymbol{1}, \ldots, N_{n_e,\alpha\beta}\,\boldsymbol{1}]~.
\eqe
Here, ``$\bullet,\alpha$'' denotes the parametric derivative $\partial{\bullet}/\partial\xi^\alpha$, and $\boldsymbol{1}$ and $N_{i}$ are the three dimensional identity tensor and the NURBS shape functions \citep{Hughes2005_01}. $\tauab$ and $M^{\alpha\beta}_{0}$ need to be specified for the finite element implementation through Eq.~\eqref{e:surf_Cauchy_compo} and \eqref{e:surf_Moment_compo}. $\tauab$ corresponds to the components of the in-plane Kirchhoff stress tensor $\btau = J\bsig$. They are equal to the components $S^{\alpha\beta}$ of the in-plane second Piola-Kirchhoff (2.PK) stress $\bS=S^{\alpha\beta}\,\bA_{\alpha}\otimes\bA_{\beta}$ since
\eqb{l}
\tauab = \ba^{\alpha}\cdot\btau\ba^{\beta} = \bA^{\alpha}\cdot\bS\bA^{\beta}=S^{\alpha\beta}~,
\label{e:tau_s_connect}
\eqe
due to $\btau = \bF\,\bS\,\bF^{\text{T}}$ and $\ba^{\alpha} = \bF^{-\text{T}} \bA^{\alpha}$.
Here $\bF=\ba_{\alpha}\otimes\bA^{\alpha}$ is the surface deformation gradient. $\bA_{\alpha}$ ($\ba_{\alpha}$) and $\bA^{\alpha}$ ($\ba^{\alpha}$) are the tangent and dual vectors in the reference (current) configuration (see \ref{s:curvilinear_desc}). Following Eq.~\eqref{e:surf_Cauchy_compo},
the 2.PK stress $\bS$ can also be written as
\eqb{lll}
\bS \is \ds 2\pa{W}{\bC}~,
\label{e:Surf_2PK_curve_con_energy}
\eqe
where $\bC=\bF^{\text{T}}\,\bF$ is the right surface Cauchy-Green deformation tensor. $\bS$ can be computed by using Eq.~\eqref{e:Surf_2PK_curve_con_energy}. However, if the model is developed based on the logarithmic strain, $\bS$ can not be directly computed and $\bS$ needs to connected to the logarithmic strain $\bE^{(0)}:=1/2\ln(\bC)$. Using Eq.~\eqref{e:Surf_2PK_curve_con_energy}, $\bS$ can be then obtained as
\eqb{lll}
\ds \bS \is \ds 2\pa{W}{\bE^{(0)}}:\mathcal{L}^{1}~,
\label{e:Surf_2PK_curve_con_log}
\eqe
where
\eqb{lll}
\ds \mathcal{L}^{1} \dis \ds \pa{\bE^{(0)}}{\bC}~.
\eqe
$\mathcal{L}^{1}$ and $\mathcal{L}^{2}:=\partial^{2}\bE^{(0)}/\partial\bC\otimes\partial\bC$ are needed for the calculation of $\bS$ and its corresponding elasticity tensor, which appears in the FE stiffness matrix\footnote{See \citet{Kumar2014_01} for $\mathcal{L}^{1}$ and $\mathcal{L}^{2}$.}. There is a high computational cost for the calculation of $\mathcal{L}^{1}$ and $\mathcal{L}^{2}$ \citep{Asghari2007_01,Germain2010_01,Miehe2011_01,Kumar2014_01,Ghaffari2017_01} due to double and quadruple contraction with the logarithmic stress $\bS^{(0)} : = \partial W/\partial \bE^{(0)}$ and its tangent $\partial\bS^{(0)}/\partial\bE^{(0)}$ (see \citet{Kumar2014_01}). This computational cost can be reduced for isotropic material models \citep{bonet_2008_01,Arghavani2011_01} but this is not possible for anisotropic material models. It is convenient to use $\bE^{(0)}$, since it simplifies the formulation of the strain energy density (see \ref{s:symmetry_group_matraial_invatriant}). But $\bS^{(0)}$ and its corresponding elasticity tensor need to be transformed to the 2.PK\footnote{The Cauchy stress tensor can also be used.} stress tensor and its corresponding elasticity tensor to be used in a classical FE formulation. This approach is used by \citet{Ghaffari2017_01}. The algebraic strain and deformation measures, like $\bC$ and $\bF$, can directly be linearized, discretized and used in a classical FE formulation. So, the linearization and implementation are more efficient if the material model can be formulated based on $\bC$. The 2.PK stress tensor and its corresponding elasticity tensor can be obtained directly as the first and second partial derivative of $W$ with respect to $\bC$. In the next section, the strain energy density $W(\bE^{(0)})$ of \citet{Kumar2014_01} is rewritten based on $\bC$ and thus the performance of the model is increased by a factor of 1.5.\\
\section{Material model}\label{s:metric_mat_model}
\begin{figure}
    \begin{subfigure}{1\linewidth}
        \centering
     \includegraphics[height=55mm]{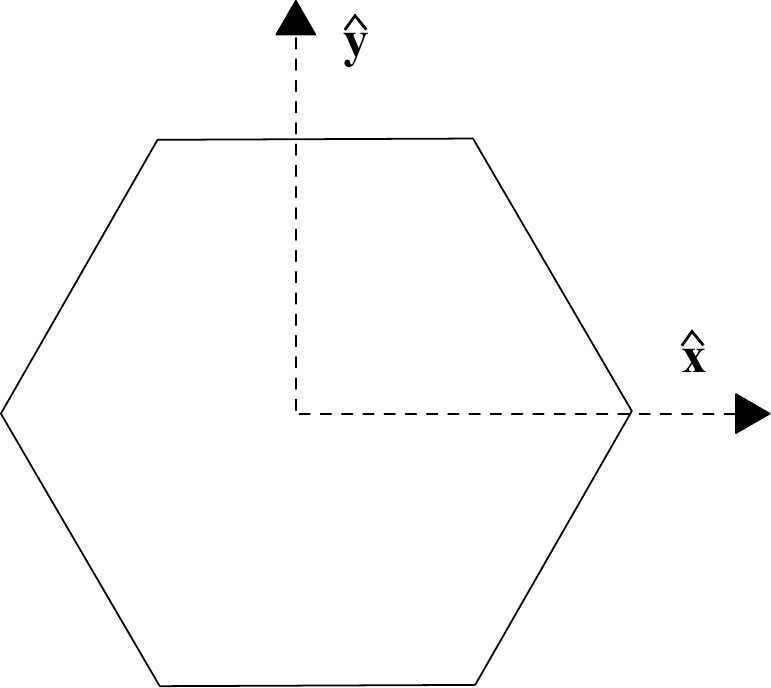}
    \end{subfigure}
    \caption{Anisotropy of the material: Orthonormal vectors characterize the graphene lattice. $\hat{\bx}$ and $\hat{\by}$ are the armchair and zigzag directions \citep{Ghaffari2017_01}.}
    \label{f:lattice}
\end{figure}

In this section, a nonlinear constitutive law for graphene is proposed. Experimentally measured strains up to $12.5\%$ \citep{Hugo2014_01}, $20\%$ \citep{Tomori2011_01} and even $25\%$ \citep{Lee2008_01,Kim2009_01} have been reported for graphene. This is consistent with atomistic simulations \citep{Lu2009_01,Lu2011_01,Cao2014_01} and first principle simulations \citep{Wei2009_01,Xu2012_01,Kumar2014_01,Shodja2017_01}. See also \citet{Galiotis2015_01} and \citet{Akinwande2017} for reviews on the matter. Up to those strains, the deformation is elastic and reversible, and so hyperelastic material models can be used. Those are based on the surface strain energy density $W$. It can be decomposed into the membrane and bending parts $W_{\text{m}}$ and  $W_{\text{b}}$ as \citep{Ghaffari2017_01}
\eqb{lll}
W \is W_{\text{m}}(\bC) + W_{\text{b}}(\bC,\,\buab)~.
\eqe
Models based on a Taylor expansion of the elasticity tensor have many parameters. Many experiments, and multidimensional optimization should be conducted in order to calibrate these parameters \citep{Delfani2013_01,Delfani2015_01,Delfani2016_01}. On the other hand, models based on a set of invariants are more simple \citep{Kumar2014_01}. This is the case for the membrane model of \citet{Kumar2014_01} and the bending model of \citet{Canham1970_01}, which are used here. A possible set of invariants for $\bC$ and the structural tensor $\bbH$ are (see \ref{s:symmetry_group_matraial_invatriant})
\eqb{lll}
 \ds \sJ_{1\bC} \dis \ds \sqrt{\det(\bC)}=J~,\\
 \ds \sJ_{2\bC} \dis \ds \frac{1}{2}\bar{\bC}^{\bot}:\bar{\bC}^{\bot} = \frac{1}{4}\left(\frac{\Lambda_1}{\Lambda_2}+\frac{\Lambda_2}{\Lambda_1}-2\right)~,\\
 \ds \sJ_{3\bC} \dis \ds \frac{1}{8}\bbH(\bar{\bC},\bar{\bC},\bar{\bC})= \frac{1}{8}\left[(\hat{\bM}:\bar{\bC})^3-3(\hat{\bM}:\bar{\bC})(\hat{\bN}:\bar{\bC})^2\right]
 =\frac{1}{8}\left(\frac{\lambda_1}{\lambda_2}-\frac{\lambda_2}{\lambda_1}\right)^3\,\cos(6\theta)~,
 \label{e:graphene_invar_C}
\eqe
where $J=\det{\bF}$, $\bar{\bC}$ is the \textcolor{cgn}{area-invariant} part of $\bC$, and $\bar{\bC}^{\bot}$ is traceless part of $\bar{\bC}$. The latter are defined based on the \textcolor{cgn}{area-invariant} surface deformation gradient $\bar{\bF}$ as
\eqb{lll}
\bar{\bF} \dis \ds J^{-\frac{1}{2}}\bF~,\\[3mm]
\bar{\bC} \dis \ds \bar{\bF}^{\text{T}}\,\bar{\bF} =\frac{1}{J}\bC~,\\[3mm]
\bar{\bC}^{\bot}\dis \ds\frac{1}{J}\left(\bC-\frac{1}{2}\tr(\bC)\,\bI\right)~.
\label{e:F_C_tlC}
\eqe
$\hat{\bM}$ and $\hat{\bN}$ are two traceless \textcolor{cgm}{tensors that} are related to the lattice direction and $\bbH$ (see Eqs.~\textcolor{cgm}{\eqref{e:tensors_M_N}} and \eqref{e:graphene_structural_tensor}). $\lambda_{\alpha}$ and $\Lambda_{\alpha}=\lambda^2_{\alpha}$ ($\alpha=1,2$) are the two eigenvalues of the right surface stretch tensor $\bU$ and $\bC$, respectively. $\theta$ is the maximum stretch angle relative to the armchair direction $\hat{\bx}$ and defined as (see Fig.~\ref{f:lattice})
\eqb{lll}
\theta \is \arccos{(\bY_{\!1}\cdot\hat{\bx})}~,
\eqe
where $\bY_{\!1}$ is the direction of the maximum stretch. Using the spectral decomposition, $\bU$ and $\bC$ can be written as
\eqb{lll}
\bU \is \ds \sum_{\alpha=1}^{2}{\lambda_{\alpha}}\,\bY_{\!\alpha}\otimes\bY_{\!\alpha}~,\\[5mm]
\bC \is \ds \sum_{\alpha=1}^{2}{\Lambda_{\alpha}}\,\bY_{\!\alpha}\otimes\bY_{\!\alpha}~.
\eqe
\noindent
The first and second invariants, $\sJ_{1\bC}$ and $\sJ_{2\bC}$, model material behavior under pure dilatation and shear, the third one, $\sJ_{3\bC}$, models anisotropic behavior. The derivative of these invariants with respect to $\bC$ can be easily determined (see Eq.~\eqref{e:derivative_metric_invariants_graphene}) and used to obtain stress and elasticity tensors without any transformation. The material model can be developed based on additive or multiplicative combinations of the invariants. This can complicate the development of material models and many combinations should be tested to find the best choice in terms of model accuracy and computational efficiency. \citet{Kumar2014_01} show that the logarithmic surface strain $\bE^{(0)}$ and its invariants can model the nonlinear hyperelastic response of graphene very well. The invariants of $\bE^{(0)}$ can be approximated by the invariants of $\bC$. This approximation is sufficient to model the material behavior in the full range of deformation for which the original material model is valid. So, the exact value of the logarithmic strain is not needed anymore. This ensures the accuracy and efficiency of the model. The second and third invariant of $\bE^{(0)}$ (see \ref{s:symmetry_group_matraial_invatriant}) can be approximated as
\eqb{lll}
\sJ_{2\bE^{(0)}} \approx f_1 = e_1\,\sJ_{2\bC}-e_2\,\sJ_{2\bC}^2~,\\[2mm]
\sJ_{3\bE^{(0)}} \approx f_2 = \sJ_{3\bC}(g_1-g_2\,\sJ_{2\bC})~,
\label{e:log_strain_appx}
\eqe
where $e_{\alpha}$ and $g_{\alpha}$ are constants (see Tab.~\ref{t:cons_taylor_log_strain}) that are independent of the material response and computed from the kinematics of the strains. \textcolor{cgn}{The error of this approximation is less than 0.02\%}.
\begin{table}
  \centering
\begin{tabular}{ c c c c }
  \hline
  $e_1$ & $e_2$ & $g_1$ & $g_2$ \\
  \hline
  0.25 & 0.0811 & 0.125 & 0.06057 \\
  \hline
\end{tabular}
\caption{Constants for the Taylor expansion of the invariants of the logarithmic strain.}\label{t:cons_taylor_log_strain}
\end{table}\noindent
$f_1$ is based on the second order Taylor series expansion of $\sJ_{2\bC}$, while $f_2$ is based on the second order expansion of $\sJ_{2\bC}$ and $\sJ_{3\bC}$ omitting the monomials $\sJ_{2\bC}$, $\sJ_{2\bC}^2$ and $\sJ_{3\bC}^2$ for the following reasons: $\sJ_{3\bC}^2$ results in $\cos(12\,\theta)$, which has too high periodicity, while $\sJ_{2\bC}$ and $\sJ_{2\bC}^2$  do not have periodicity of $60^{\circ}$.\\
Using Eq.~\eqref{e:log_strain_appx}, the membrane energy of \citet{Kumar2014_01} can be modified into
\eqb{lll}
W_{\text{m}}(\sJ_{1\bC},\sJ_{2\bC},\sJ_{3\bC}) \is \varepsilon\big[1 - (1+\hat{\alpha}\,\ln(\sJ_{1\bC}))\,\exp(-\hat{\alpha}\,\ln(\sJ_{1\bC}))\big]
+ 2\,\mu\,f_1+\eta\,f_2~,
\label{e:membrane_strain_energy_metric}
\eqe
where $\mu$ and $\eta$ are defined as \citep{Kumar2014_01}
\eqb{lll}
\mu \dis \mu_{0} - \mu_{1}\,(\sJ_{1\bC})^{\hat{\beta}} ~,\\[2mm]
\eta \dis \eta_{0} - \eta_{1}\,(\ln\,\sJ_{1\bC})^2~.
\eqe\noindent
The Canham bending strain energy density is \citep{Canham1970_01}
\eqb{lll}
\ds W_{\mathrm{b}} \dis \ds J\frac{c}{2}\left(\kappa_1^2+\kappa_2^2\right)~,
\label{e:canham_energy_k1_k2}
\eqe
where $\kappa_{\alpha}$ are the principal surface curvatures (see \ref{s:curvilinear_desc}). The membrane and bending material parameters are given in Tabs.~\ref{t:Graphene_cons} and \ref{t:graphene_bending_material_cons}.\noindent
\begin{table}
  \centering
    \begin{tabular}{c c c c c c c c }
      \hline
        & $\hat{\alpha}$ & $\varepsilon~[\textnormal{N/m}]$  &  $\mu_0~[\textnormal{N/m}]$ & $\mu_1~ [\textnormal{N/m}]$ & $\hat{\beta}$ & $\eta_0~[\textnormal{N/m}]$ & $\eta_1~[\textnormal{N/m}]$ \\
      \hline
      GGA  & 1.53 & 93.84  & 172.18 & 27.03 & 5.16 & 94.65 & 4393.26 \\
      LDA  & 1.38 & 116.43 & 164.17 & 17.31 & 6.22 & $86.9{}^{\text{a}}$ & $3611.5{}^{\text{a}}$ \\
      \hline
    \end{tabular}
      \caption{Material constants of graphene \citep{Kumar2014_01}. GGA = generalized gradient approximation; LDA = local density approximation. GGA and LDA are two approximations in the density functional theory. \textcolor{cgn}{These material parameters can be obtained by fitting $W_{\text{m}}$ to the quantum data of pure dilatation and uniaxial tension.} $^{\text{a}}$~See correction of \citet{Kumar2016_01}.}\label{t:Graphene_cons}
\end{table}\noindent
\begin{table}
  \centering
     \begin{tabular}{c c c c}
     \hline
     & FGBP & SGBP & QM \\
     \hline
    $c$~[nN$\cdot$nm] & 0.133 & 0.225 & 0.238 \\
    \hline
  \end{tabular}
  \caption{Bending stiffness according to various atomistic models \citep{Lu2009_01,Kudin2001_01}. FGBP = first generation Brenner potential; SGBP = second generation Brenner potential; QM = quantum mechanics. The QM parameter is used in all simulations. \textcolor{cgn}{It is obtained by fitting $W_{\text{b}}$ to the quantum data of bending.}}
  \label{t:graphene_bending_material_cons}
\end{table}
The second Piola-Kirchhoff stress tensor (related to the membrane strain energy density) follows from Eq.~\eqref{e:membrane_strain_energy_metric} as
\eqb{lll}
\bS_{\text{m}} \dis \ds 2\pa{W_{\text{m}}}{\bC} = 2\pa{W_{\text{m}}}{\sJ_{1\bC}}\pa{\sJ_{1\bC}}{\bC}+2\pa{W_{\text{m}}}{\sJ_{2\bC}}\pa{\sJ_{2\bC}}{\bC} +2\pa{W_{\text{m}}}{\sJ_{3\bC}}\pa{\sJ_{3\bC}}{\bC}~,
\eqe
which becomes
\eqb{lll}
\bS_{\text{m}} \is \ds H_1\bC^{-1}+ \frac{H_2}{J}\bar{\bC}^{\bot}
+ \frac{H_3}{4J}(a_{\hat{\bM}}\hat{\bM}+ a_{\hat{\bN}}\hat{\bN})~,
\label{e:mem_2PK_tensor_metric}
\eqe
where $\partial\sJ_{i\bC}\,/\partial\bC$, $H_i$, $a_{\hat{\bM}}$ and $a_{\hat{\bN}}$ are given in \ref{s:some_coef}.
The elasticity tensor (related to the membrane strain energy density) is defined as
\eqb{lll}
\ds \mathbb{C}_{\text{m}} \dis \ds \frac{\partial^2{W_{\text{m}}}}{\partial{\bC}\otimes\partial{\bC}}= \frac{\partial^2{W_{\text{m}}}}{\partial{C_{\alpha\beta}}\,\partial{C_{\gamma\delta}}} \,\bA_{\alpha}\otimes\bA_{\beta}\otimes\bA_{\gamma}\otimes\bA_{\delta}~,
\label{e:metric_2PK_tensorial}
\eqe
which, for the proposed membrane strain energy density, is (see \ref{s:some_coef})
\eqb{lll}
\ds \mathbb{C}_{\text{m}} \is 2\biggl\{ \ds \left(\frac{J}{2}\pa{H_1}{\sJ_{1\bC}}-\sJ_{2\bC}\pa{H_1}{\sJ_{2\bC}} -\frac{3}{2}\sJ_{3\bC}\,\pa{H_1}{\sJ_{3\bC}}\right)\,\bC^{-1} \otimes \bC^{-1}+\frac{1}{J^2}\pa{H_2}{\sJ_2}\,\bar{\bC}^{\bot}\otimes\,\bar{\bC}^{\bot}\\[4mm]
\plus \ds \frac{2}{J}\pa{H_1}{\sJ_2}\,\left[\bC^{-1} \otimes \bar{\bC}^{\bot}\right]^{\text{S}}
+  \frac{1}{4J}\,\pa{H_1}{\sJ_3}\,\left[\bC^{-1} \otimes \bZ \right]^{\text{S}}
+ \ds \frac{1}{2J^2}\pa{H_3}{\sJ_2}\left[\bZ\otimes \bar{\bC}^{\bot} \right]^{\text{S}}\\[4mm]
\mi \ds \frac{1}{2}H_1\,(\bC^{-1} \boxtimes \bC^{-1} + \bC^{-1} \oplus \bC^{-1})
+ \ds \frac{H_2}{2J^2}\,(\bI \boxtimes \bI + \bI \oplus \bI-\bI \otimes \bI)\\[4mm]
\plus \ds \frac{3H_3}{2J^2}\,\left[(\hat{\bM}:\bar{\bC})\,(\hat{\bM} \otimes \hat{\bM}-\hat{\bN}\otimes\hat{\bN})-(\hat{\bN}:\bar{\bC})(\hat{\bM}\otimes\,\hat{\bN}+\hat{\bN}\otimes\,\hat{\bM})\right] \biggr\}~,
\label{e:mem_2PK_elastiicty_tensor_metric}
\eqe
with
\eqb{lll}
\bZ \dis \ds a_{\hat{\bM}}\,\hat{\bM}+a_{\hat{\bN}}\,\hat{\bN}~,
\eqe
\eqb{lll}
\ds (\bA\otimes\bB)^{\text{S}} \dis \ds \frac{1}{2}(\bA\otimes\bB+\bB\otimes\bA)~.
\eqe
The multiplication operators\footnote{\citet{Kintzel_2006_01} and \citet{Kintzel_2006_02} use $\times$ instead of $\oplus$.} $\otimes$~, $\oplus$ and $\boxtimes$ are defined for two second order tensors of $\bA$ and $\bB$ as
\eqb{lll}
\bA\otimes\bB \is A^{\alpha\beta}\,B^{\gamma\delta}\,\bA_{\alpha}\otimes\bA_{\beta}\otimes\bA_{\gamma}\otimes\bA_{\delta}~,\\
\bA\oplus\bB \is A^{\alpha\beta}\,B^{\gamma\delta}\,\bA_{\alpha}\otimes\bA_{\gamma}\otimes\bA_{\delta}\otimes\bA_{\beta} =A^{\alpha\delta}\,B^{\beta\gamma}\,\bA_{\alpha}\otimes\bA_{\beta}\otimes\bA_{\gamma}\otimes\bA_{\delta}~,\\
\bA\boxtimes\bB \is A^{\alpha\beta}\,B^{\gamma\delta}\,\bA_{\alpha}\otimes\bA_{\gamma}\otimes\bA_{\beta}\otimes\bA_{\delta} =A^{\alpha\gamma}\,B^{\beta\delta}\,\bA_{\alpha}\otimes\bA_{\beta}\otimes\bA_{\gamma}\otimes\bA_{\delta}~.
\eqe
The tensorial form of $\bS$ and $\mathbb{C}_{\text{m}}$ in Eqs.~\eqref{e:mem_2PK_tensor_metric} and \eqref{e:mem_2PK_elastiicty_tensor_metric} can also be used in non-curvilinear, e.g. Cartesian, shell formulations. The constitutive law needs to be written in curvilinear coordinates to be used in the shell formulation of  \citet{Duong2016_01}. $\bC$, $\bC^{-1}$, $\hat{\bM}$, $\hat{\bN}$ and $\bI$ can be written in the curvilinear coordinate basis as
\eqb{lll}
\bC \is \ds A^{\alpha\gamma}\,\augd\,A^{\delta\beta}\,\bA_{\alpha}\otimes\bA_{\beta}~,\\
\bC^{-1} \is \ds \aab\,\bA_{\alpha}\otimes\bA_{\beta}~,\\
\bI \is \Aab\,\bA_{\alpha}\otimes\bA_{\beta}~,\\
\hat{\bM} \is \hat{M}_{\alpha\beta}\,\bA^{\alpha}\otimes\bA^{\beta}=\hat{M}^{\alpha\beta}\,\bA_{\alpha}\otimes\bA_{\beta}~,\\
\hat{\bN} \is \hat{N}_{\alpha\beta}\,\bA^{\alpha}\otimes\bA^{\beta}=\hat{N}^{\alpha\beta}\,\bA_{\alpha}\otimes\bA_{\beta}~,
\label{e:curvlinear_connect_C_Cinv_M_N}
\eqe
where $\Auab$ ($\auab$) and $\Aab$ ($\aab$) are the covariant and contra-variant components of the metric tensors in the reference configuration (current configuration), see \ref{s:curvilinear_desc}. Here, $\hat{M}_{\alpha\beta}$~, $\hat{M}^{\alpha\beta}$, $\hat{N}_{\alpha\beta}$ and $\hat{N}^{\alpha\beta}$ are given by
\eqb{lll}
\hat{M}_{\alpha\beta} \is \bA_{\alpha}\cdot\hat{\bM}\cdot\bA_{\beta}~;~\hat{M}^{\alpha\beta} = \bA^{\alpha}\cdot\hat{\bM}\cdot\bA^{\beta},\\
\hat{N}_{\alpha\beta} \is \bA_{\alpha}\cdot\hat{\bN}\cdot\bA_{\beta}~;~\hat{N}^{\alpha\beta} = \bA^{\alpha}\cdot\hat{\bN}\cdot\bA^{\beta}~.
\eqe
In addition, $\mathbb{C}_{\text{m}}$ can be written as
\eqb{lll}
\mathbb{C}_{\text{m}} \dis \ds C^{\alpha\beta\gamma\delta}_{\text{m}}\,\bA_{\alpha}\otimes\bA_{\beta}\otimes\bA_{\gamma}\otimes\bA_{\delta}~,
\eqe
where $C^{\alpha\gamma\delta\beta}_{\text{m}}$ is given by
\eqb{lll}
C^{\alpha\beta\gamma\delta}_{\text{m}} \is \bA^{\alpha}\otimes\bA^{\beta}:\mathbb{C}_{\text{m}}: \bA^{\gamma}\otimes\bA^{\delta}~.
\eqe
$S_{\text{m}}^{\alpha\beta}$ and $C^{\alpha\beta\gamma\delta}_{\text{m}}$ can be obtained analytically by substitution of Eq.~\eqref{e:curvlinear_connect_C_Cinv_M_N} into Eqs.~\eqref{e:mem_2PK_tensor_metric} and \eqref{e:mem_2PK_elastiicty_tensor_metric}, and factorization of $\bA_{\alpha}\otimes\bA_{\beta}$ and $\bA_{\alpha}\otimes\bA_{\beta}\otimes\bA_{\gamma}\otimes\bA_{\delta}$. The 2.PK stress components $S^{\alpha\beta}$ are equal to the Kirchhoff stress components $\tau^{\alpha\beta}$, as was shown in Sec.~\ref{s:finite_element}, Eq.~\eqref{e:tau_s_connect}. Likewise $c^{\alpha\beta\gamma\delta}_{\text{m}}=\textcolor{cgm}{\vaa\otimes\vab : \mathbb{c}_{\text{m}}: \vag\otimes\vad} = C^{\alpha\beta\gamma\delta}_{\text{m}}$, where $\mathbb{c}$ is the material tangent corresponding to $\btau$ and is used in the FE formulation of \citet{Duong2016_01}. For $\tau^{\alpha\beta}_{\text{m}}$ thus follows
\eqb{lll}
\tauab_{\text{m}} \is \ds \ds H_1\aab+ \frac{H_2}{J^2}\left(A^{\alpha\gamma}\,\augd\,A^{\delta\beta}-\frac{1}{2}\tr(\bC)\Aab\right)
+ \frac{H_3}{4J}\left(a_{\hat{\bM}}\hat{M}^{\alpha\beta}+ a_{\hat{\bN}}\hat{N}^{\alpha\beta}\right)~,
\eqe
where $\tr(\bC)$ can be written in curvilinear coordinates as
\eqb{lll}
\tr(\bC) \is \bC:\bI=\auab\,\bA^{\alpha}\otimes\bA^{\beta}:\Agd\,\bA_{\gamma}\otimes\bA_{\delta}=\auab\,\Aab~.
\eqe
The proposed relation for $\tauab_{\text{m}}$ is simpler and has lower computational cost then the one by \citet{Ghaffari2017_01}.
For the bending energy given in Eq.~\eqref{e:canham_energy_k1_k2}, the stress and moment components are derived in \citep{Ghaffari2017_01,Sauer2017_01}, i.e.
\eqb{lll}
\tauab_{\text{b}} \dis \ds 2\pa{W_{\text{b}}}{\auab}=J\,\left[c\,(2H^2+\kappa)\aab-4c\,H\,\bab\right]~,
\eqe
\eqb{lll}
\Mab_{0} \dis \ds \pa{W_{\text{b}}}{\buab}=c\,J\,\bab~.
\eqe
Here, $\bab$ is the contra-variant components of the curvature tensor (see \ref{s:curvilinear_desc}).
In addition, the elasticity tensors for bending are given as \citep{Sauer2017_01}
\eqb{lll}
\cabgd_{\text{b}} \dis \ds 4\paqq{W_{\text{b}}}{\auab}{\augd}=c_{aa}\,\aab\,\agd+c_{a}\,\aabgd+c_{bb}\,\bab\,\bgd+c_{ab}\left(\aab\,\bgd+\bab\,\agd\right)~,\\[3mm]
\dabgd \dis \ds 2\paqq{W_{\text{b}}}{\auab}{\bugd}=d_{aa}\,\aab\,\agd+d_{a}\,\aabgd+d_{ab}\,\aab\,\bgd+d_{ba}\bab\,\agd~,\\[3mm]
\eabgd \dis \ds 2\paqq{W_{\text{b}}}{\buab}{\augd}=d^{\gamma\delta\alpha\beta}~,\\[3mm]
\fabgd \dis \ds \paqq{W_{\text{b}}}{\buab}{\bugd}=f_{a}\,\aabgd~,
\eqe
with
\eqb{lll}
c_{aa}\is -J\left(14H^{2}+c\kappa\right)~,\\
c_{a}\is 2J\left(-6c\,H^{2}+c\,\kappa\right)~,\\
c_{bb}\is 4c\,J~,\\
c_{ab}\is c_{ba}=4c\,J\,H~,\\
d_{aa}\is 4J\,c\,H~,\\
d_{a}\is 4J\,c\,H~,\\
d_{ab}\is -J\,c~,\\
d_{ba}\is -2J\,c~,\\
f_{a}\is -J\,c~.
\eqe
\section{Elementary model behavior}\label{s:elementary_behav}
\begin{figure}
    \begin{subfigure}{0.49\textwidth}
        \centering
     \includegraphics[height=58mm]{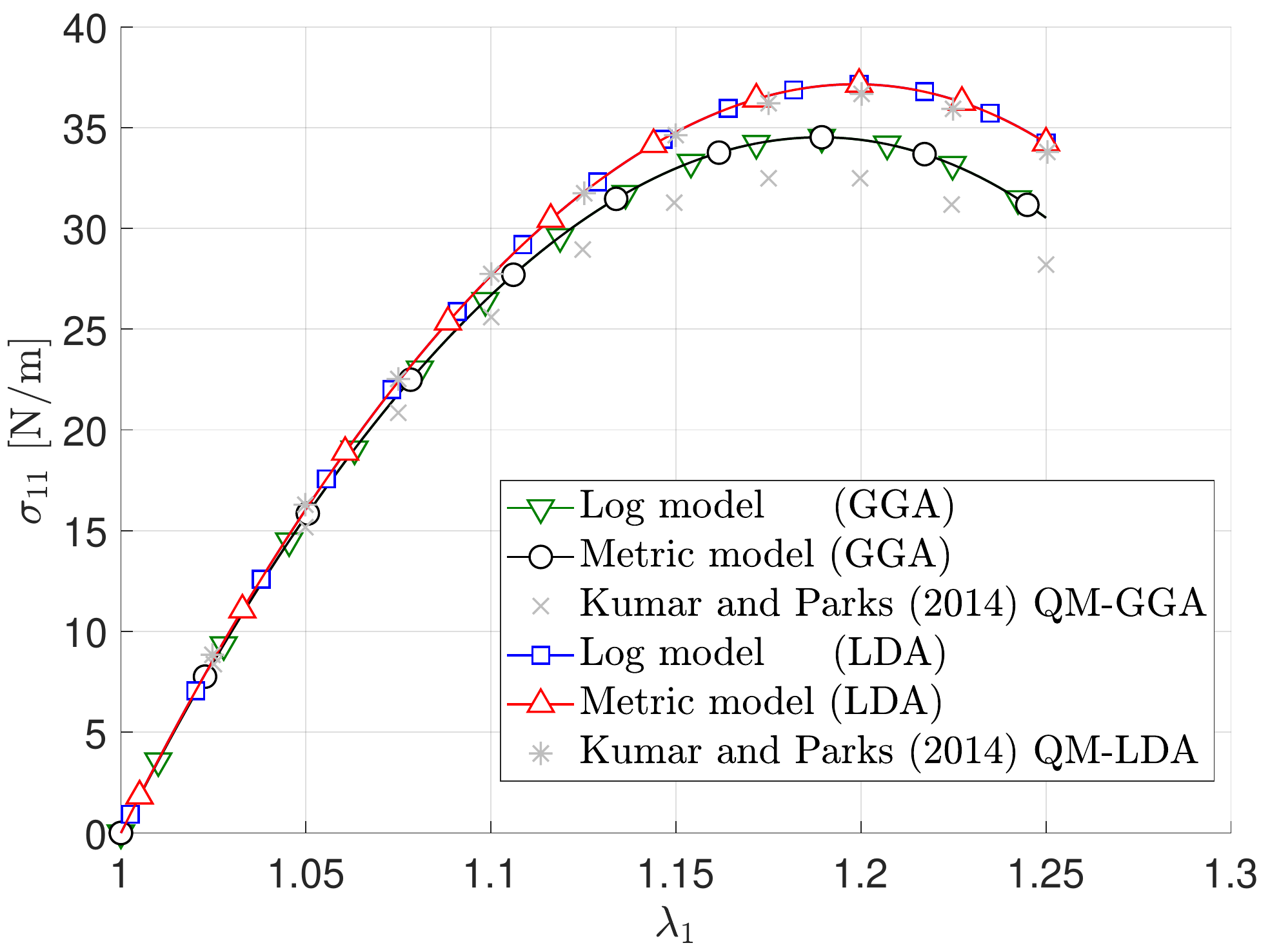}
        \vspace{-7mm}
        \subcaption[l]{}
        \label{f:sigma11_uniaxial_arm_chair_Metric_Model}
    \end{subfigure}
    \begin{subfigure}{0.49\textwidth}
        \centering
    \includegraphics[height=58mm]{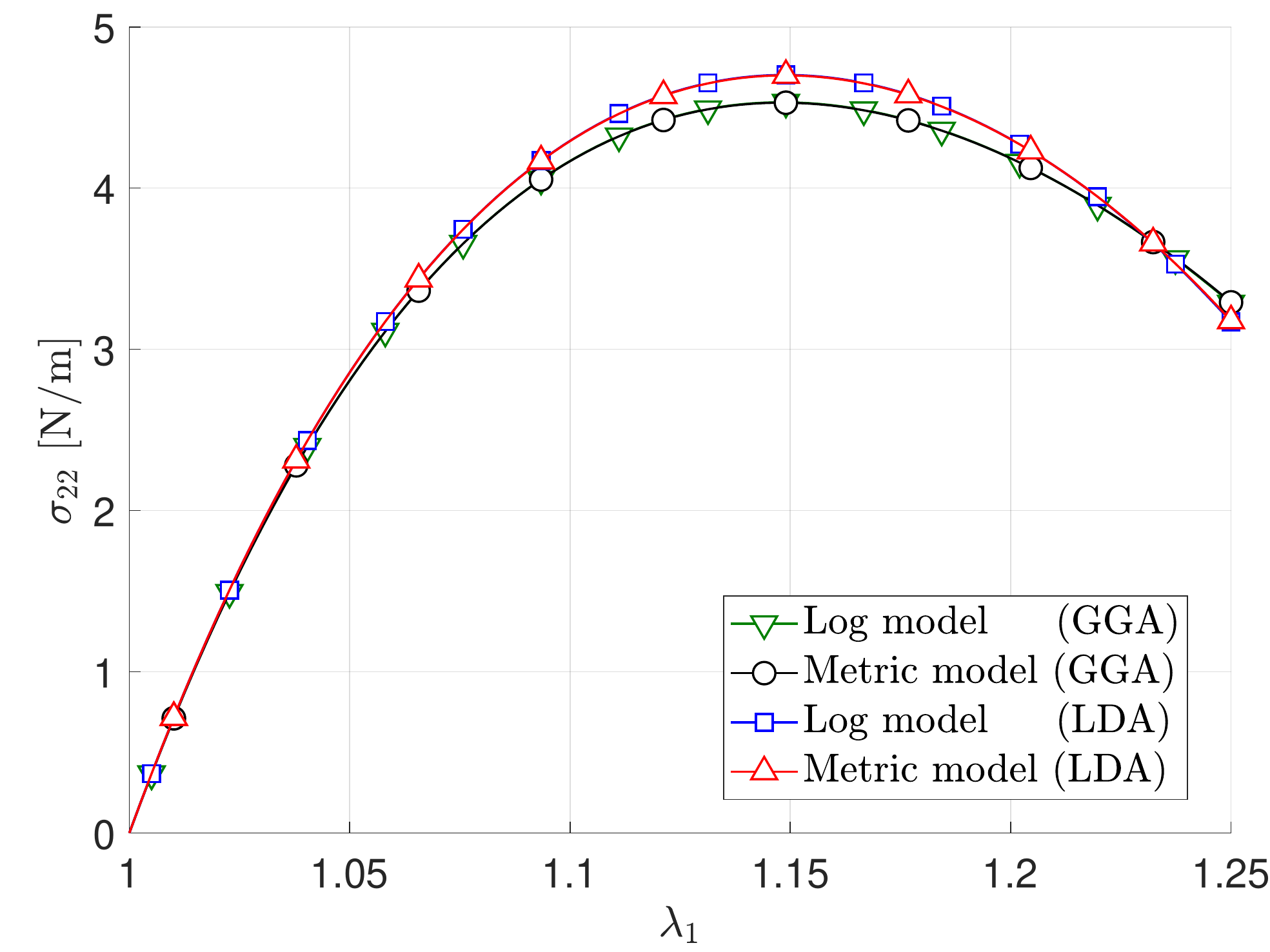}
        \vspace{-7mm}
        \subcaption{}
        \label{f:sigma22_uniaxial_arm_chair_Metric_Model}
    \end{subfigure}
    \vspace{-3.5mm}
    \caption{Uniaxial stretch in the armchair direction: Comparison of the log model \citep{Ghaffari2017_01} and the metric model (proposed here): (\subref{f:sigma11_uniaxial_arm_chair_Metric_Model}) Stress in the stretched direction;
(\subref{f:sigma22_uniaxial_arm_chair_Metric_Model}) stress in the perpendicular direction. The maximum relative error for $\sigma_{11}$ and $\sigma_{22}$ are 0.019 and 0.199 percent. \textcolor{cgn}{The quantum mechanical results are taken from \citet{Kumar2014_01}.}}
\label{f:sigma_uniaxial_arm_chair_Metric_Model}
\end{figure}

\begin{figure}
    \begin{subfigure}{0.49\textwidth}
        \centering
 \includegraphics[height=58mm]{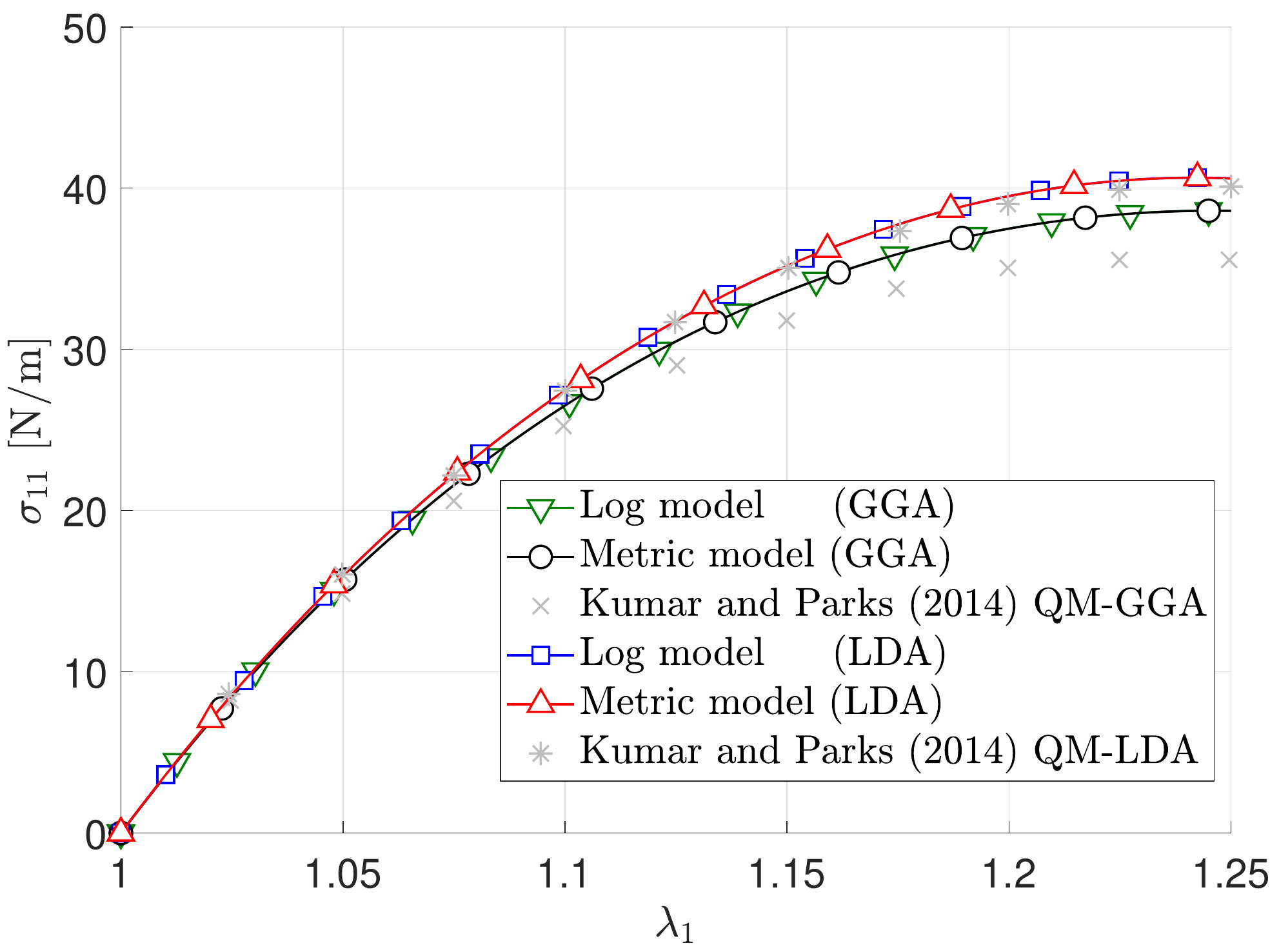}
        \vspace{-7mm}
        \subcaption{}
        \label{f:sigma11_uniaxial_zig_zag_Metric_Model}
    \end{subfigure}
    \begin{subfigure}{0.49\textwidth}
        \centering
    \includegraphics[height=58mm]{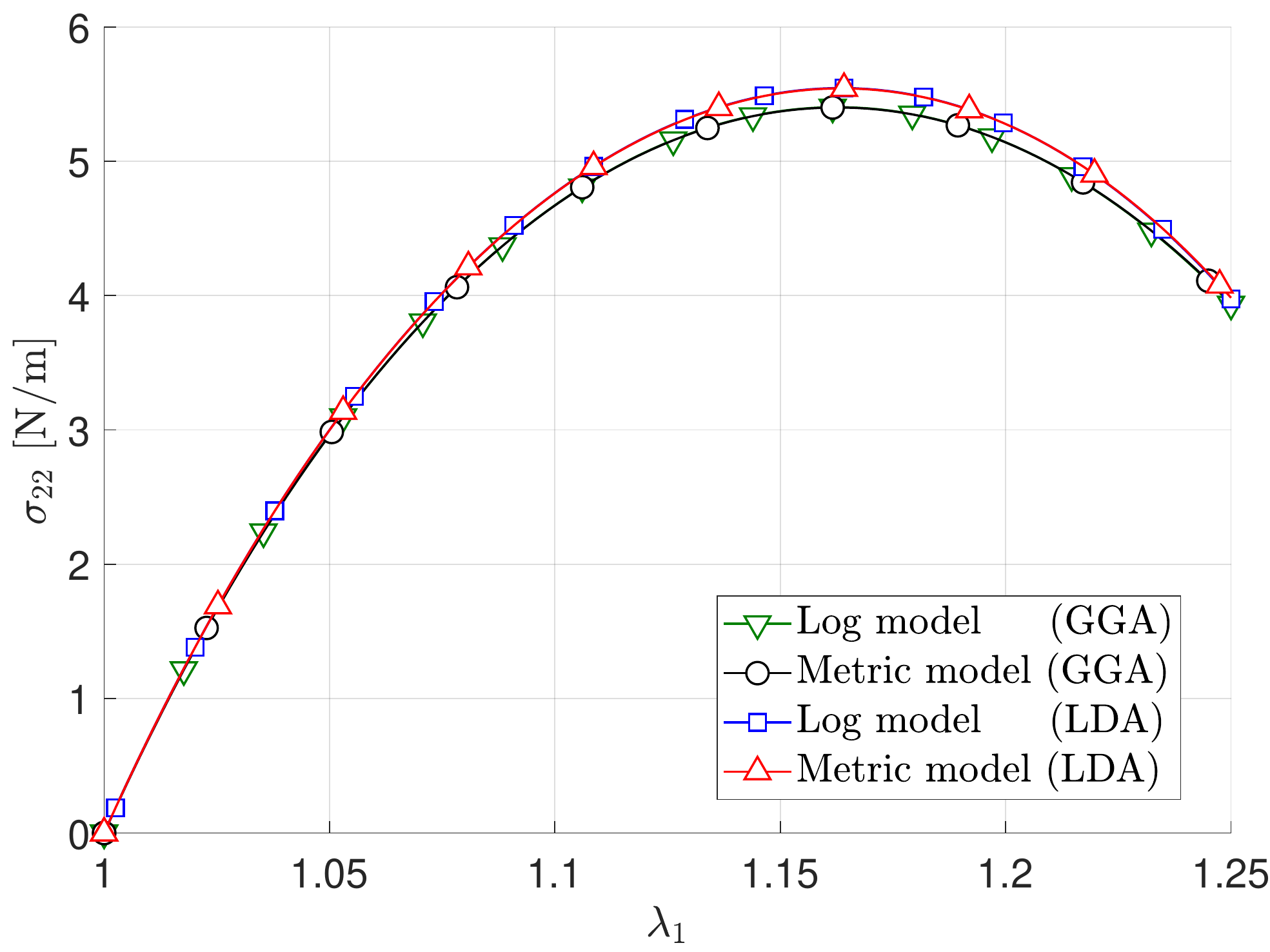}
        \vspace{-7mm}
        \subcaption{}
        \label{f:sigma22 uniaxial zig zag Metric Model}
    \end{subfigure}
    \vspace{-3.5mm}
    \caption{Uniaxial stretch in the zigzag direction: Comparison of the log model \citep{Ghaffari2017_01} and the metric model (proposed here): (\subref{f:sigma11_uniaxial_zig_zag_Metric_Model}) Stress in the stretched direction; (\subref{f:sigma22 uniaxial zig zag Metric Model}) stress in the perpendicular direction. The maximum relative error for $\sigma_{11}$ and $\sigma_{22}$ are 0.019 and 0.194 percent. \textcolor{cgn}{The quantum mechanical results are taken from \citet{Kumar2014_01}.} }
\label{f:sigma uniaxial zig zag Metric Model}
\end{figure}

\begin{figure}
    \begin{subfigure}{0.49\textwidth}
        \centering
     \includegraphics[height=58mm]{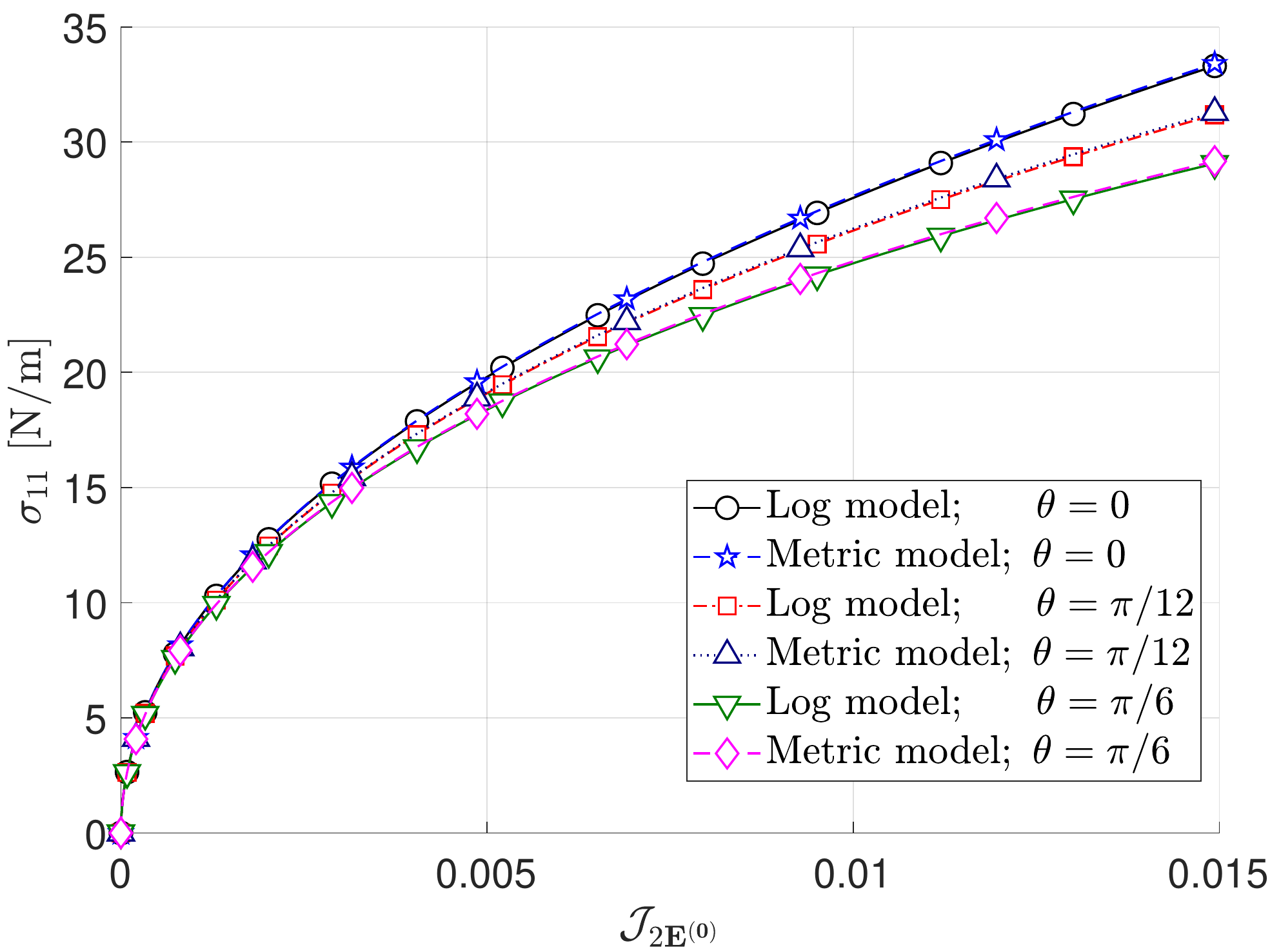}
        \vspace{-6mm}
        \subcaption{}
        \label{f:Analytical_solution_pure_shear_GGA_sig11_Metric_Model}
    \end{subfigure}
    \begin{subfigure}{0.49\textwidth}
        \centering
 \includegraphics[height=58mm]{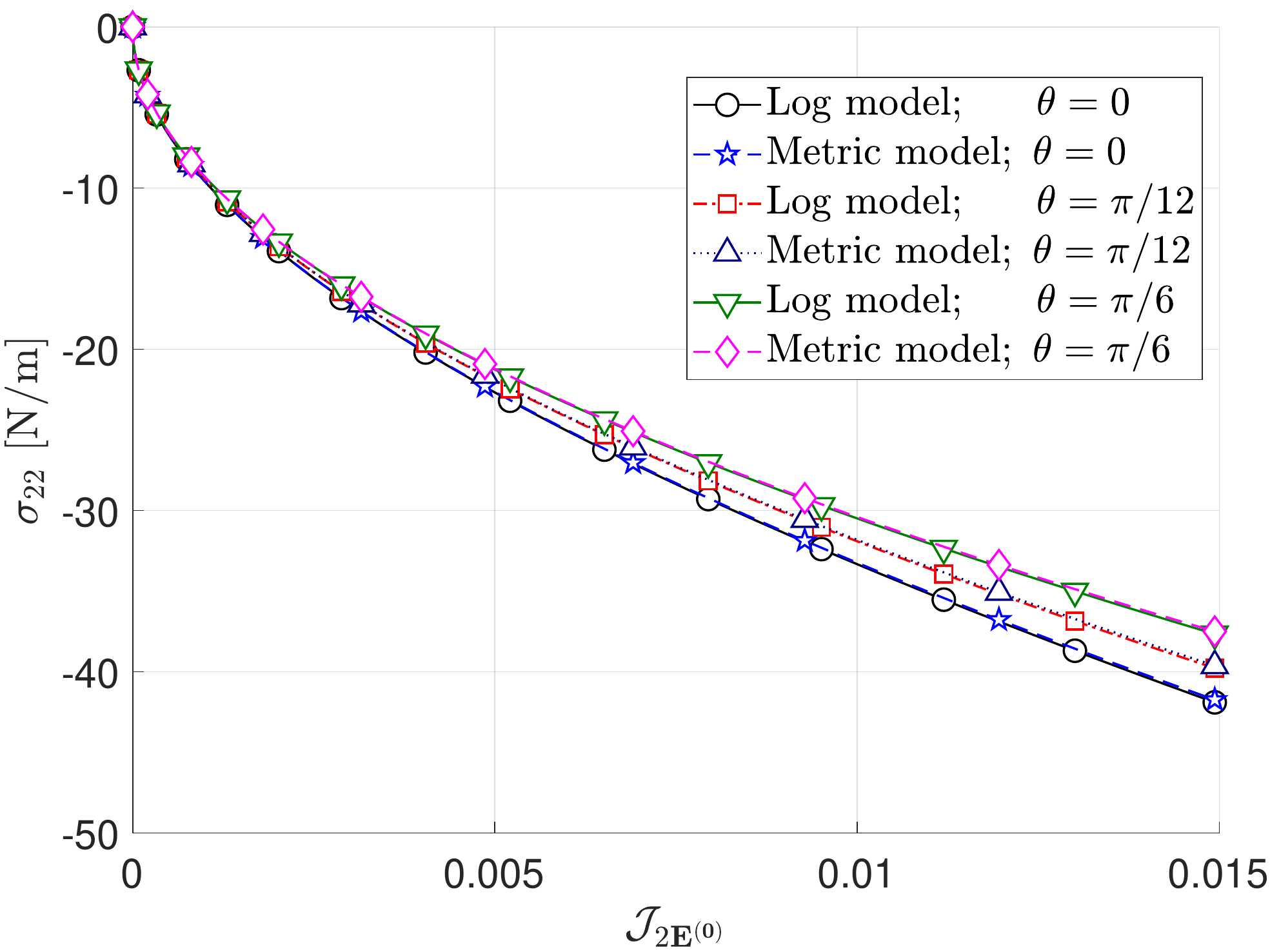}
        \vspace{-6mm}
        \subcaption{}
        \label{f:Analytical_solution_pure_shear_GGA_sig22_Metric_Model}
    \end{subfigure}
    \vspace{-3.5mm}
    \caption{Pure shear based on the GGA material parameters (see Tab.~\ref{t:Graphene_cons}): Comparison of the log model \citep{Ghaffari2017_01} and the metric model (proposed here): (\subref{f:Analytical_solution_pure_shear_GGA_sig11_Metric_Model}) Stress in the pull direction; (\subref{f:Analytical_solution_pure_shear_GGA_sig22_Metric_Model}) stress in the compression direction. The maximum relative error for $\sigma_{11}$ and $\sigma_{22}$ are 0.35 and 0.42 percent. $\theta$ denotes the direction of pulling relative to the armchair direction.}
    \label{f:Analytical_solution_pure_shear_GGA_sig_Metric_Model}
\end{figure}

\begin{figure}
    \begin{subfigure}{0.49\textwidth}
        \centering
    \includegraphics[height=58mm]{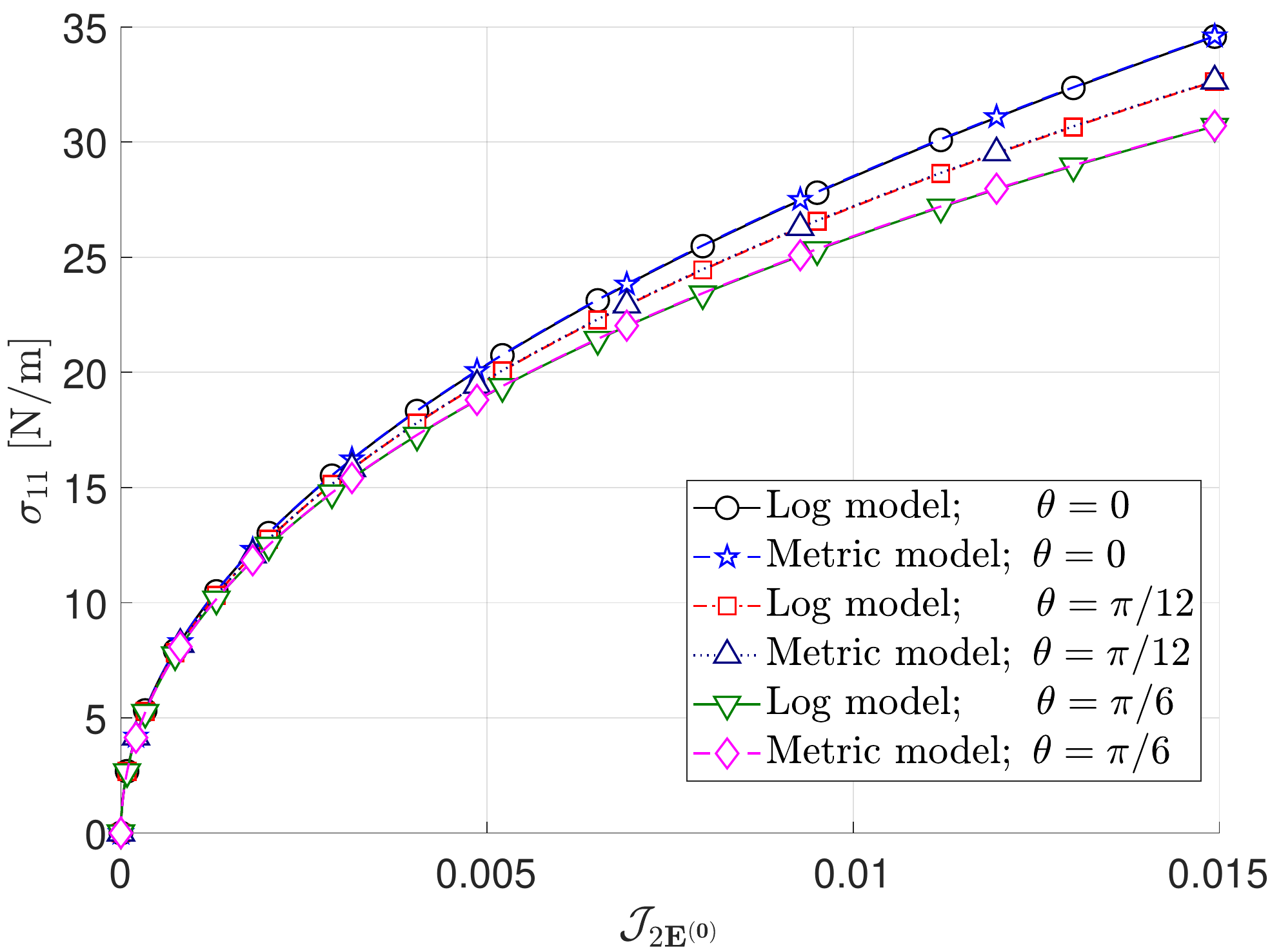}
    \vspace{-6mm}
        \subcaption{}
        \label{f:Analytical_solution_pure_shear_LDA_sig11_Metric_Model}
    \end{subfigure}
    \begin{subfigure}{0.49\textwidth}
        \centering
    \includegraphics[height=58mm]{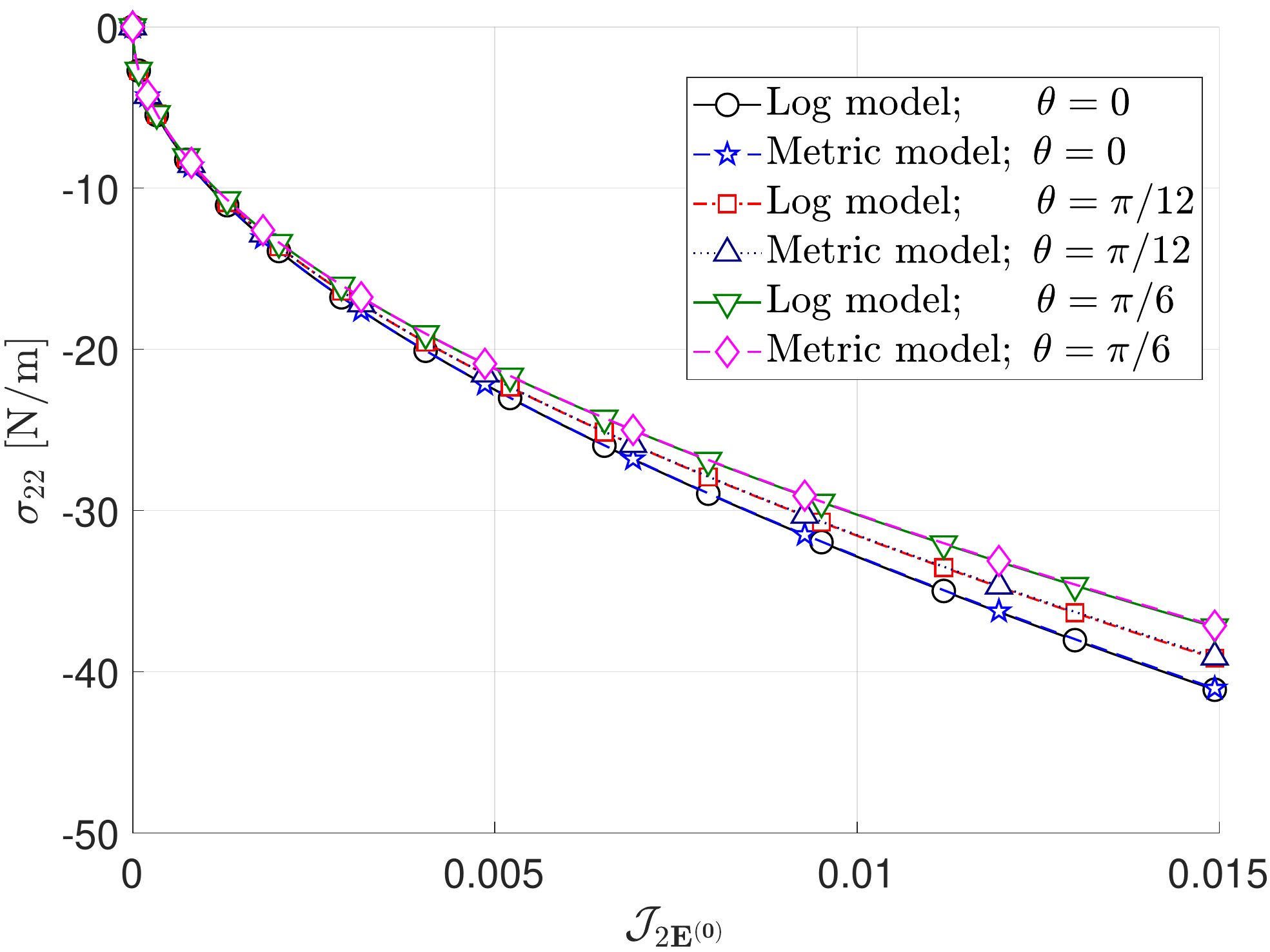}
    \vspace{-6mm}
        \subcaption{}
        \label{f:Analytical_solution_pure_shear_LDA_sig22_Metric_Model}
    \end{subfigure}
    \vspace{-3.5mm}
    \caption{Pure shear based on the LDA material parameters (see Tab.~\ref{t:Graphene_cons}): Comparison of the log model \citep{Ghaffari2017_01} and the metric model (proposed here): (\subref{f:Analytical_solution_pure_shear_LDA_sig11_Metric_Model}) Stress in the pull direction; (\subref{f:Analytical_solution_pure_shear_LDA_sig22_Metric_Model}) stress in the compression direction. The maximum relative error for $\sigma_{11}$ and $\sigma_{22}$ are 0.12 and 0.22 percent. $\theta$ denotes the direction of pulling relative to the armchair direction.}
    \label{f:Analytical_solution_pure_shear_LDA_sig_Metric_Model}
\end{figure}

In this section, the new material model and its FE implementation are verified by testing the behavior of a graphene sheet under uniaxial stretch and pure shear. The performance of the proposed metric model is investigated and compared with the logarithmic model (log model) of \citet{Ghaffari2017_01}.\\
For uniaxial stretch, the sheet is stretched in the armchair and zigzag direction and fixed in the perpendicular direction. The Cartesian components of the stresses in the pulled direction, $\sigma_{11}$, and perpendicular direction, $\sigma_{22}$, are presented in Figs.~\ref{f:sigma11_uniaxial_arm_chair_Metric_Model} and \ref{f:sigma22_uniaxial_arm_chair_Metric_Model} for pulling along the armchair direction, and  \ref{f:sigma11_uniaxial_zig_zag_Metric_Model} and \ref{f:sigma22 uniaxial zig zag Metric Model} for pulling in the zigzag direction. The new results are compared with \citet{Ghaffari2017_01} for both parameter sets following from GGA and LDA, which are two approximations of density functional theory. The stresses are nonlinear and have a distinct maximum. The maximum stress is larger if it is stretched along the zigzag direction. Also the stiffness is higher in this direction.
  \noindent
For pure shear, the sheet is pulled in one direction and compressed in the perpendicular direction. The Cartesian components of the stress in the pulled direction, $\sigma_{11}$, and compressed direction, $\sigma_{22}$, are presented in Figs.~\ref{f:Analytical_solution_pure_shear_GGA_sig11_Metric_Model} and \ref{f:Analytical_solution_pure_shear_GGA_sig22_Metric_Model} for the GGA parameter set, and in Figs.~\ref{f:Analytical_solution_pure_shear_LDA_sig11_Metric_Model} and \ref{f:Analytical_solution_pure_shear_LDA_sig22_Metric_Model} for the LDA parameter set. The stress in the pulled and compressed directions are monolithically increasing or decreasing. The boundary conditions and loads are discussed in more detail in \citet{Ghaffari2017_01}.\\
The results of the metric model for all tests in Figs.~\ref{f:sigma_uniaxial_arm_chair_Metric_Model}-\ref{f:Analytical_solution_pure_shear_LDA_sig_Metric_Model} are in excellent agreement with the log model of \citet{Ghaffari2017_01}. The metric model has the same pure dilatation and bending strain energy density terms as the log model of \citet{Ghaffari2017_01}, so there is no need to verify pure dilatation and bending. \textcolor{cgn}{Note that in general, graphene sheets can wrinkle under uniaxial stretching and shearing, which is avoided here by constraining the out-of-plane deformation.}
\section{Numerical examples}\label{s:metric_num_result}
In this section, the performance of the model is investigated by several examples. The examples are contact of a CNT and CNC with a Lennard-Jones wall and bending and twisting of CNTs and CNCs. Before applying these deformations, the CNT and CNC need to be relaxed, since they can contain residual stresses coming from the rolling of graphene. The strain energy and internal stresses of CNTs and CNCs are minimized initially, before applying the loading. In all examples, the buckling and post-buckling behavior of the structures is computed and the point of buckling can be either determined by examining the ratio of the membrane energy to the total energy or it can be determined from sharp variations in the reaction forces. The modified arc-length method of \citet{Ghaffari2015} and a line-search \citep{Ferran1996,bonet_2008_01,Souza2008,wriggers2008} are used to obtain convergence around the buckling point and capture the jump in the energy and force. The simulations will not converge without these methods even when using very small load steps. In the following examples the error is defined by
\eqb{lll}
\text{error} \is \ds \frac{\|q-q_{\text{ref}}\|}{\|q_{\text{ref}}\|}~
\label{e:error_dif}
\eqe
where $q$ can be a force or an energy, and $q_{\text{ref}}$ is a corresponding reference value. For comparison of log and metric models, the log model is considered as $q_{\text{ref}}$, and for other examples the metric model with the finest mesh is used as reference. \textcolor{cgn}{The current formulation does not use any special treatment against locking.
Instead, sufficiently fine quadratic NURBS meshes are used.}
\subsection{Carbon nanotubes}
CNT($n,m$) can be generated by rolling a graphene sheet perpendicular to the lattice vector of ($n,m$), where $n$ and $m$ are the chirality parameters \citep{Lee2012}. In this section, bending, twisting of CNTs and contact a CNT with a Lennard-Lones wall is considered, and their buckling and postbuckling behavior are simulated. The buckling point can be determined accurately by examining the ratio of the membrane energy to the total energy. At the bucking point membrane energy is converted to bending energy. These points have been obtained for bending and twisting of CNTs in \citet{Ghaffari2017_01}.
\subsubsection{CNT bending}
\begin{figure}[h]
    \begin{subfigure}{0.49\textwidth}
        \centering
     \includegraphics[height=55mm,trim=1.5cm 1cm 0cm 1cm,clip]{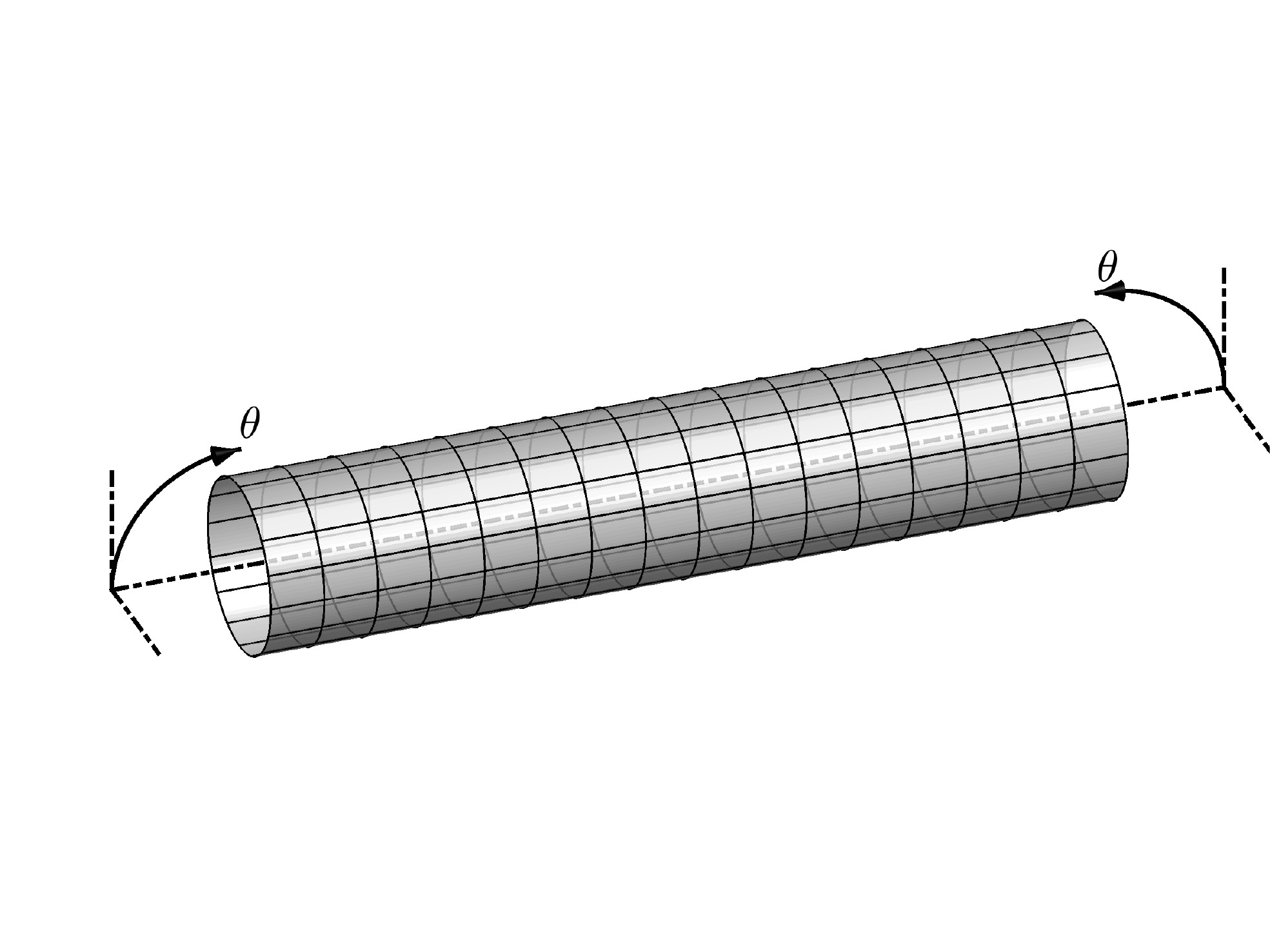}
        \vspace{-5mm}
        \subcaption{}
        \label{f:CNT_Loading_bending}
    \end{subfigure}\hspace{4mm}
     \begin{subfigure}{0.49\textwidth}
        \centering
     \includegraphics[height=55mm]{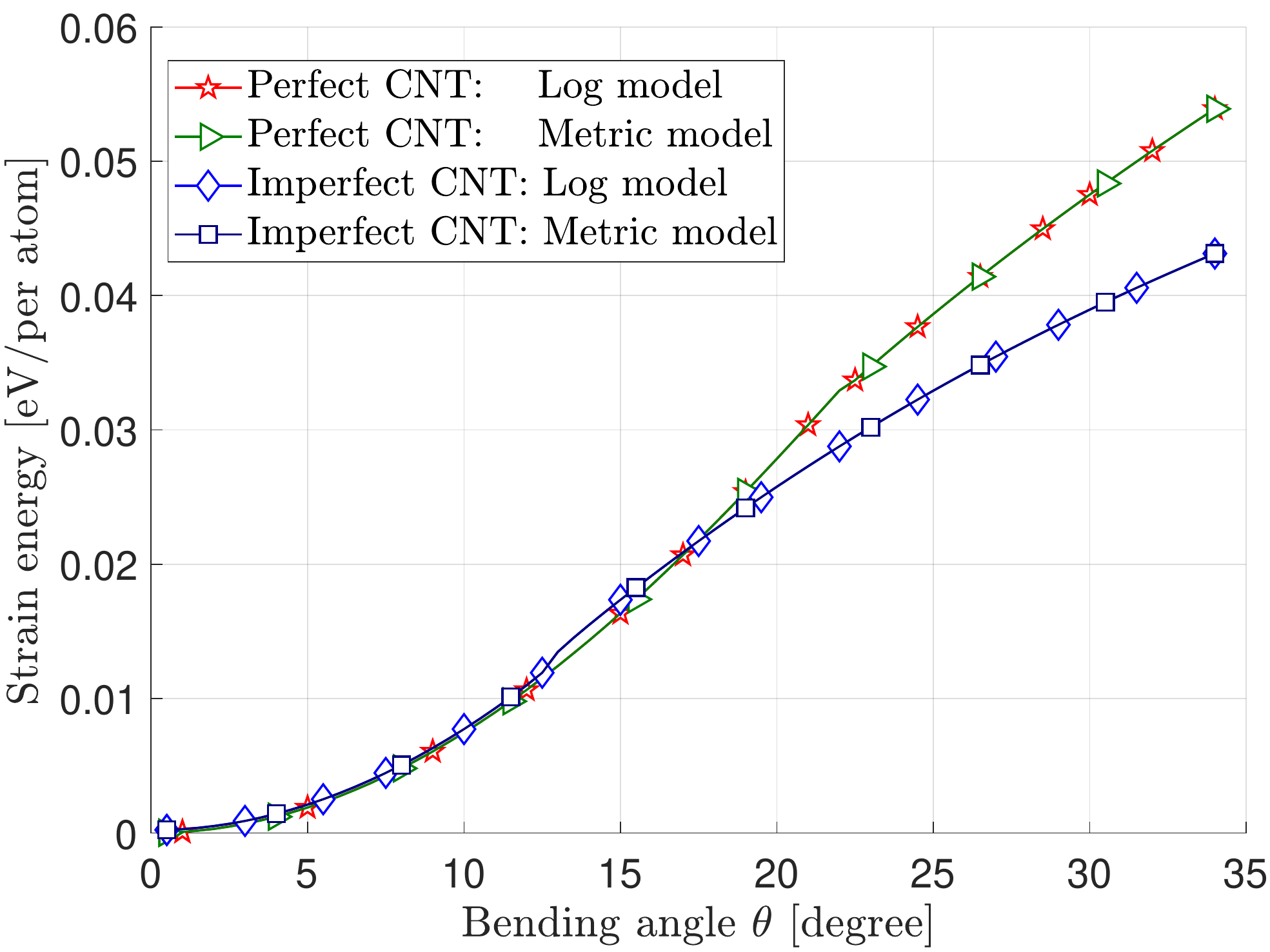}
        \vspace{-1mm}
        \subcaption{}
        \label{f:bending_Energy_central_n10_m10_L10_QI80_80_L_M_models}
    \end{subfigure}
    \vspace{-4mm}
\caption{CNT bending: (\subref{f:CNT_Loading_bending}) boundary conditions; (\subref{f:bending_Energy_central_n10_m10_L10_QI80_80_L_M_models}) strain energy per atom. CNT(10,10) with the length 10 nm is selected. Using Eq.~\eqref{e:error_dif}, the maximum error relative to the logarithmic model is $7.6500\times 10^{-4}$ percent.}
\end{figure}

First, bending of a CNT is considered as shown in Fig.~\ref{f:CNT_Loading_bending}. The end faces of the CNT are assumed to be rigid and remain planar, and the CNT is allowed to deform in the axial direction in order to avoid a net axial force. The bending angle $\theta$ is applied at both faces of the CNT equally. The variation of the strain energy per atom with the bending angle is shown in Fig.~\ref{f:bending_Energy_central_n10_m10_L10_QI80_80_L_M_models} and compared with the results of the log model of \citet{Ghaffari2017_01} for perfect and imperfect structures. The imperfection is applied as a small torque\footnote{$T=2\text{nN}\,R$ where $R$ is initial radius of the CNT.} in the middle of the CNT. The results of the metric and log models match perfectly. \textcolor{cgn}{$200\times200$ quadratic NURBS elements are used in this study. This discretization has an energy error of less than 0.1\% as the convergence study in \citet{Ghaffari2017_01} shows.}\\
\noindent
\subsubsection{CNT twisting}
\begin{figure}[h]
    \begin{subfigure}{0.49\textwidth}
        \centering
     \includegraphics[height=50mm,trim=2cm 0cm 1cm 0cm,clip]{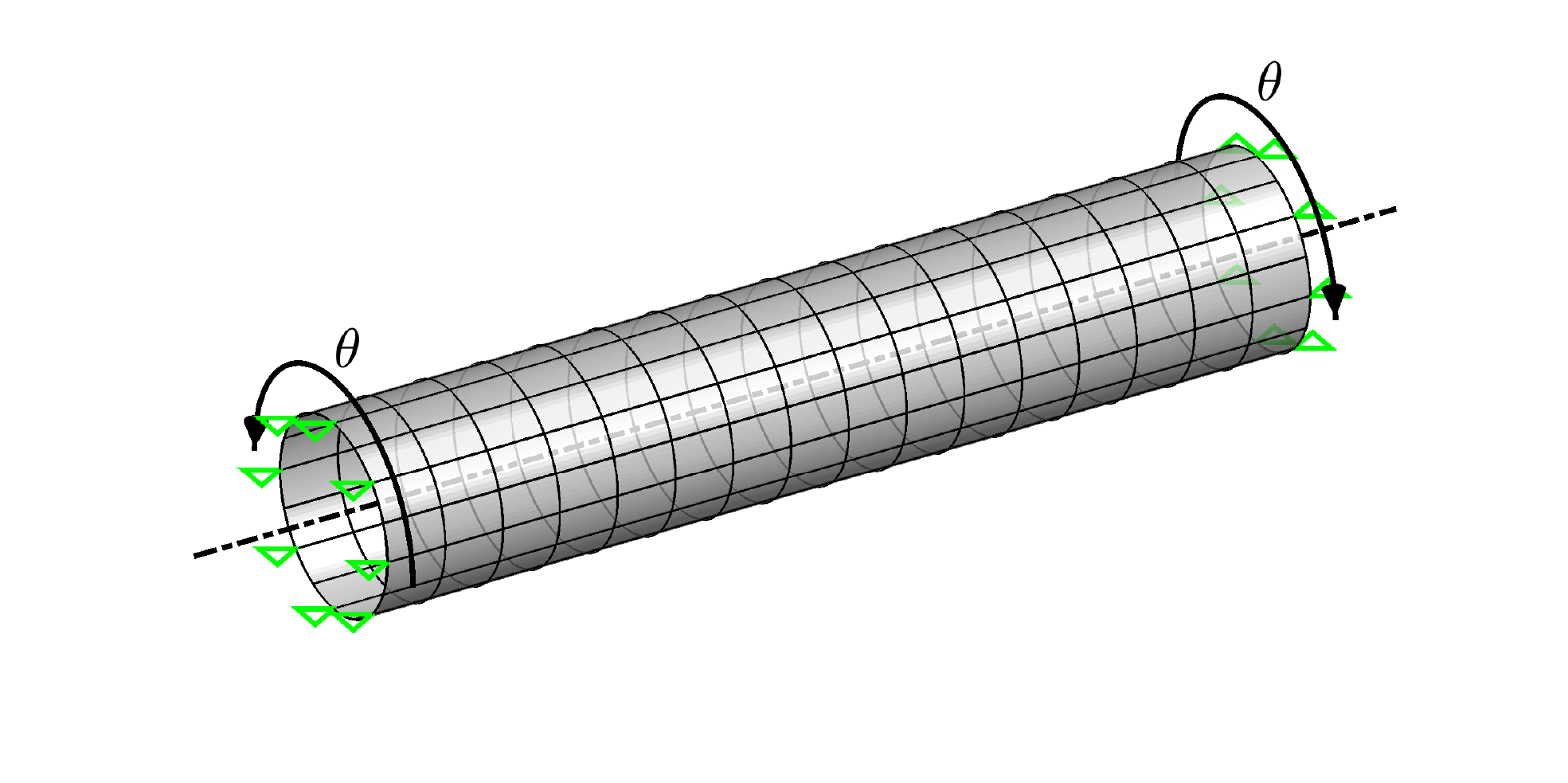}
        \vspace{0mm}
        \subcaption{}
        \label{f:CNT_Loading_torsion}
    \end{subfigure}\hspace{4mm}
     \begin{subfigure}{0.49\textwidth}
        \centering
     \includegraphics[height=55mm]{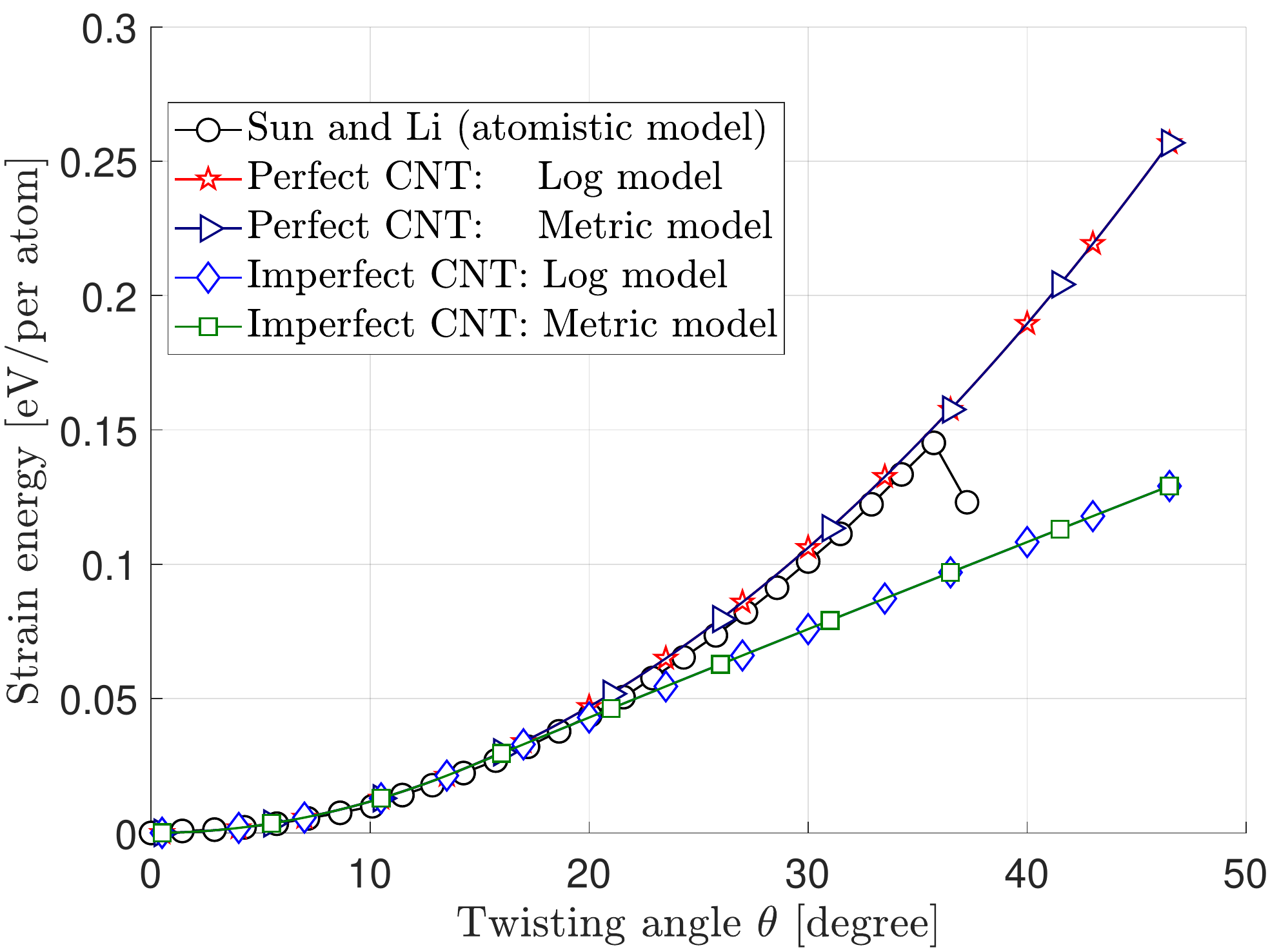}
        \vspace{0mm}
        \subcaption{}
        \label{f:torsion_Energy_torsion_n12_m6_L6_74_L_M_models}
    \end{subfigure}
    \vspace{-4mm}
\caption{CNT twisting: (\subref{f:CNT_Loading_torsion}) boundary conditions (length in the axially direction is fixed); (\subref{f:torsion_Energy_torsion_n12_m6_L6_74_L_M_models}) strain energy per atom. CNT(12,6) with the length 6.74 nm is used. The atomistic results are taken from \citet{Sun2011}. Using Eq.~\eqref{e:error_dif}, the maximum error relative to logarithmic model is 0.002 percent.}
\end{figure}

Next, twisting of a CNT is considered by applying a twisting angle at both faces of the CNT (Fig.~\ref{f:CNT_Loading_torsion}). The variation of the strain energy per atom with the twisting angle is shown in Fig.~\ref{f:torsion_Energy_torsion_n12_m6_L6_74_L_M_models} and compared with the results of the log model of \citet{Ghaffari2017_01}.
 The results of the metric and log models match up to 0.002\%. \textcolor{cgn}{$100\times100$ quadratic NURBS elements are used in this study. This discretization has an energy error of less than 0.01\% as the convergence study in \citep{Ghaffari2017_01} shows.}\\
\subsubsection{CNT contact}\label{s:CNT_contact_wall}
\begin{figure}
    \begin{subfigure}{0.49\textwidth}
        \centering
    \includegraphics[width=75mm,trim=3cm 7.5cm 0cm 0cm,clip]{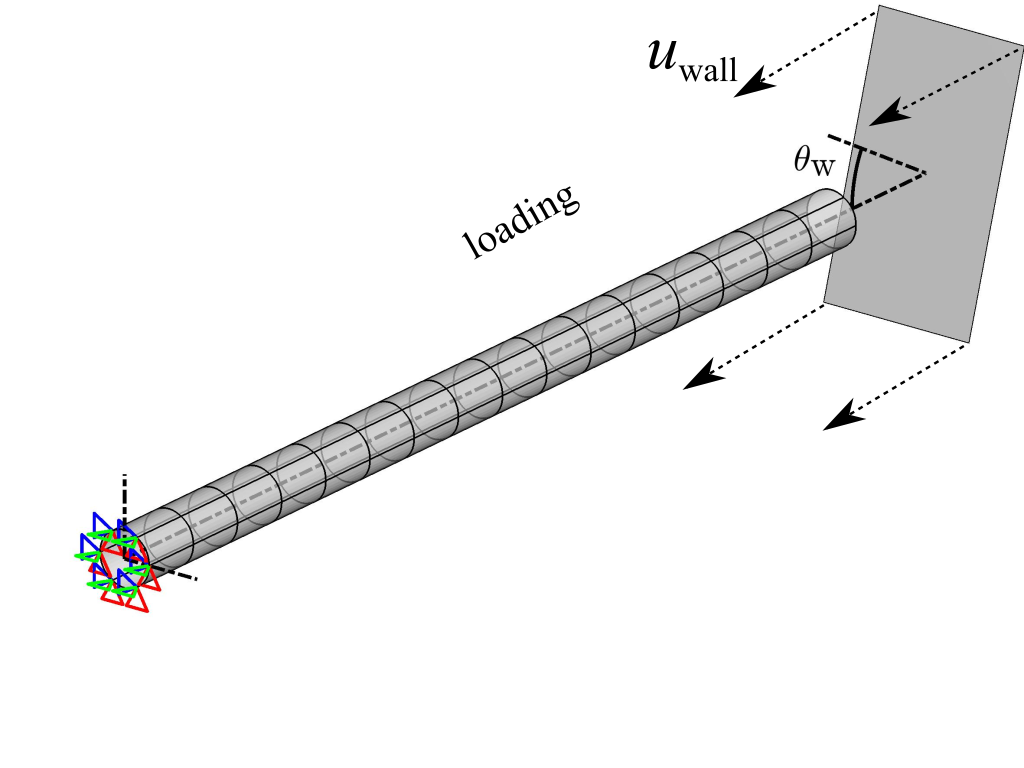}
        \vspace{-7mm}
        \subcaption{}
        \label{f:BC_CNT_LJ_wall}
    \end{subfigure}
    \begin{subfigure}{0.49\textwidth}
        \centering
    \includegraphics[width=80mm]{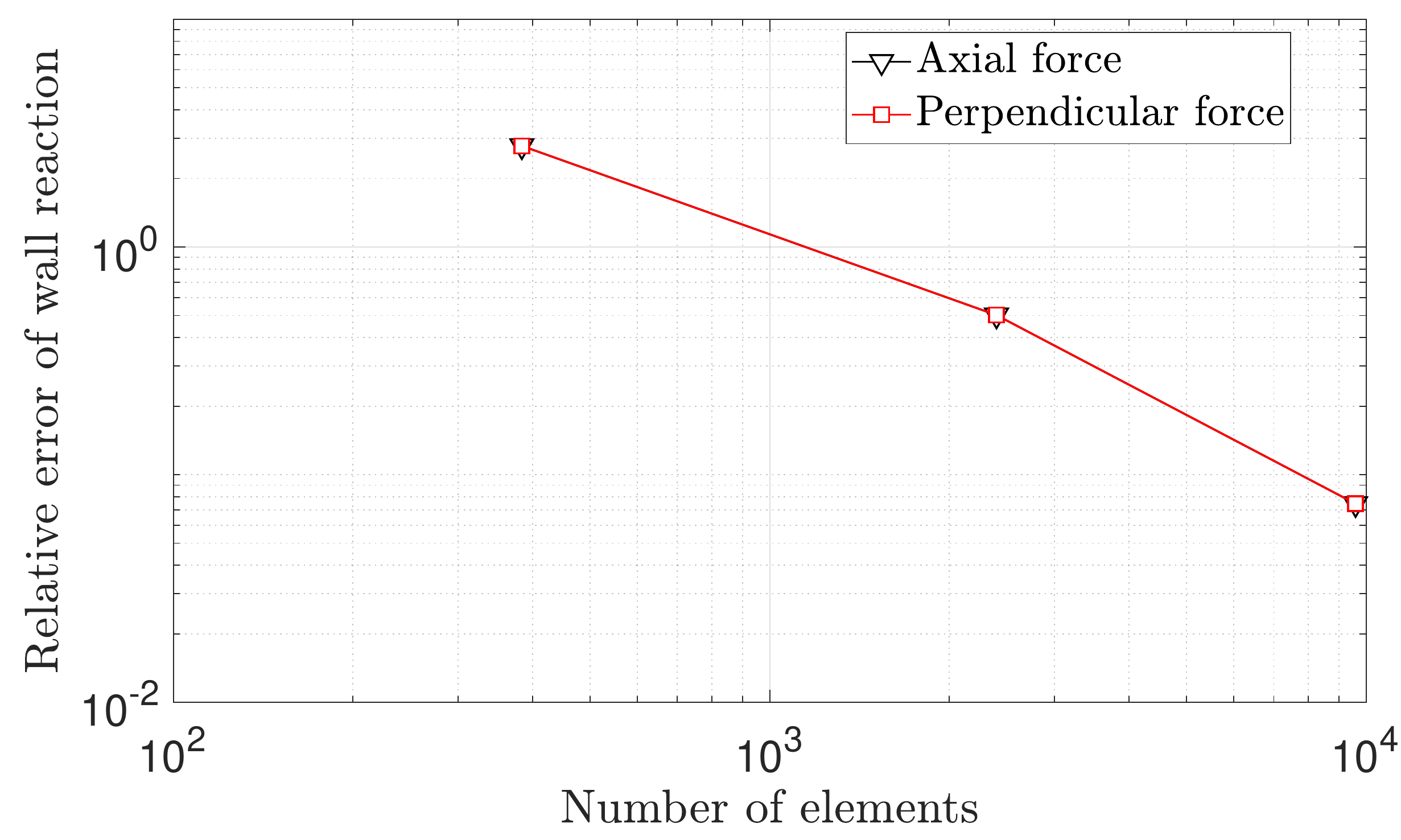}
        \vspace{-7mm}
        \subcaption{}
        \label{f:LJ_wall_reaction_conv}
    \end{subfigure}
    \vspace{-3.5mm}
    \caption{Contact of a CNT with a Lennard-Jones wall: (\subref{f:BC_CNT_LJ_wall}) boundary conditions; (\subref{f:LJ_wall_reaction_conv}) Error of the reaction force relative to the finest mesh ($320\times120$ quadratic NURBS elements). CNT(15,15) with the length 38.19 nm and the contact angle $\theta_{\text{w}}=17.45^{\circ}$ is used.}
\end{figure}

\begin{figure}
    \begin{subfigure}{0.49\textwidth}
        \centering
    \includegraphics[height=58mm]{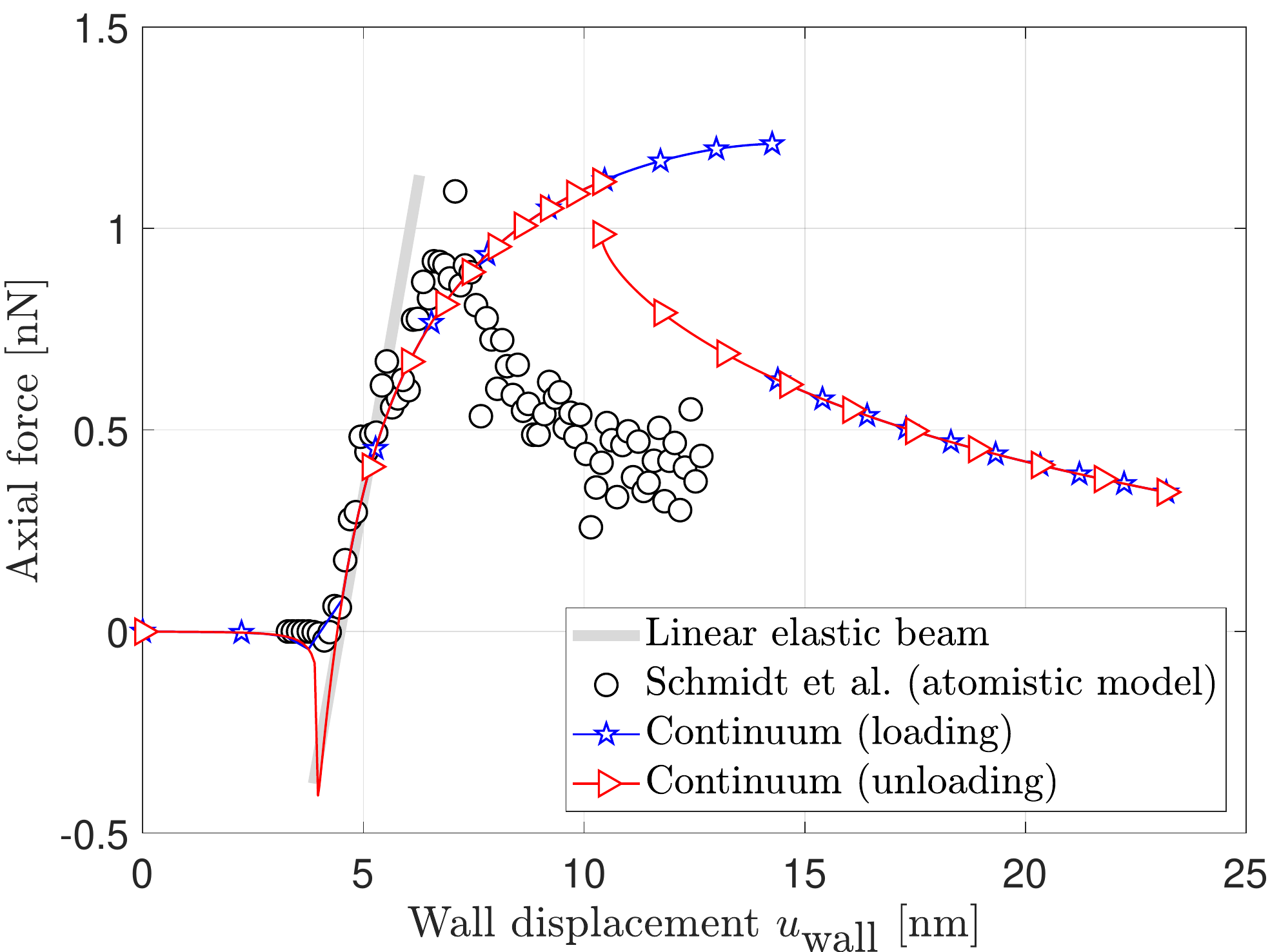}
    \vspace{-6mm}
        \subcaption{}\label{f:LJ_wall_reaction_AxLoad}
    \end{subfigure}
    \begin{subfigure}{0.49\textwidth}
        \centering
    \includegraphics[height=58mm]{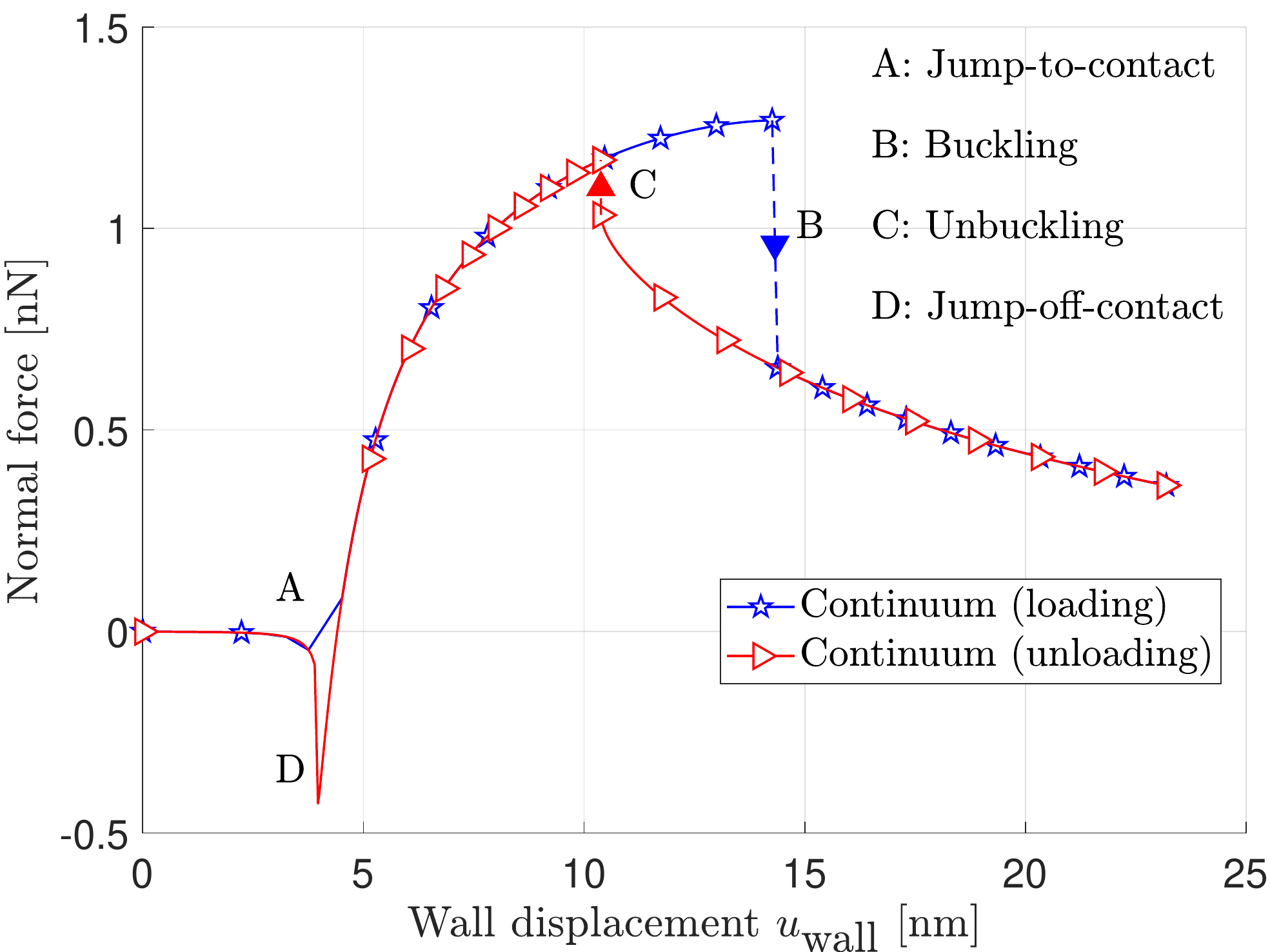}
    \vspace{-6mm}
        \subcaption{}\label{f:LJ_wall_reaction_normal}
    \end{subfigure}
    \vspace{-2.5mm}
    \caption{Contact of a CNT with a the Lennard-Jones wall: Reaction force of the wall in the (\subref{f:LJ_wall_reaction_AxLoad}) undeformed axial direction of the CNT  (\subref{f:LJ_wall_reaction_normal}) normal direction of the wall. CNT(15,15) with the length 38.19 nm and the contact angle $\theta_{\text{w}}=17.45^{\circ}$ is used. The atomistic results are taken from \citet{Schmidt2015}$^{\text{a}}$. The atomistic simulation is conducted at finite temperature but thermal effects are not considered in the continuum model. The relation between axial force $F_{\text{A}}$ and normal force $F_{\text{N}}$ is $F_{\text{N}}=F_{\text{A}}/\cos(\theta_{\text{w}})$. The data between markers is continuous. $^{\text{a}}$~The force unit in \citet{Schmidt2015} should be nN instead of eV/$\AA$.}\label{f:LJ_wall_reaction}
\end{figure}

\begin{figure}
    \begin{subfigure}{1\textwidth}
        \centering
    \includegraphics[height=80mm]{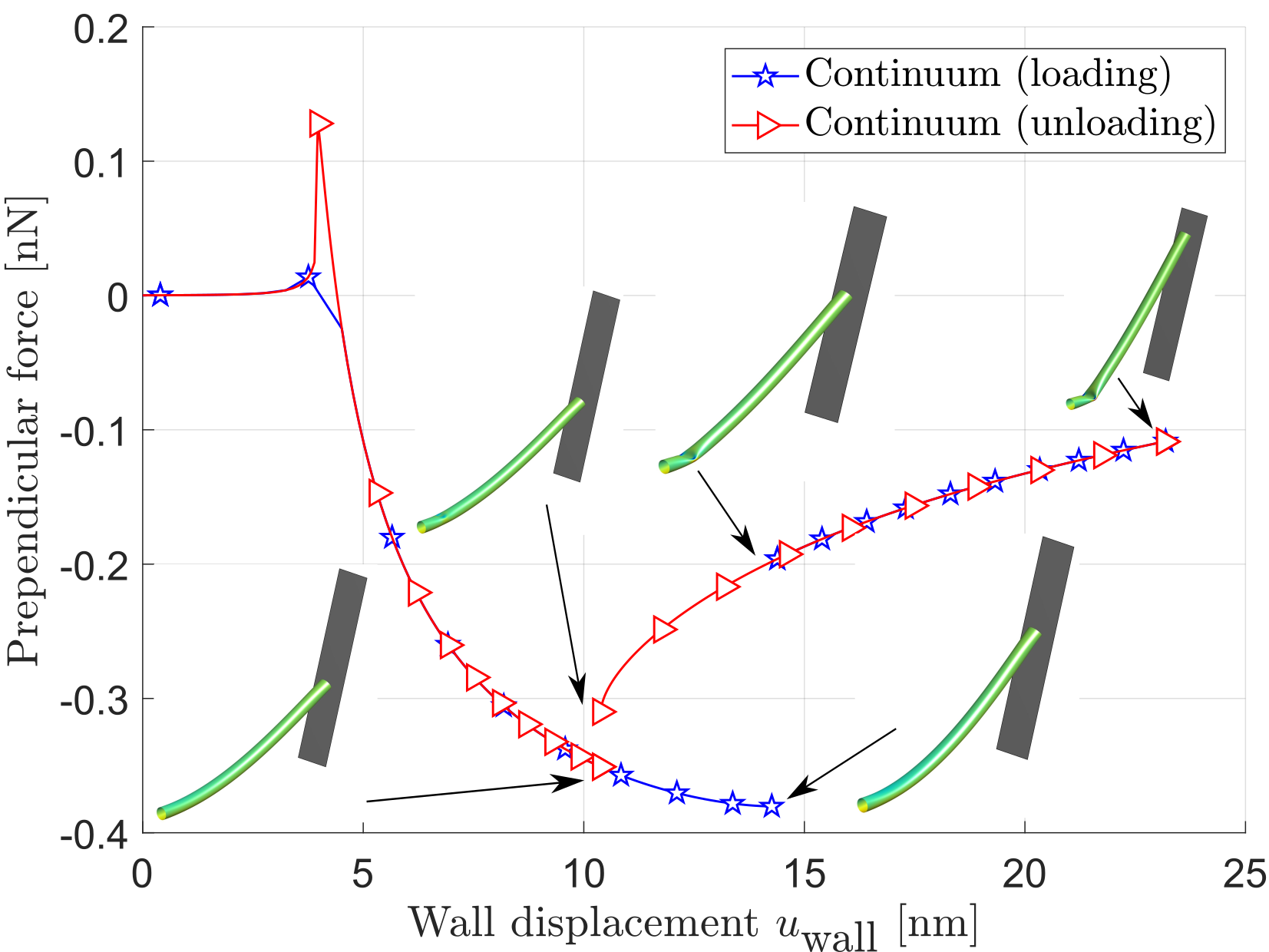}
    \end{subfigure}
    \vspace{-3.5mm}
    \caption{Contact of a CNT with a Lennard-Jones wall: Deformed geometries during loading and unloading.}
    \label{f:LJ_wall_reaction_PreLoad}
\end{figure}

Finally, the contact of a CNT with a Lennard-Jones wall is simulated. This is interesting, since a CNT can be used as a tip of an atomic force microscope (AFM). CNTs can have a set of discrete chiralities and radii so they can be mass-produced with a precise radius and length while silicon and silicon nitride tips can not be produced with an identical geometry \citep{Cheung2000}. This unique feature of CNT-based AFMs guarantees the reproducibility of measurements and experiments with different AFMs \citep{Dai1996,Hafner2001_01,Hafner2001_02,Stevens2009,Wilson2009}.
In addition, CNTs-based AFMs have a higher resolution relative to AFMs with silicon or silicon nitride tips \citep{Choi2016}. However, the measurement with the AFM is not reliable after buckling and hence buckling should be avoided. A CNT with a larger radius is more stable, but the precision and resolution of the AFM decrease.\\
Here we study contact and buckling of a CNT with a rigid wall using the setup shown in Fig.~\ref{f:BC_CNT_LJ_wall}. The wall is modeled with a coarse grained contact model (CGCM) \citep{Sauer2006_01,sauer2007_01,sauer2007_02,sauer2009_01,Ghaffari2015}. Within this model, an equivalent half space potential is used at each contact point. This potential can be written as \citep{Ghaffari2017_01}
\eqb{lll}
\Psi_{\text{(VdW)h}} \is \ds -\Gamma\left[\frac{3}{2}\left(\frac{h_0}{r}\right)^3-\frac{1}{2}\left(\frac{h_0}{r}\right)^9\right]~,
\eqe
where $h_0=0.34~\text{nm}$, $\Gamma=0.14~\text{N/m}$ and $r$ are the equilibrium distance, the interfacial adhesion energy per unit area and the normal distance of a surface point to the wall. The wall is moved in the axial direction toward the CNT (during loading) and away from it (during unloading). The results converge with mesh refinement as the error plot based on Eq.~\eqref{e:error_dif} in Fig.~\ref{f:LJ_wall_reaction_conv} shows. \textcolor{cgn}{Quadratic NURBS meshes with $32\times12$, $80\times30$, $160\times60$ and $320\times120$ elements are used for the convergence study. The difference in the wall reaction between the finest and second finest mesh is $7.45\%$.}
The contact force is compared with the atomistic results of \citet{Schmidt2015} in Fig.~\ref{f:LJ_wall_reaction}a. The \textcolor{cgm}{contact} force is \textcolor{cgm}{also} compared to an analytical solution (see \ref{s:circ_beam_analy}). The extremum of the axial and perpendicular forces are $1.21$ nN and --0.38 nN, respectively. The CNT buckles during loading which leads to a sharp drop in the contact force. This discontinuity is captured by using the arc-length method of \citet{Ghaffari2015} in conjugation with a line-search method. During unloading, the reaction force is different than during loading.
Note that there are two instabilities 1.~buckling / unbuckling (at point B \& C in Fig.~\ref{f:LJ_wall_reaction_normal}) 2.~jump-to- / jump-off-contact (at point A \& D in Fig.~\ref{f:LJ_wall_reaction_normal}). The second is also common to other adhesive systems at small length scales \citep{Sauer2009_02,Raj2018_01}. The deformed CNT is shown before and after buckling and jump-to / jump-off-contact in Fig.~\ref{f:LJ_wall_reaction_PreLoad}.\\
\subsection{Carbon nanocones}
\begin{figure}
     \begin{subfigure}{.32\textwidth}
    \includegraphics[angle=90,origin=c,width=50mm]{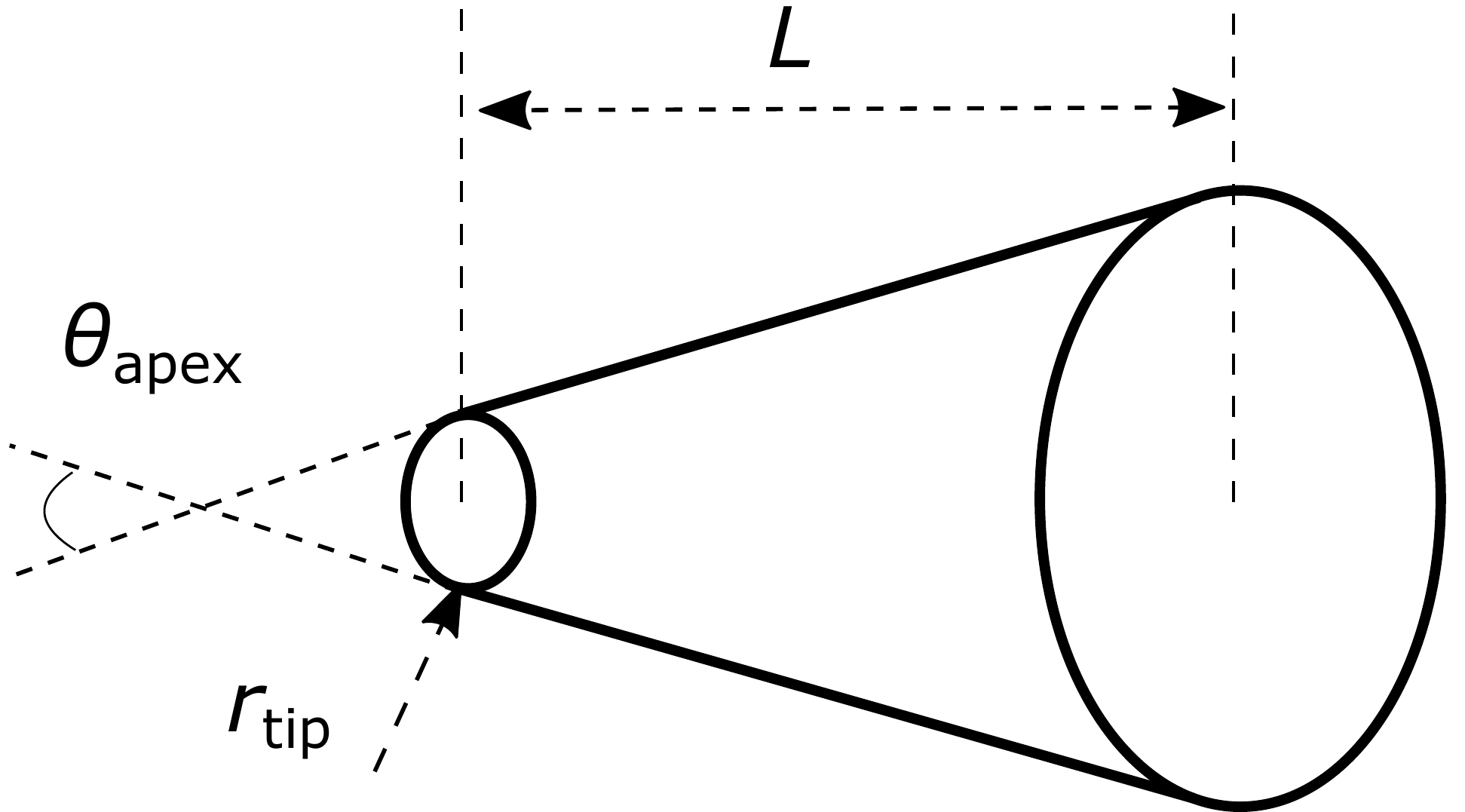}
    \subcaption{}
    \label{f:nanocone_sketch}
    \end{subfigure}
    \begin{subfigure}{0.32\textwidth}
        \centering
    \includegraphics[width=50mm,trim=2.4cm 4.5cm 1.1cm 2.5cm,clip]{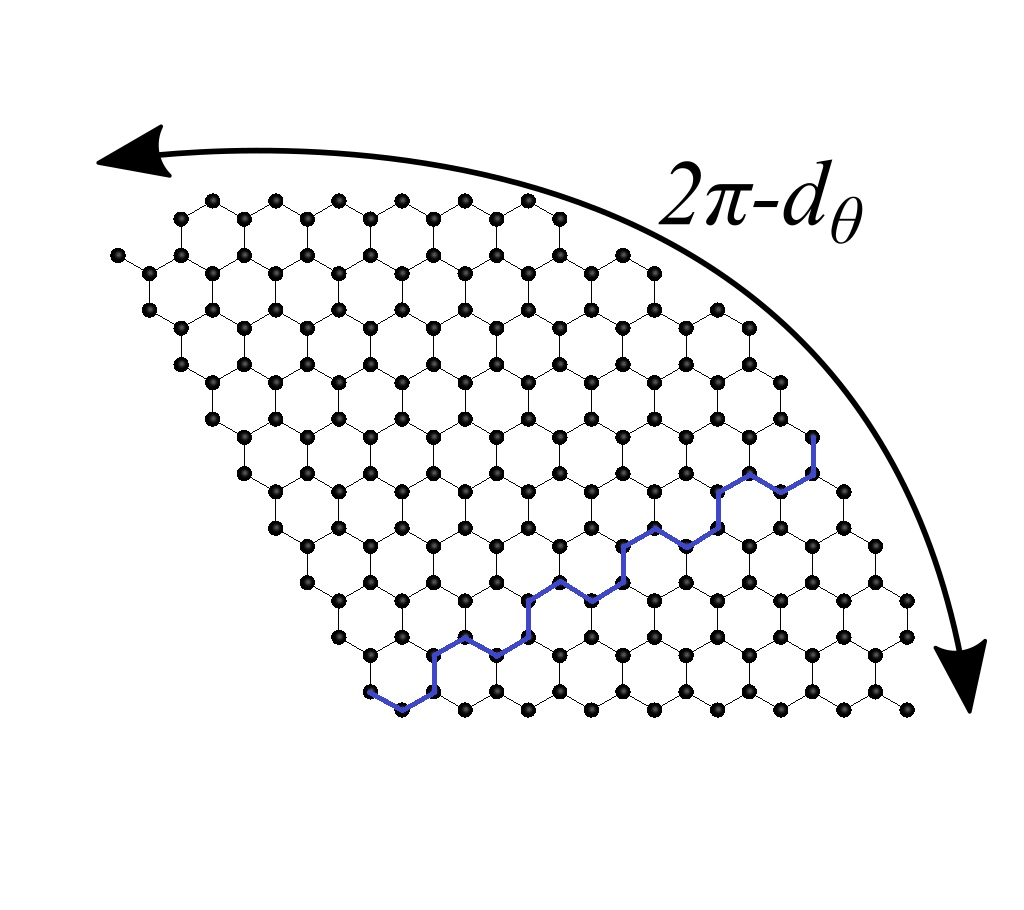}
        \subcaption{}
       \label{f:cone_sheet_armchair_line}
    \end{subfigure}
    \begin{subfigure}{0.32\textwidth}
        \centering
    \includegraphics[width=50mm,trim=3.5cm 8.5cm 1.cm 2.5cm,clip]{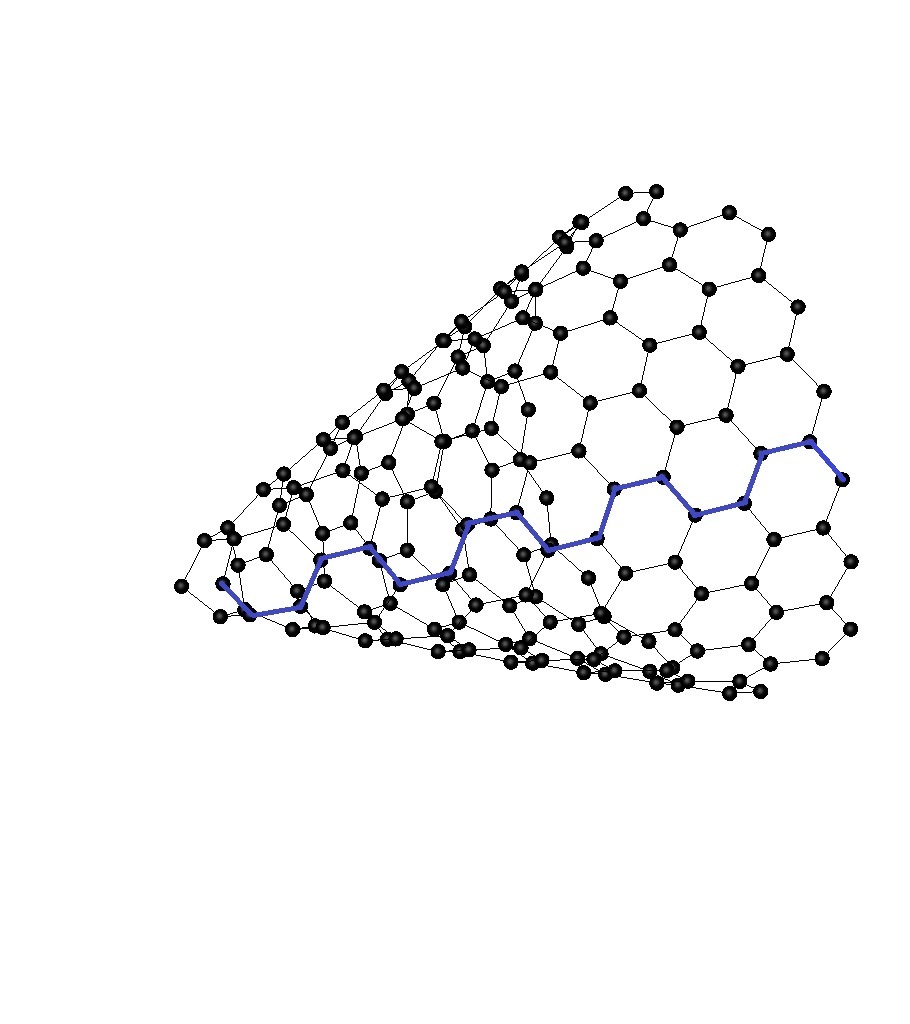}
        \subcaption{}
        \label{f:cone_with_armchair_line}
    \end{subfigure}
    \vspace{-3.5mm}
    \caption{CNC geometry and lattice: (\subref{f:nanocone_sketch}) CNC dimensions; (\subref{f:cone_sheet_armchair_line}) flat graphene sheet; (\subref{f:cone_with_armchair_line}) rolled graphene sheet. The armchair direction is shown by a blue line.}
    \label{f:cone_armchair_line}
\end{figure}

A CNC($\theta_{\text{apex}}$) can be generated by rolling a sector of a graphene sheet. It is described with an apex angle $\theta_{\text{apex}}$ and length $L$ (see Fig.~\ref{f:nanocone_sketch}). A zigzag (or armchair) line can only be matched with another zigzag (or armchair) line to create a CNC (see Fig.~\ref{f:cone_armchair_line}b). A CNC can be generated from a sector of a graphene sheet \textcolor{cgm}{that} is cut with the declination angle $d_{\theta}$\footnote{I.e. the angle of the removed sector.} of $60^{\circ}$, $120^{\circ}$, $180^{\circ}$, $240^{\circ}$ and $300^{\circ}$. Thus $\theta_{\text{apex}}$ can have the discrete values \citep{Lee2012,Ansari2014_01}
\eqb{lll}
\theta_{\text{apex}} \is 2\,\arcsin(1-d_{\theta}/360)~.
\eqe\noindent
\noindent
Similar to the preceding CNT examples, bending, twisting and wall contact of CNCs are considered in the following. It is seen that CNCs buckle without applying an imperfection due to
the variation of the chirality along the different tangential coordinates.\\
\subsubsection{CNC bending}
\begin{figure}
    \begin{subfigure}{0.49\textwidth}
        \centering
    \includegraphics[height=60mm,trim=4cm 0cm 2cm 0cm,clip]{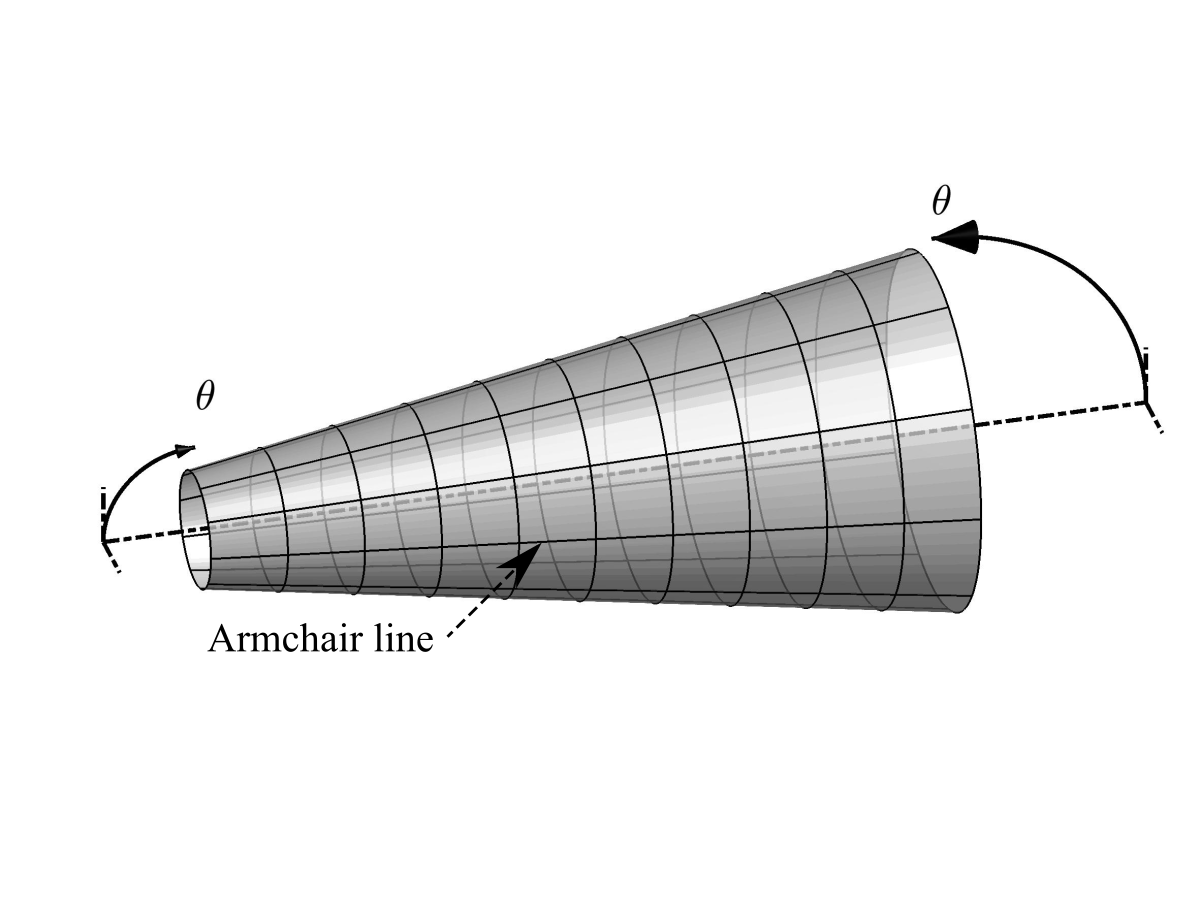}
        \vspace{-2mm}
        \subcaption{}
        \label{f:CNC_bending_BC}
    \end{subfigure}
    \begin{subfigure}{0.49\textwidth}
        \centering
    \includegraphics[height=60mm]{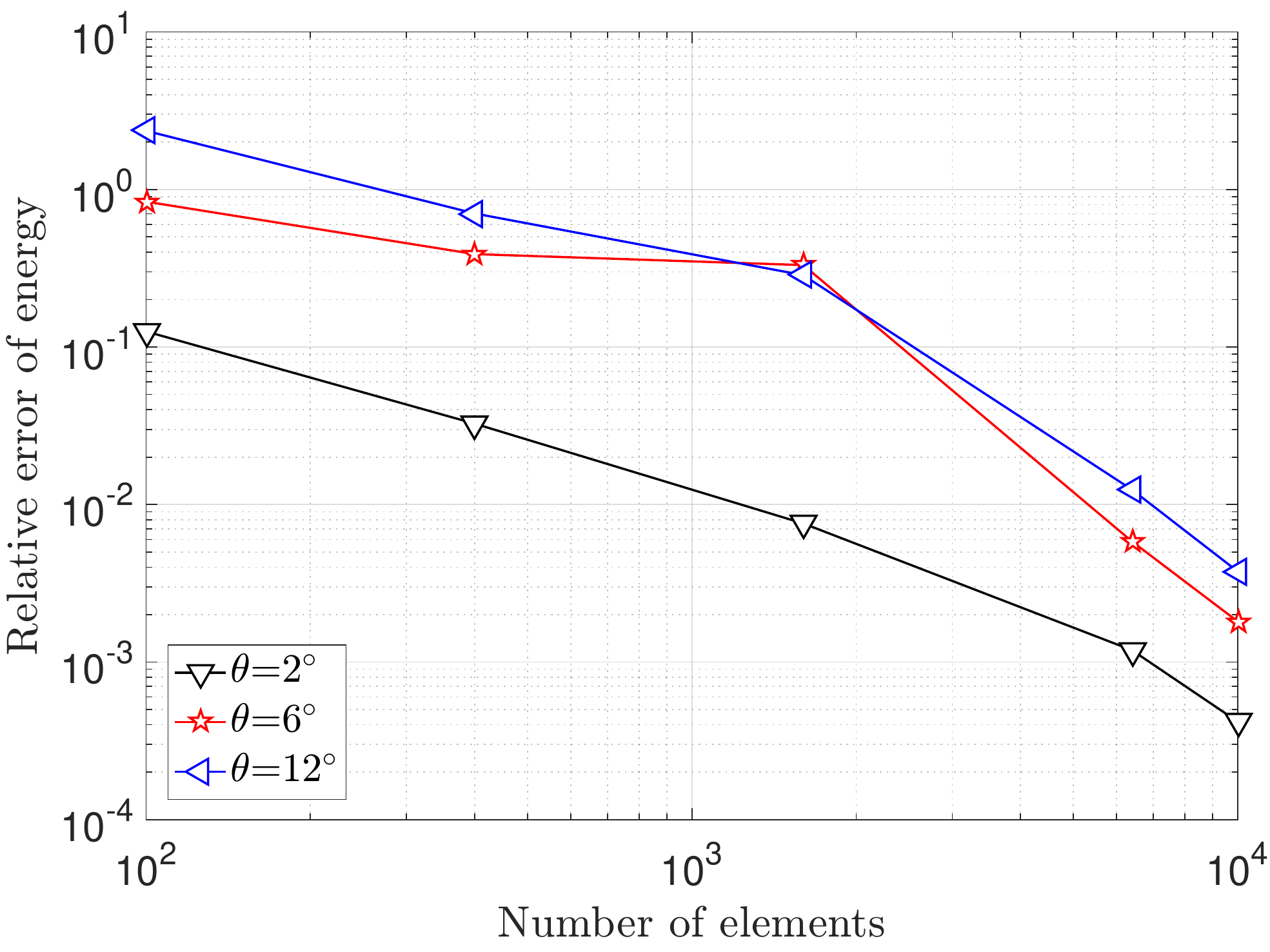}
        \vspace{-7mm}
        \subcaption{}
        \label{f:Bending_CNC_L12_A60_strainEnergy_conv}
    \end{subfigure}
    \vspace{-3.5mm}
    \caption{CNC bending: (\subref{f:CNC_bending_BC}) boundary conditions; (\subref{f:Bending_CNC_L12_A60_strainEnergy_conv}) Error of strain energy relative to the finest mesh (using $120\times120$ quadratic NURBS elements). CNC($19.2^{\circ}$) with the length and tip radius 12.04 nm and 1 nm is used.}
\end{figure}

\begin{figure}
    \begin{subfigure}{0.49\textwidth}
        \centering
    \includegraphics[height=58mm]{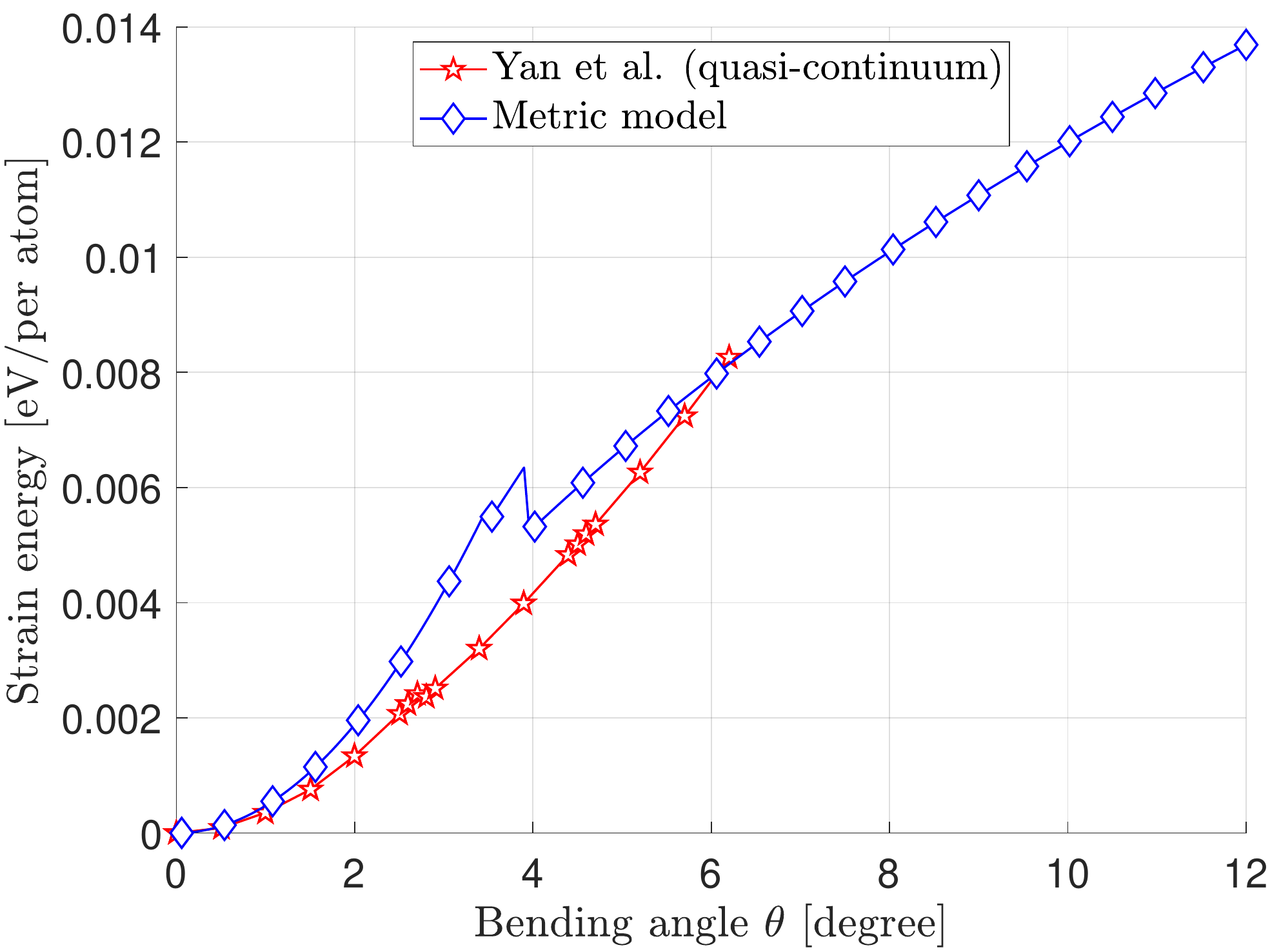}
    \vspace{-7mm}
        \subcaption{}
        \label{f:bending_CNC_L12_A60_Energy}
    \end{subfigure}
    \begin{subfigure}{0.49\textwidth}
        \centering
    \includegraphics[height=58mm]{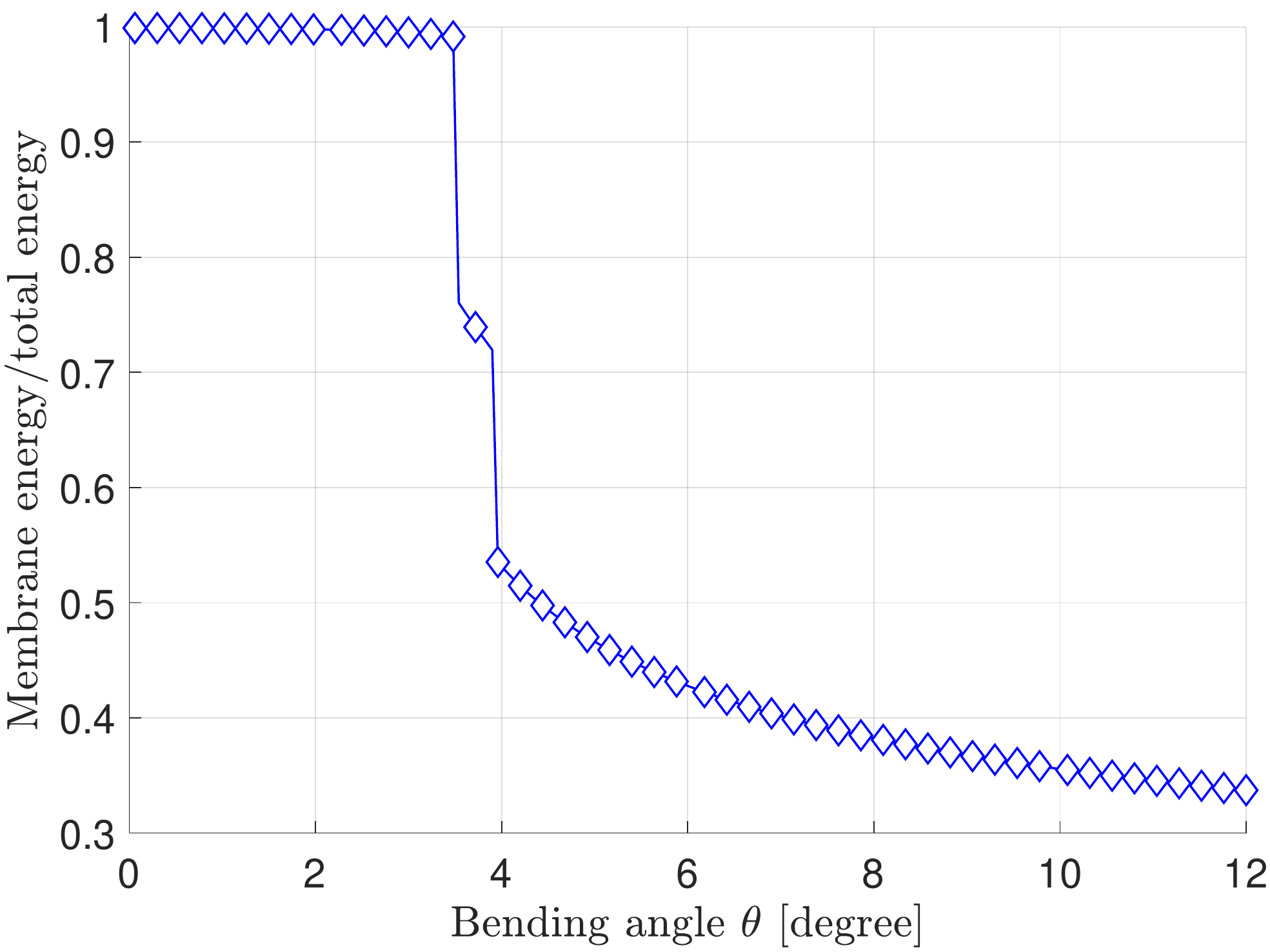}
    \vspace{-7mm}
        \subcaption{}
        \label{f:bending_CNC_L12_A60_Energy_ratio}
    \end{subfigure}
    \vspace{-3.5mm}
    \caption{CNC bending: (\subref{f:bending_CNC_L12_A60_Energy}) Comparison of strain energy per atom \textcolor{cgm}{for the proposed metric model and the} quasi-continuum \textcolor{cgm}{model} of \citet{Yan_2013_01}; (\subref{f:bending_CNC_L12_A60_Energy_ratio}) the ratio of the membrane energy to the total energy \textcolor{cgm}{for the metric model}.  CNC($19.2^{\circ}$) with the length and tip radius 12.04 nm and 1 nm is used.}
\end{figure}

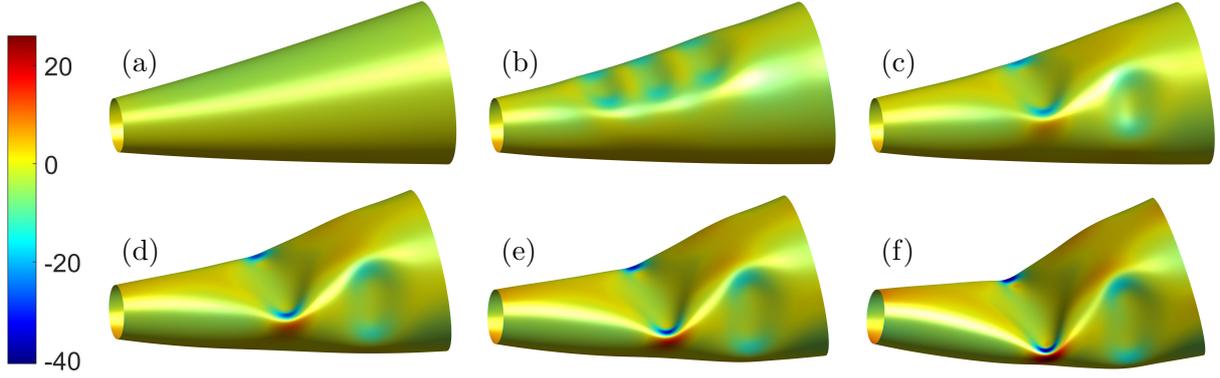
\begin{figure}
\begin{center} \unitlength1cm
\begin{picture}(18,5)
\put(1,2.5){\includegraphics[width=50mm,trim=4cm 17cm 4cm 8cm,clip]{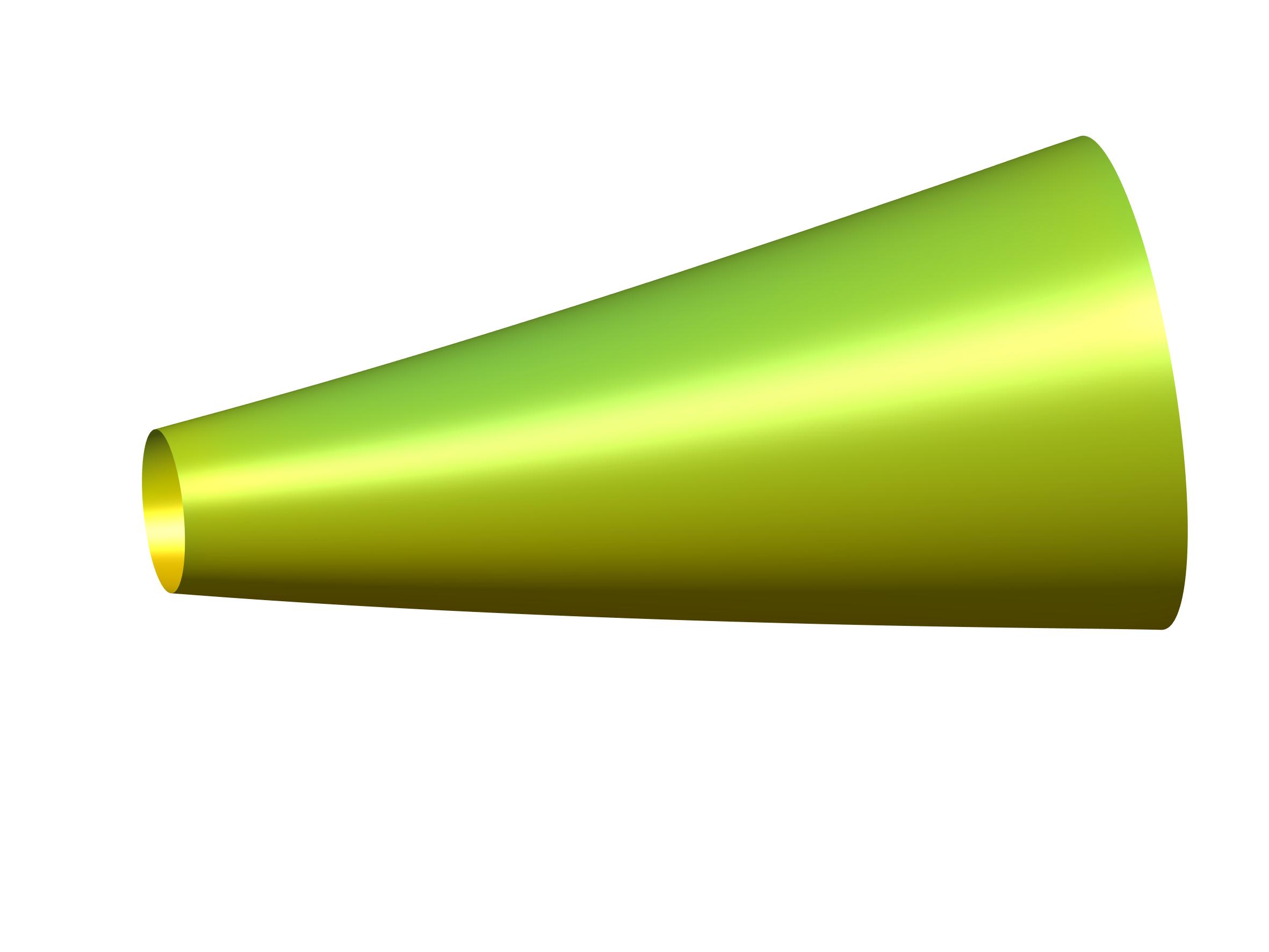}}
\put(6,2.5){\includegraphics[width=50mm,trim=4cm 17cm 4cm 8cm,clip]{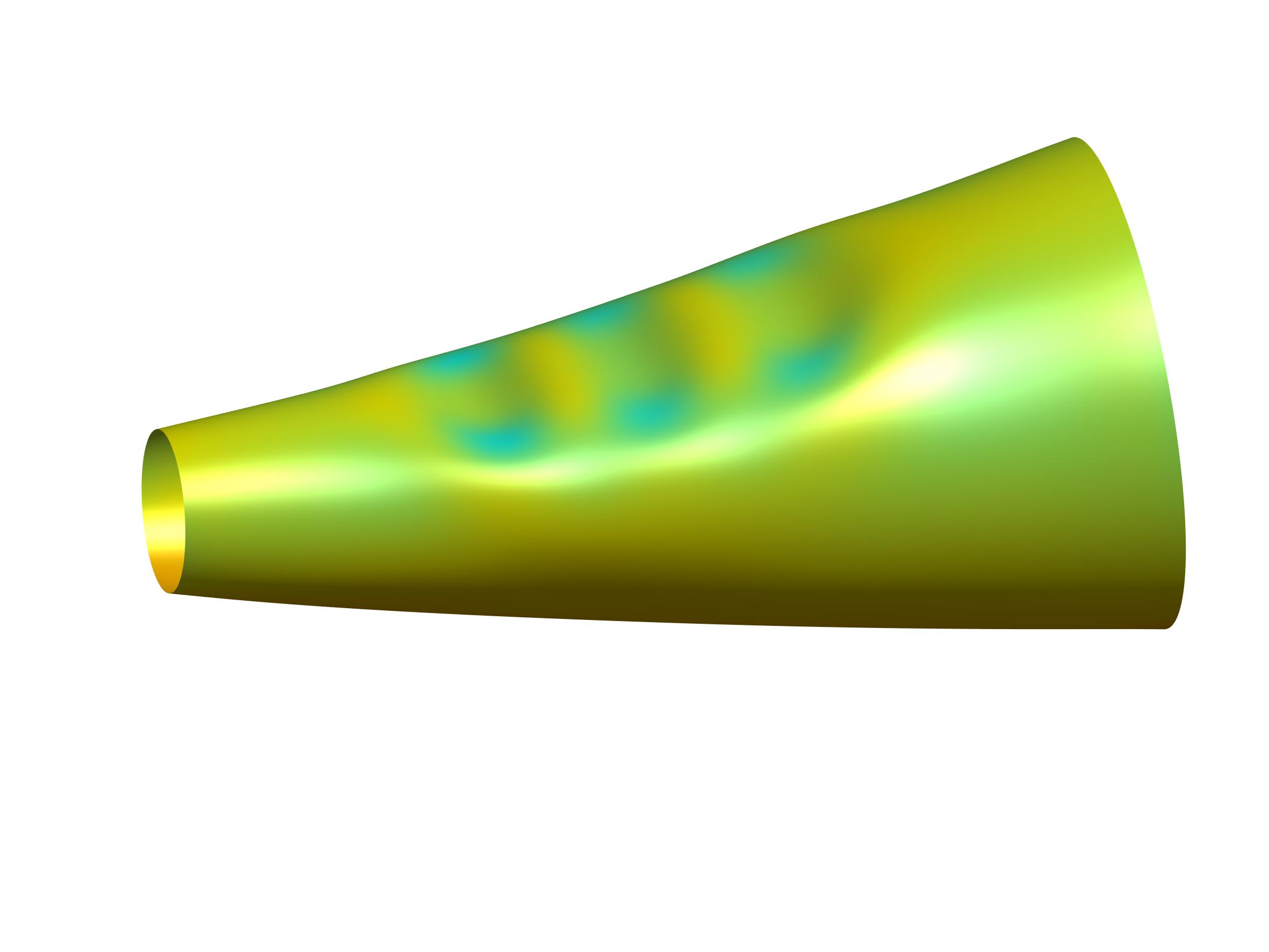}}
\put(11,2.5){\includegraphics[width=50mm,trim=4cm 17cm 4cm 8cm,clip]{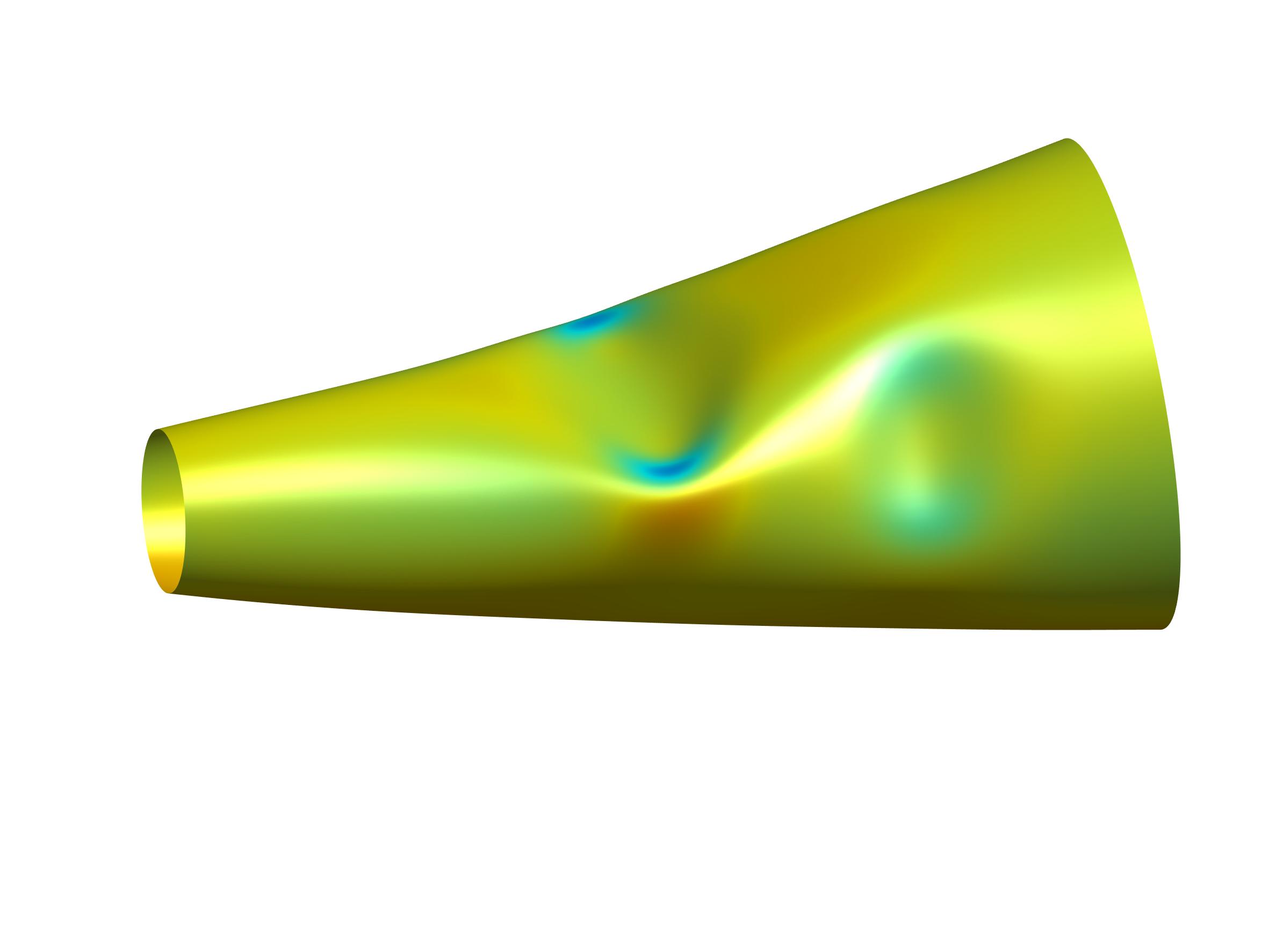}}

\put(1,0){\includegraphics[width=50mm,trim=4cm 17cm 4cm 8cm,clip]{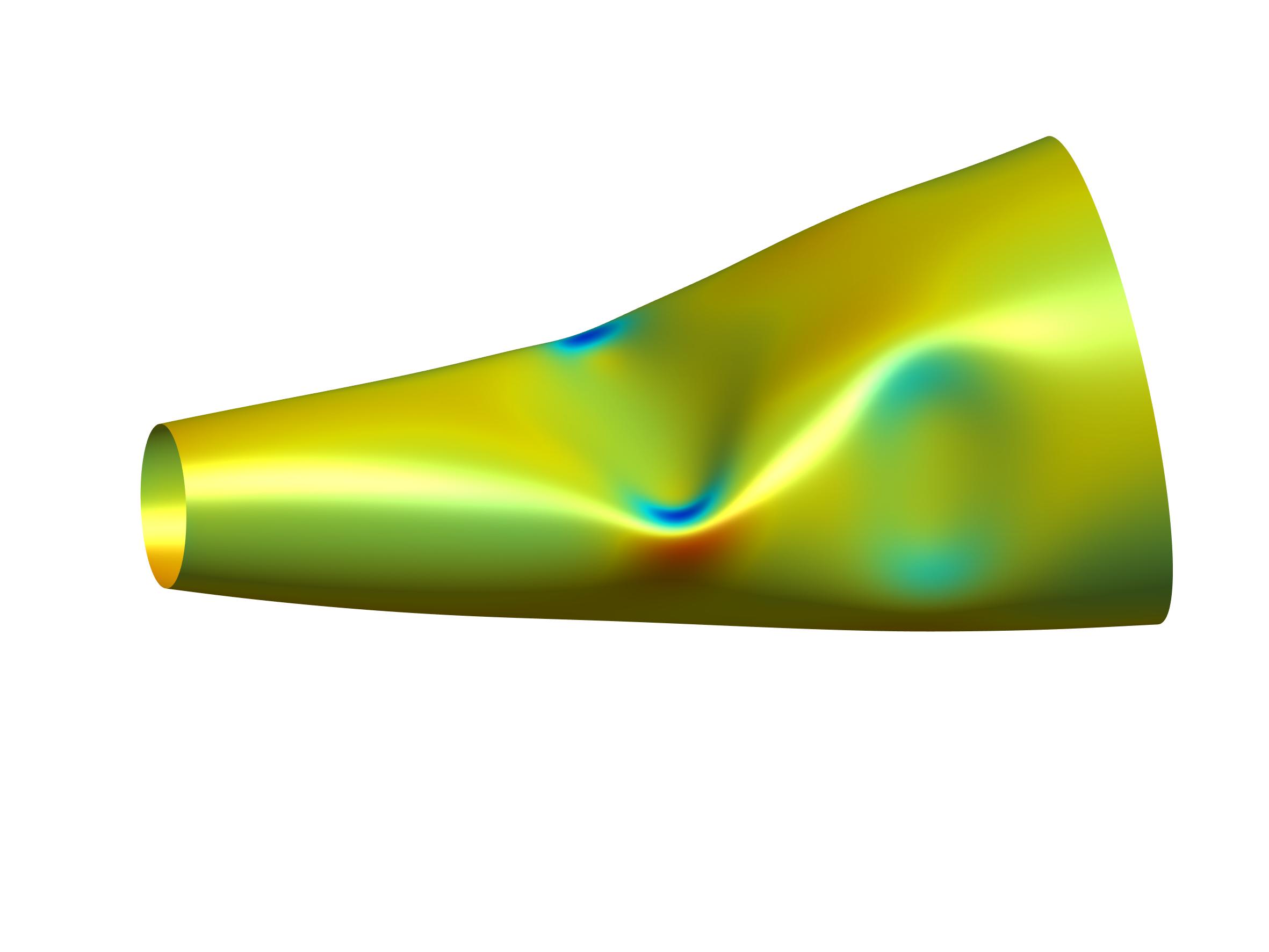}}
\put(6,0){\includegraphics[width=50mm,trim=4cm 17cm 4cm 8cm,clip]{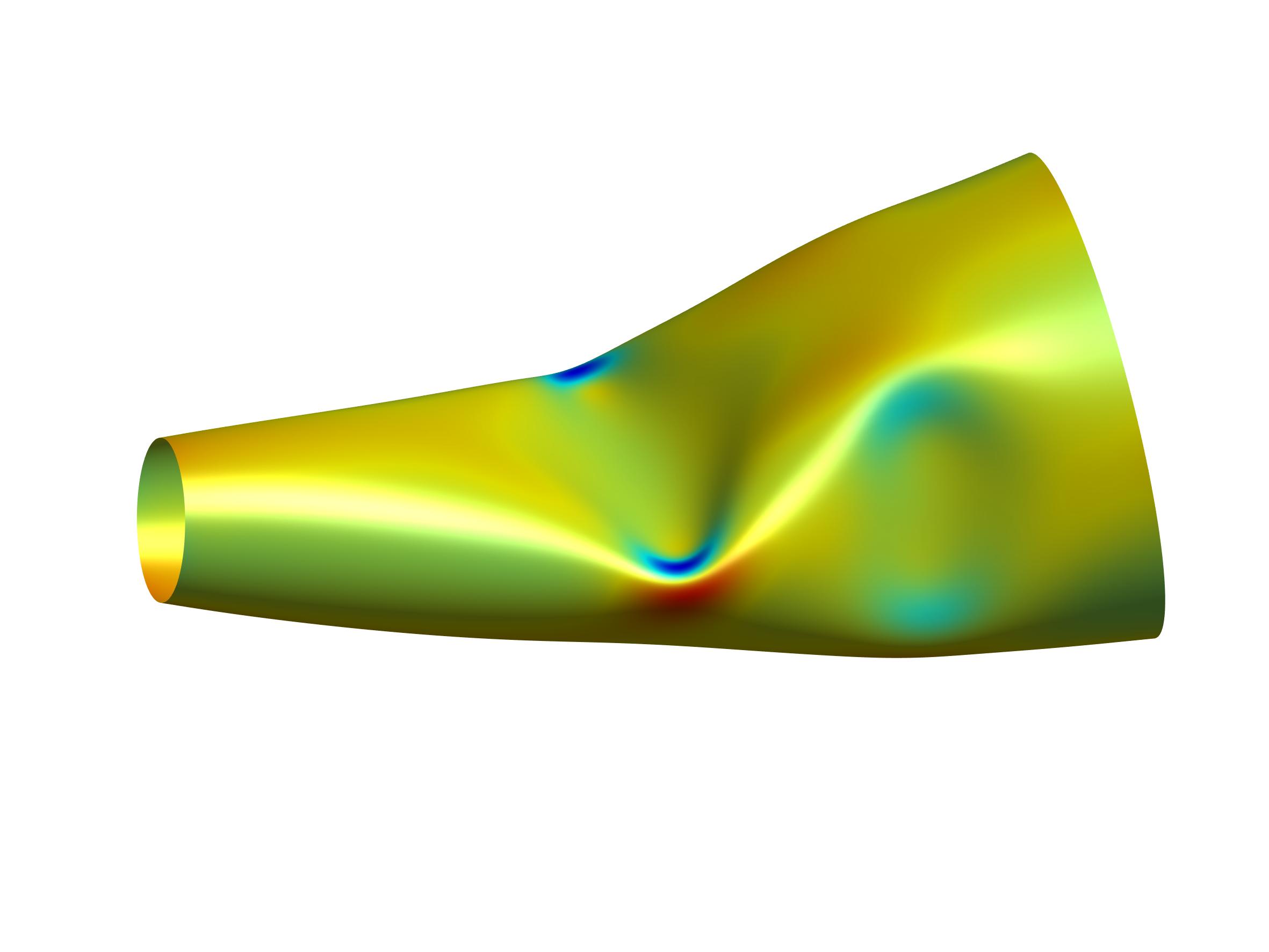}}
\put(11,0){\includegraphics[width=50mm,trim=4cm 17cm 4cm 8cm,clip]{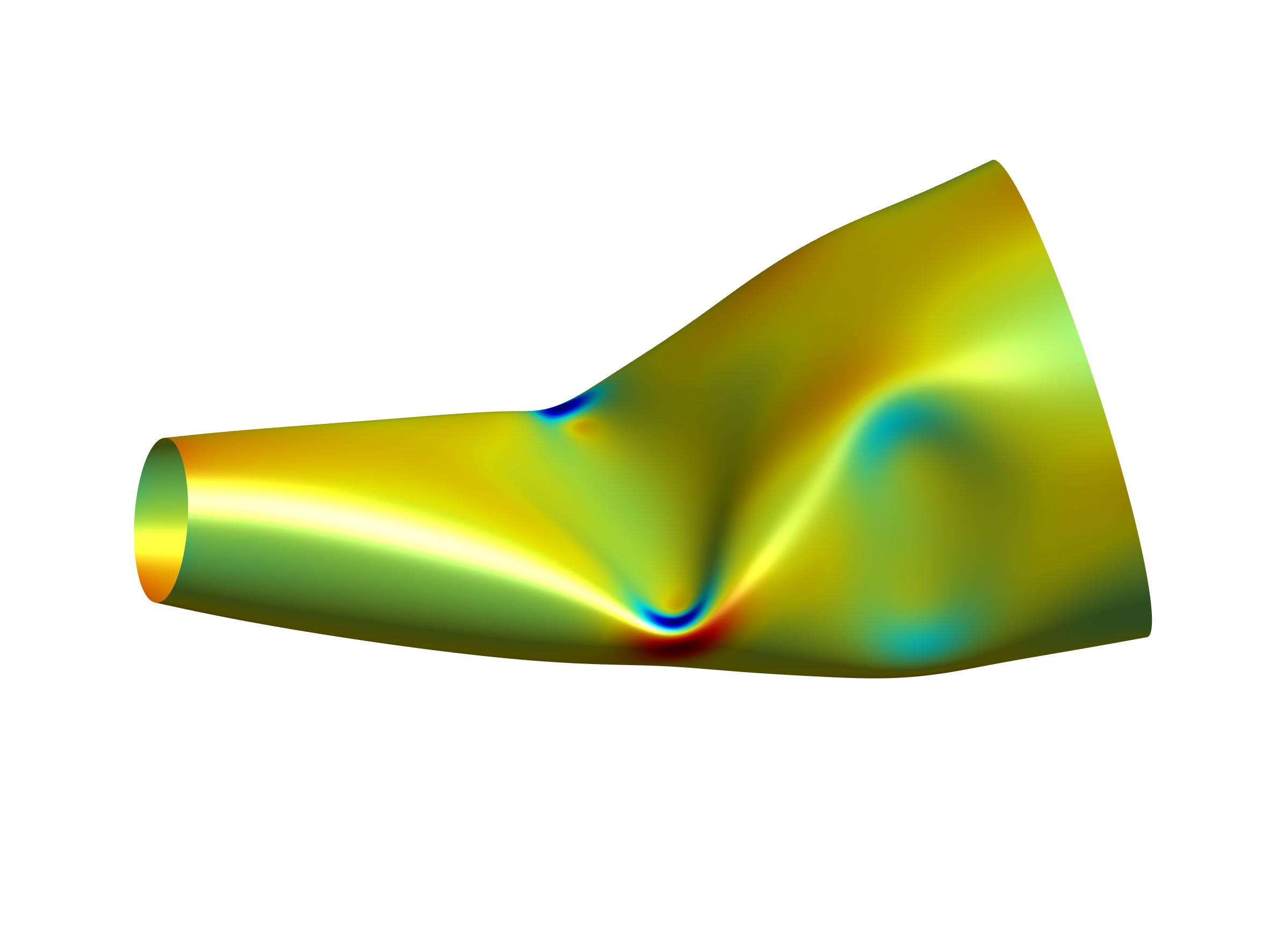}}

\put(1.5,4){(a)}
\put(6.5,4){(b)}
\put(11.5,4){(c)}
\put(1.5,1.5){(d)}
\put(6.5,1.5){(e)}
\put(11.5,1.5){(f)}
\put(0,0){\includegraphics[height=45mm]{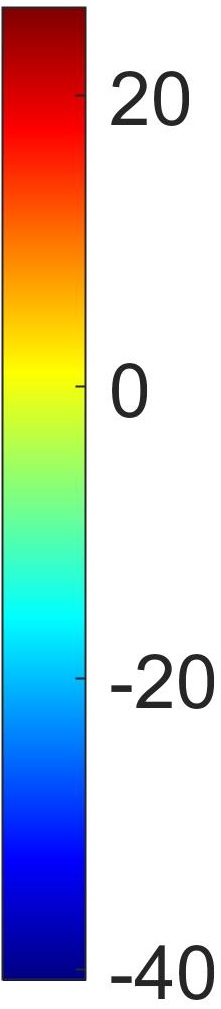}}

\end{picture}
\vspace{-7mm}
\caption{CNC bending: Comparison of $\tr(\bsig_{\text{KL}})$ [$\text{N/m}$] at the bending angles
(a) $\theta=2^{\circ}$,
(b) $\theta=3.5^{\circ}$,
(c) $\theta=4^{\circ}$ and
(d) $\theta=6^{\circ}$,
(e) $\theta=8^{\circ}$,
(f) $\theta=12^{\circ}$.  CNC($19.2^{\circ}$) with the length and tip radius 12.04 nm and 1 nm is used.}
\label{f:CNT_bending_energy_contours_ISO}
\end{center}
\end{figure}

The first example considers CNC bending. The boundary conditions for CNC bending are shown in Fig.~\ref{f:CNC_bending_BC}. The end faces of the CNC are kept rigid and the bending angle is applied to them equally. \textcolor{cgm}{Here, t}he CNC can deform in the axial direction to avoid net axial loading. Fig.~\ref{f:Bending_CNC_L12_A60_strainEnergy_conv} demonstrates the FE convergence under mesh refinement by examining the error from Eq.~\eqref{e:error_dif}, where $q_{\text{ref}}$ is the result from the finest mesh. \textcolor{cgn}{Quadratic NURBS meshes with $m\times m$ elements, for $m$ = 10, 20, 40, 60, 80, 100, 120, are used for the convergence study. The energy difference between the finest and second finest mesh is below $0.38\%$.}
\noindent
The strain energy per atom and the ratio of the membrane energy to the total energy as a function of the bending angle are given in Figs.~\ref{f:bending_CNC_L12_A60_Energy} and \ref{f:bending_CNC_L12_A60_Energy_ratio}. The structure buckles at two loading levels: At $\theta=3.48^{\circ}$ the CNC buckles at the tip, and at $\theta=3.9^{\circ}$ the CNC buckles at end. These buckling points can be precisely obtained from the ratio of the membrane energy to the total energy (see Fig.~\ref{f:bending_CNC_L12_A60_Energy_ratio}). Fig.~\ref{f:CNT_bending_energy_contours_ISO} shows the deformation and stress invariant $\tr(\bsig_{\text{KL}})=\bsig_{\text{KL}}:\bolds{1}$ following from Eq.~\eqref{e:shell_stress} at different bending angles.\\
\noindent
\noindent
\subsubsection{CNC twisting}
\begin{figure}[h]
    \begin{subfigure}{0.49\textwidth}
        \centering
    \includegraphics[height=60mm,trim=9cm 0cm 4cm 0cm,clip]{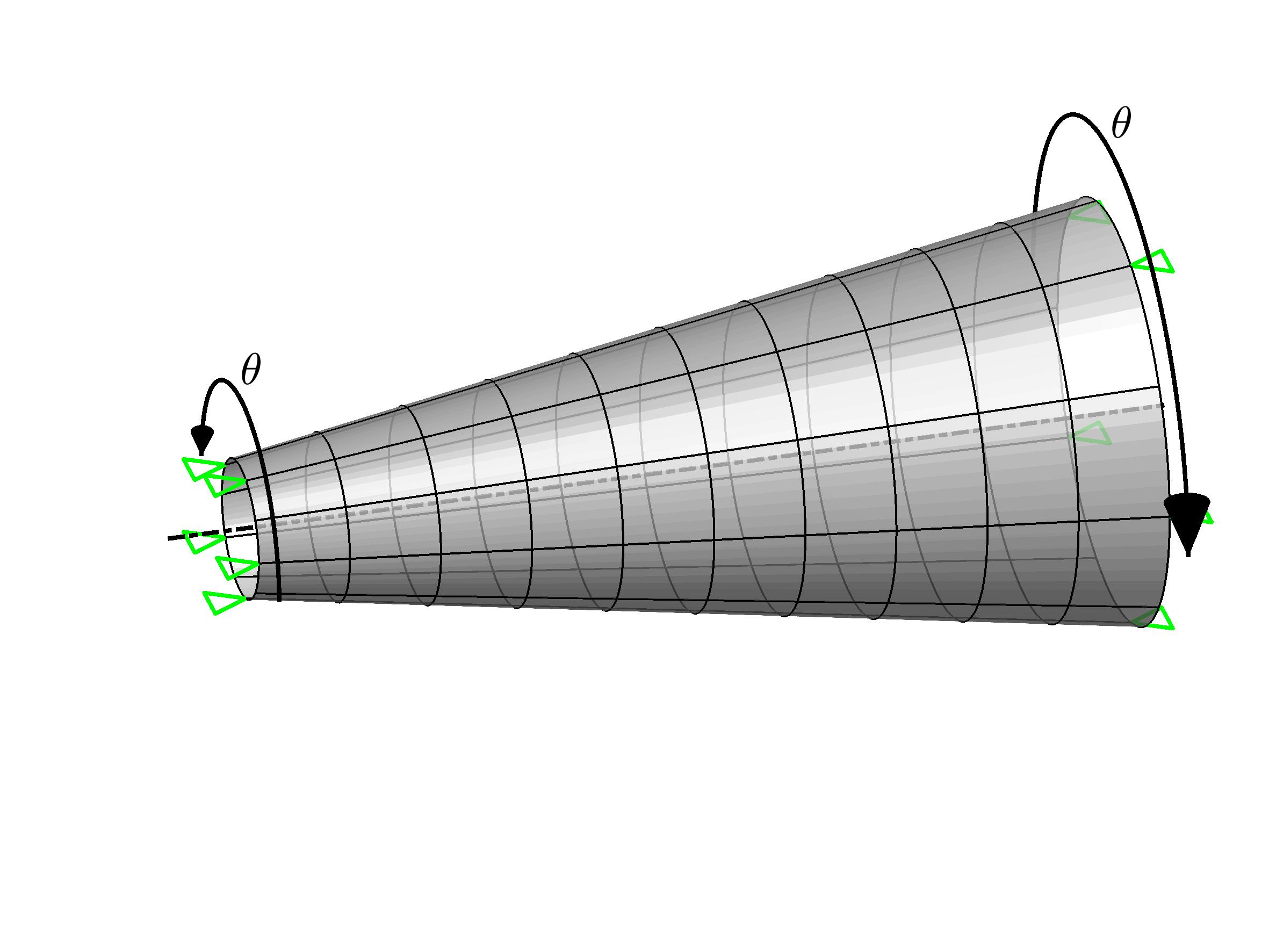}
        \vspace{-2mm}
        \subcaption{}
        \label{f:CNC_torsion_BC}
    \end{subfigure}
    \begin{subfigure}{0.49\textwidth}
        \centering
    \includegraphics[height=60mm]{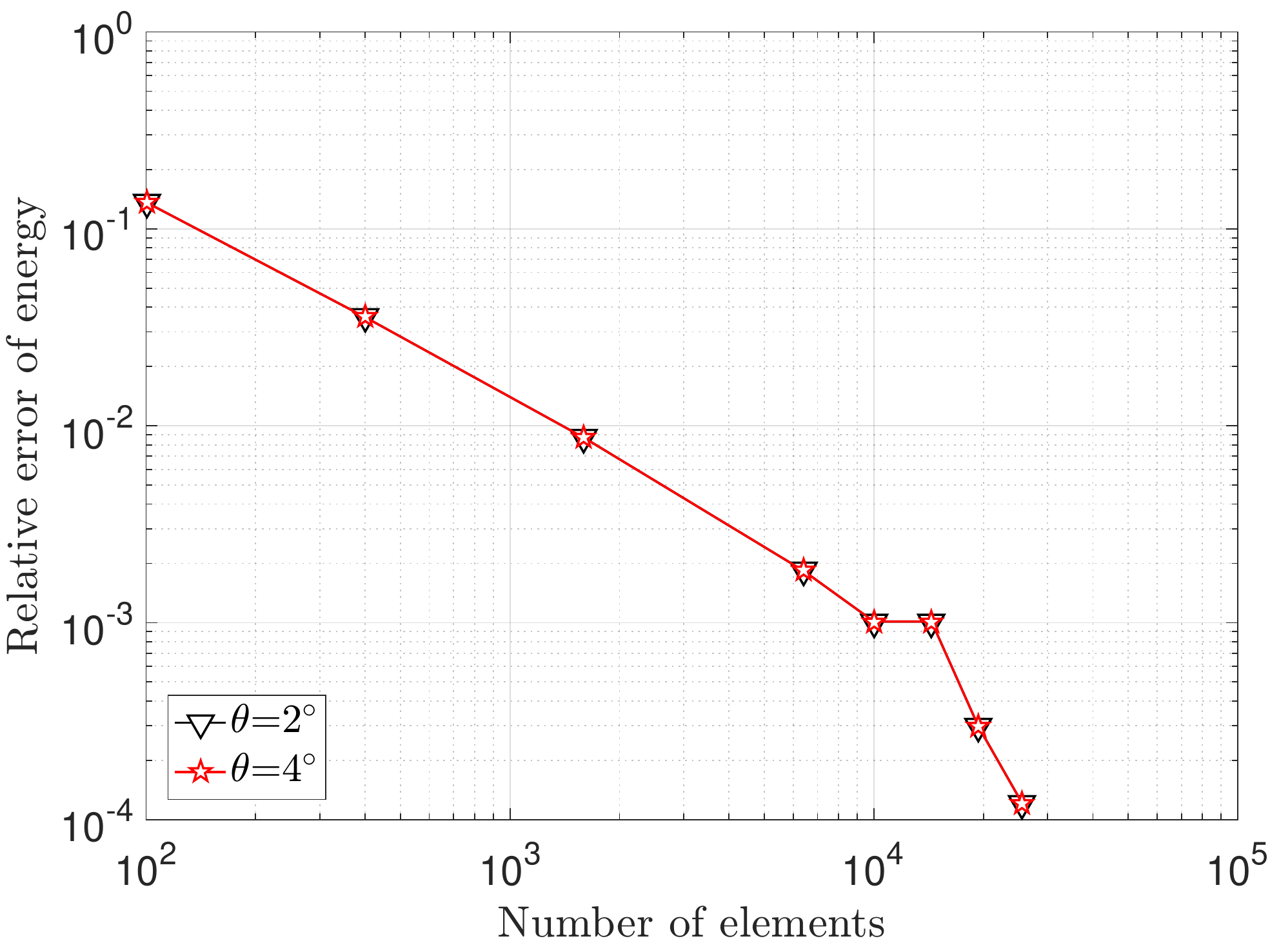}
        \vspace{-7mm}
        \subcaption{}
        \label{f:Torsion_CNC_L12_A60_strainEnergy_conv}
    \end{subfigure}
    \vspace{-3.5mm}
    \caption{CNC twisting: (\subref{f:CNC_torsion_BC}) boundary conditions; (\subref{f:Torsion_CNC_L12_A60_strainEnergy_conv}) Error of strain energy relative to the finest mesh (using $180\times180$ quadratic NURBS elements). CNC($19.2^{\circ}$) with the length and tip radius 12.04 nm and 1 nm is used.}
\end{figure}

\begin{figure}
    \begin{subfigure}{0.49\textwidth}
        \centering
    \includegraphics[height=58mm]{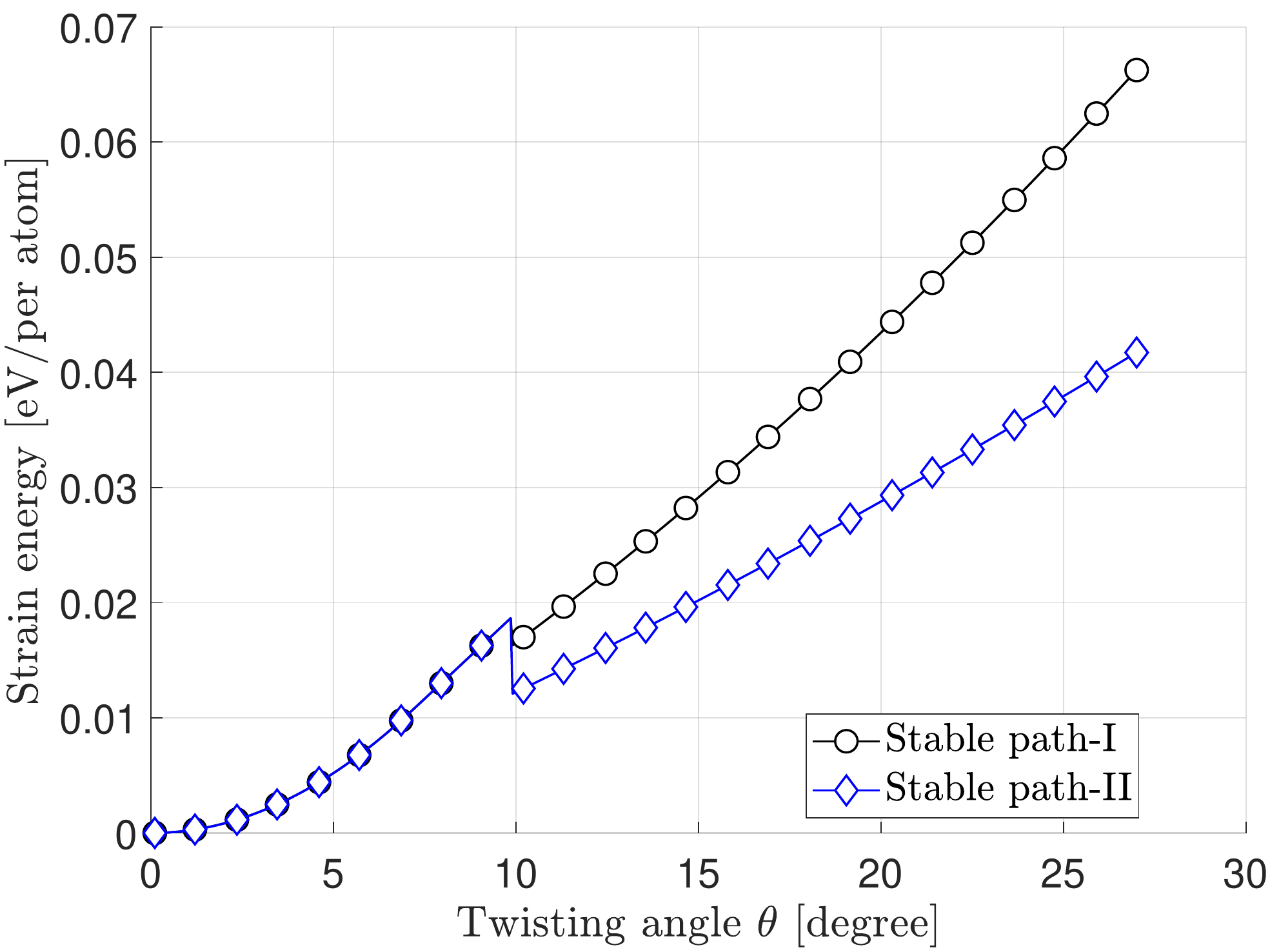}
    \vspace{-7mm}
        \subcaption{}
        \label{f:Torsion_CNC_L12_A60_Energy}
    \end{subfigure}
    \begin{subfigure}{0.49\textwidth}
        \centering
    \includegraphics[height=58mm]{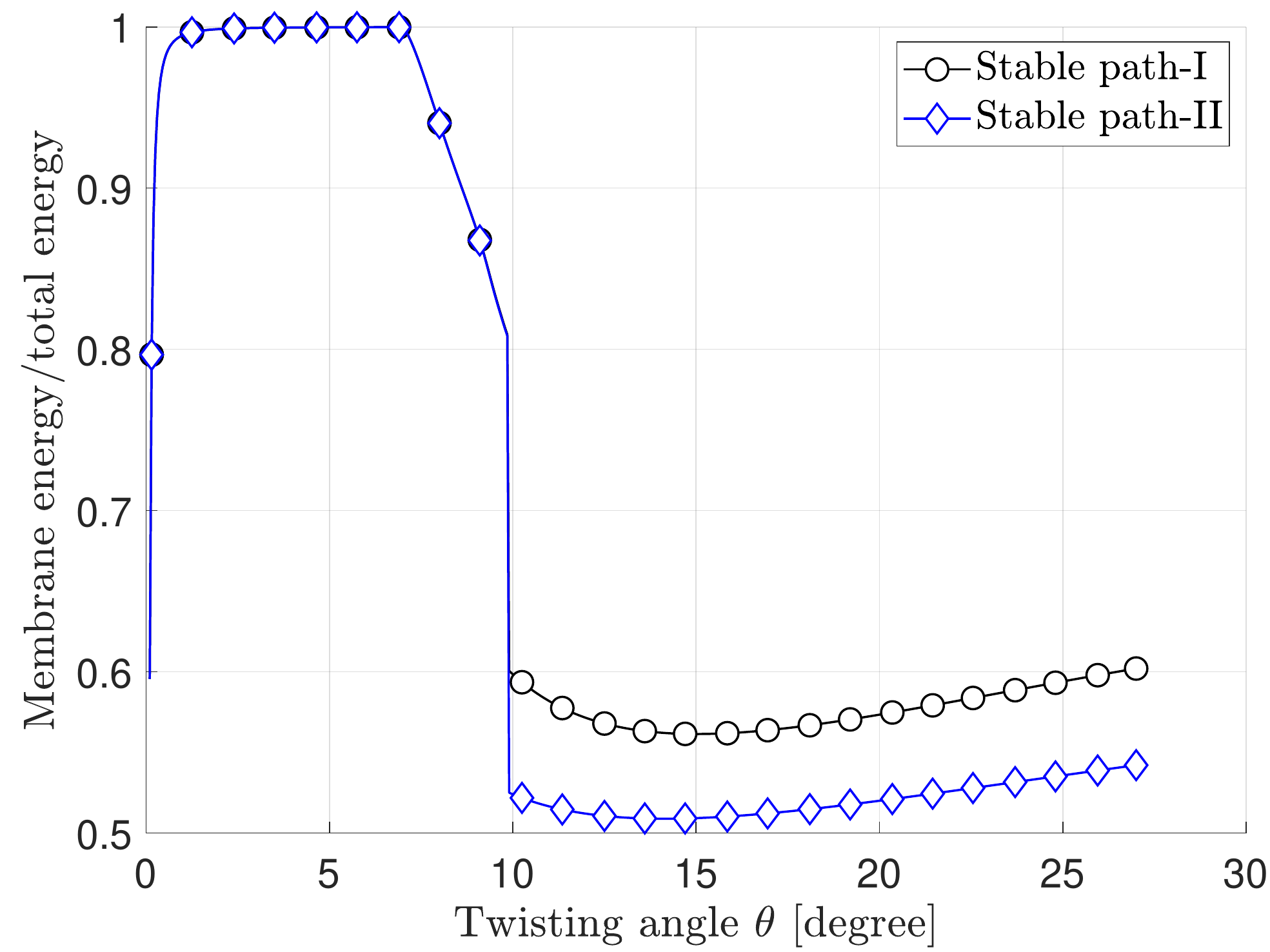}
    \vspace{-7mm}
        \subcaption{}
        \label{f:Torsion_CNC_L12_A60_Energy_ratio}
    \end{subfigure}
    \vspace{-3.5mm}
    \caption{CNC twisting: (\subref{f:Torsion_CNC_L12_A60_Energy}) Strain energy per atom; (\subref{f:Torsion_CNC_L12_A60_Energy_ratio}) the ratio of the membrane energy to the total energy.  CNC($19.2^{\circ}$) with the length and tip radius 12.04 nm and 1 nm is used.}
\end{figure}

\begin{figure}
\begin{center} \unitlength1cm
\begin{picture}(18,5)
\put(1,2.75){\includegraphics[width=50mm,trim=4cm 20cm 2cm 6cm,clip]{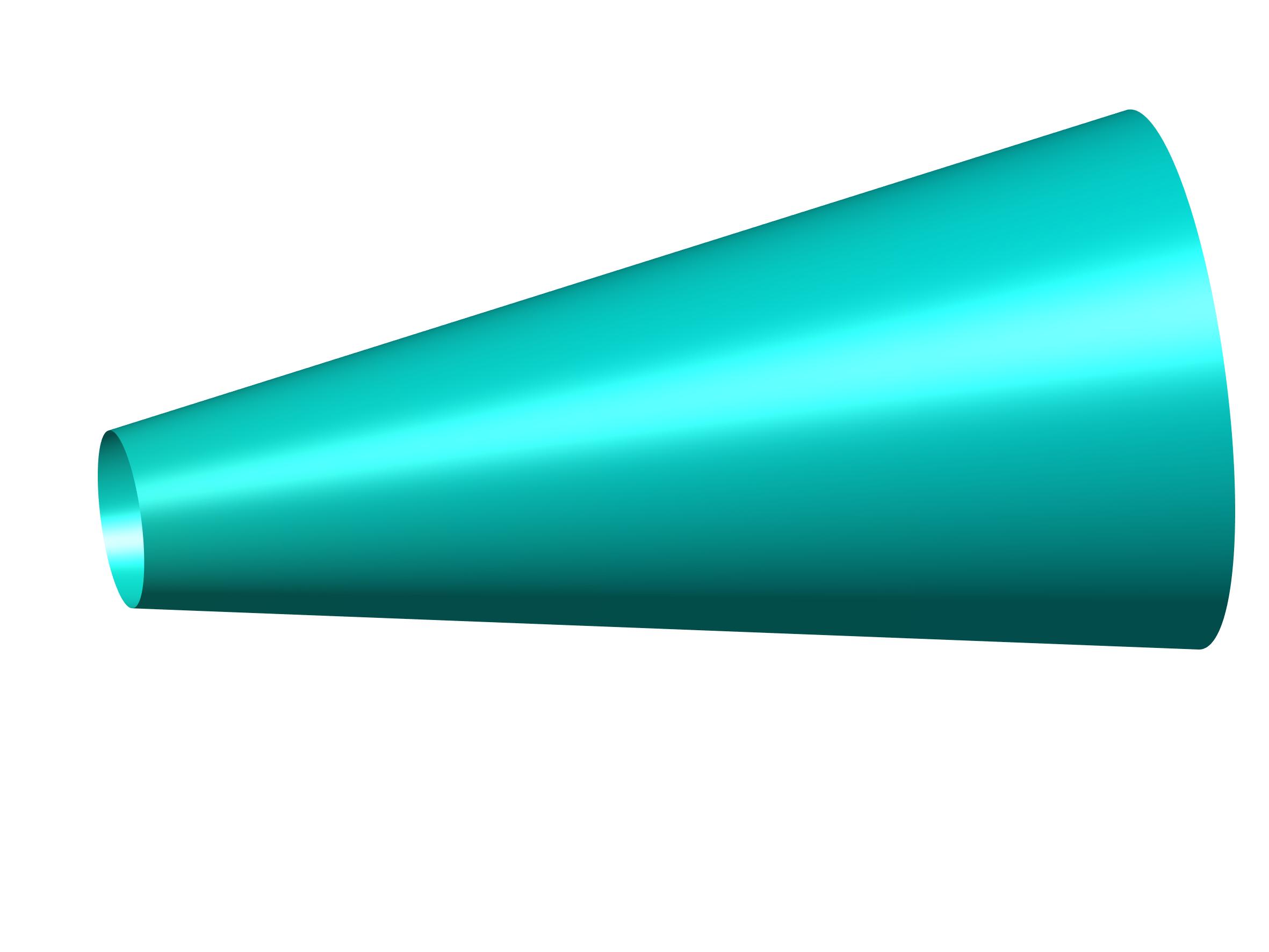}}
\put(6,2.75){\includegraphics[width=50mm,trim=4cm 20cm 2cm 6cm,clip]{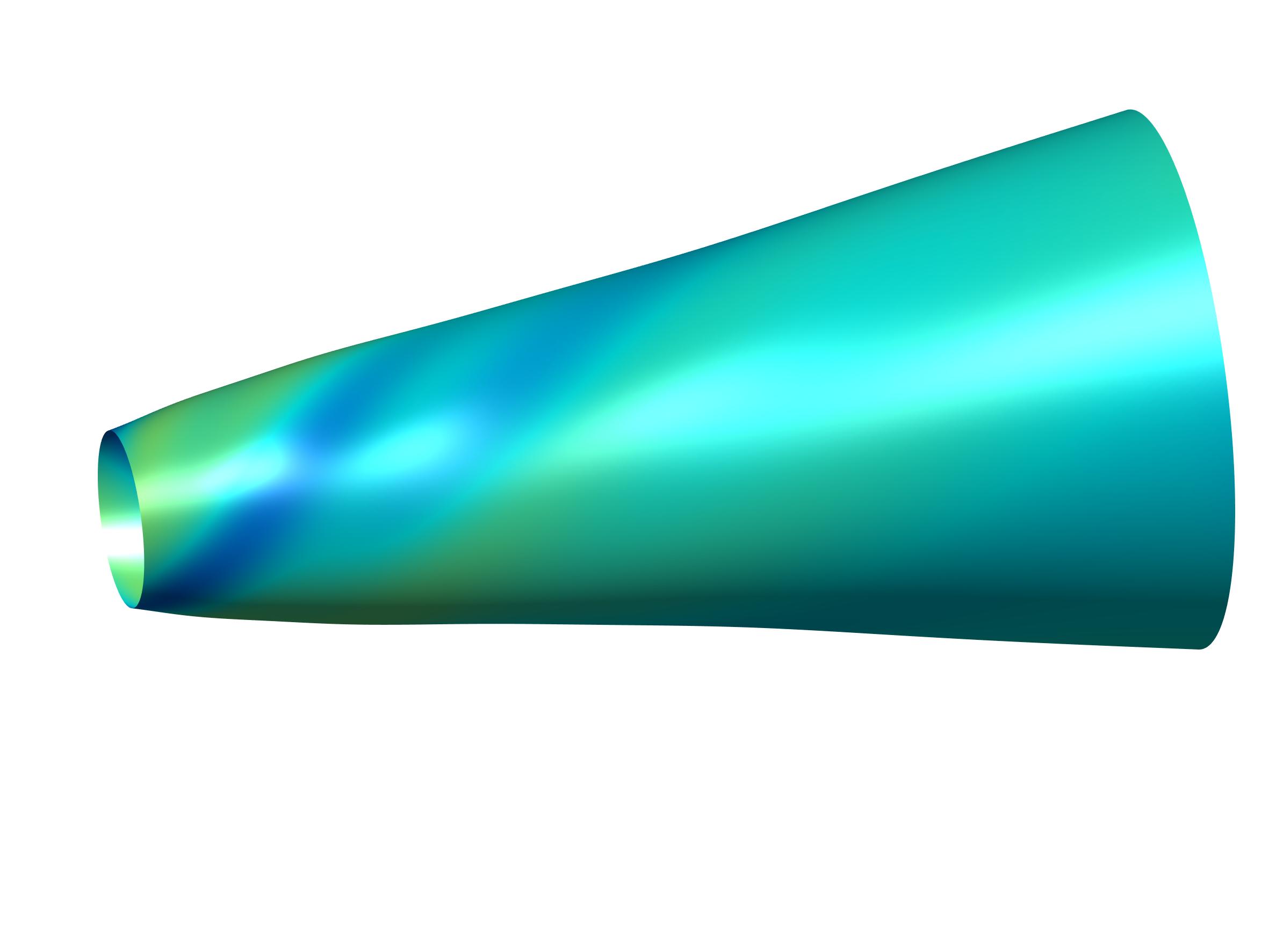}}
\put(11,2.75){\includegraphics[width=50mm,trim=4cm 20cm 2cm 6cm,clip]{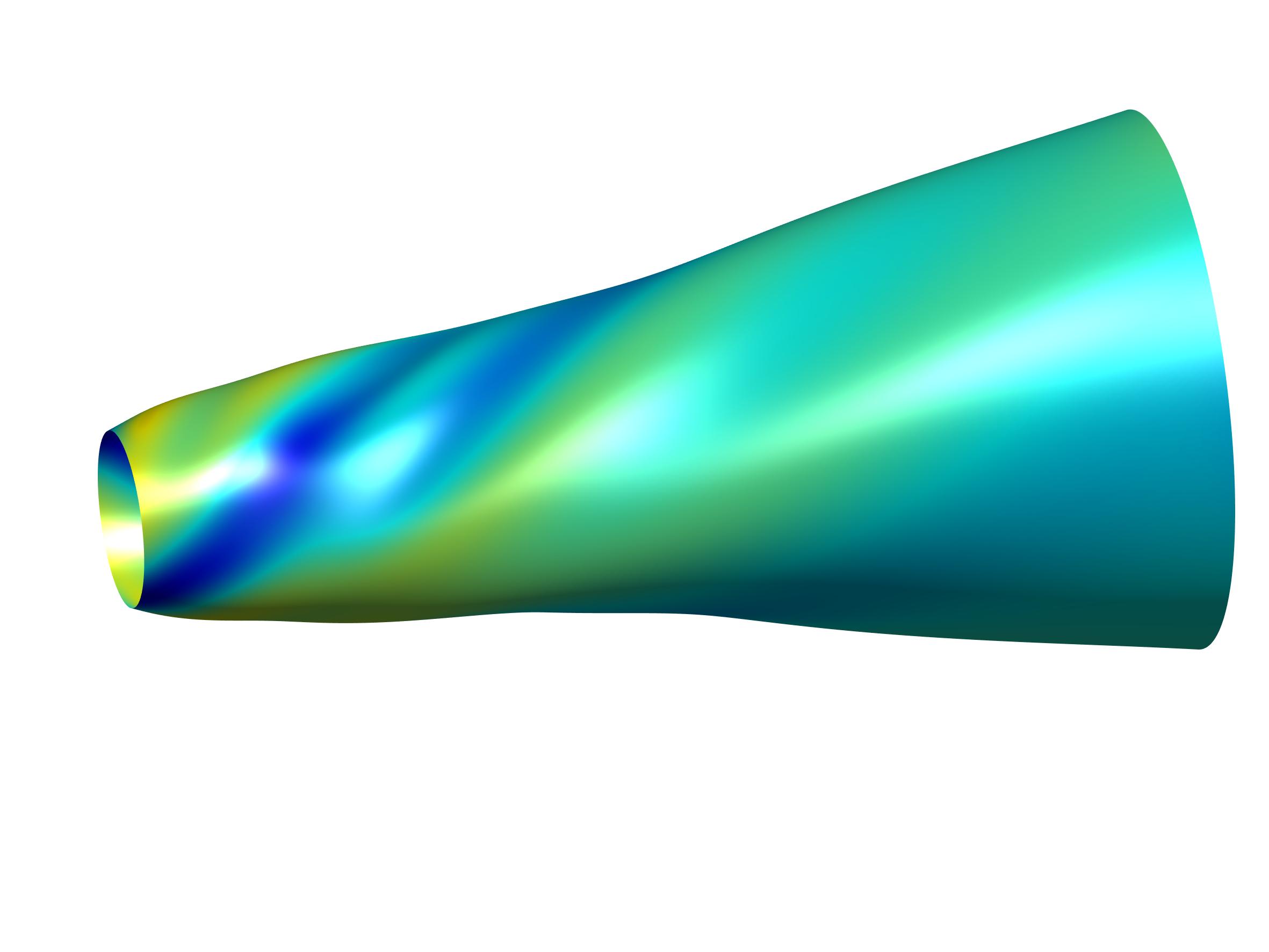}}

\put(1,0){\includegraphics[width=50mm,trim=4cm 20cm 2cm 6cm,clip]{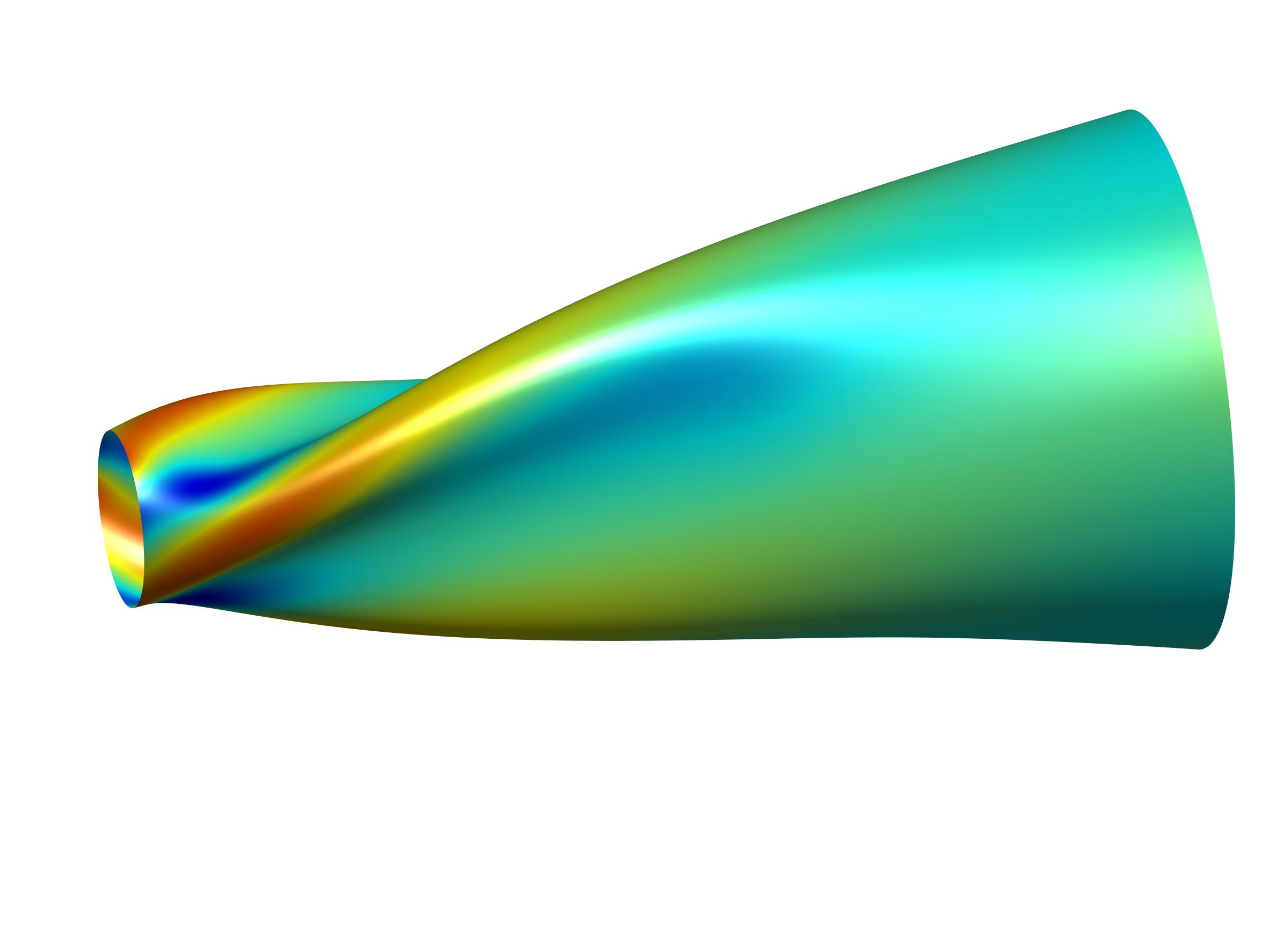}}
\put(6,0){\includegraphics[width=50mm,trim=4cm 20cm 2cm 6cm,clip]{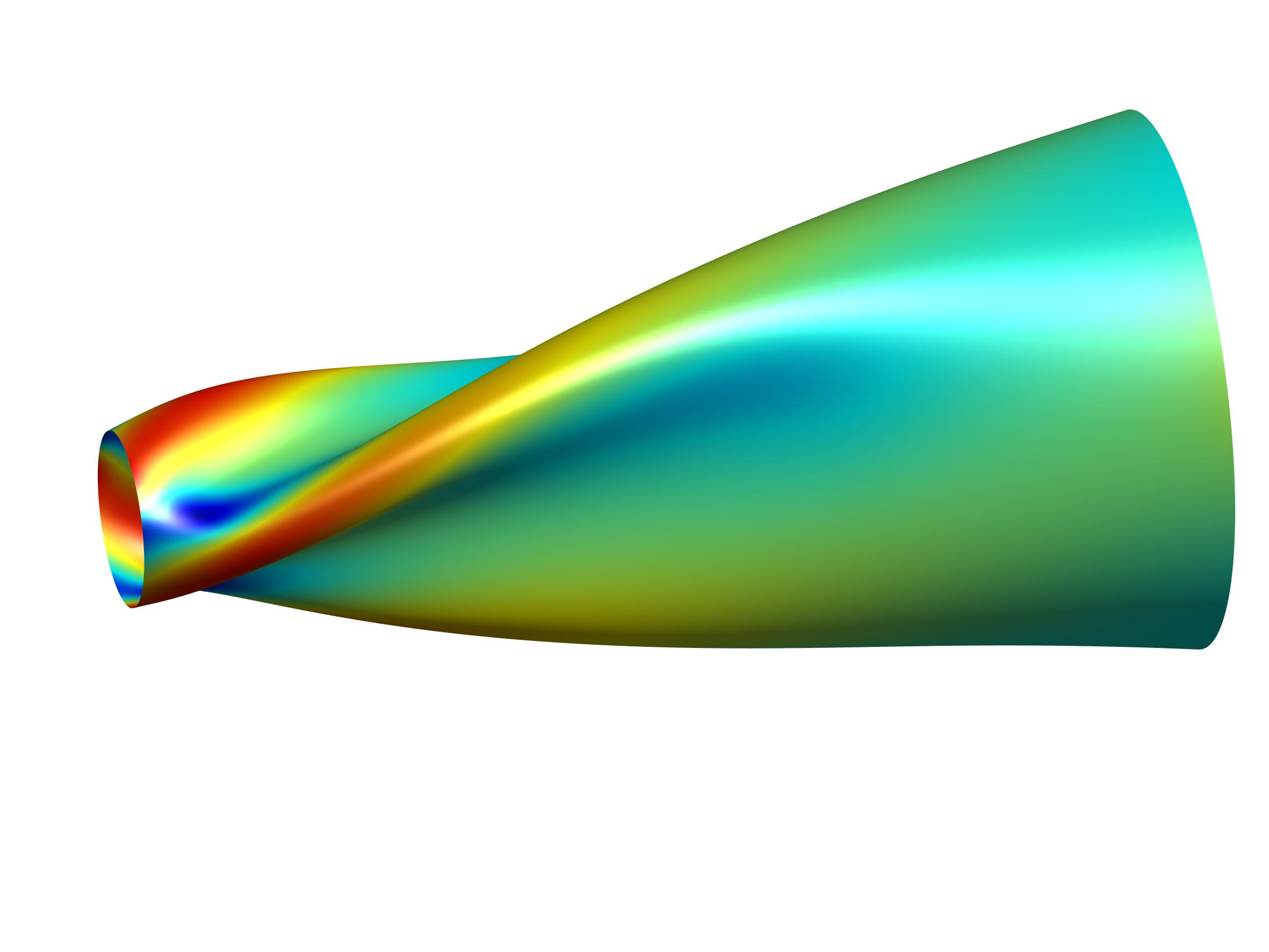}}
\put(11,0){\includegraphics[width=50mm,trim=4cm 20cm 2cm 6cm,clip]{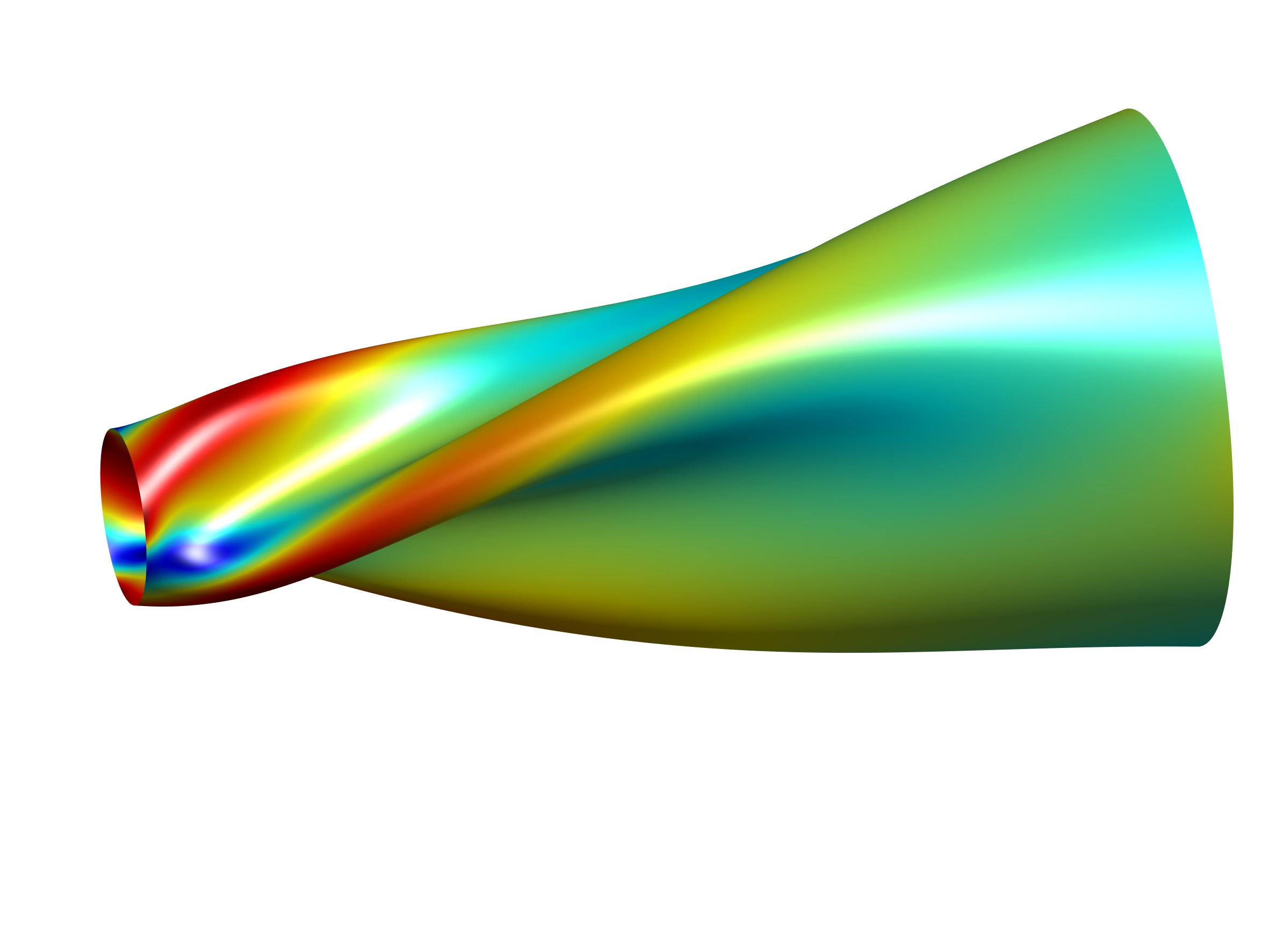}}

\put(1.5,4.25){(a)}
\put(6.5,4.25){(b)}
\put(11.5,4.25){(c)}
\put(1.5,1.5){(d)}
\put(6.5,1.5){(e)}
\put(11.5,1.5){(f)}

\put(0,0){\includegraphics[height=47.5mm]{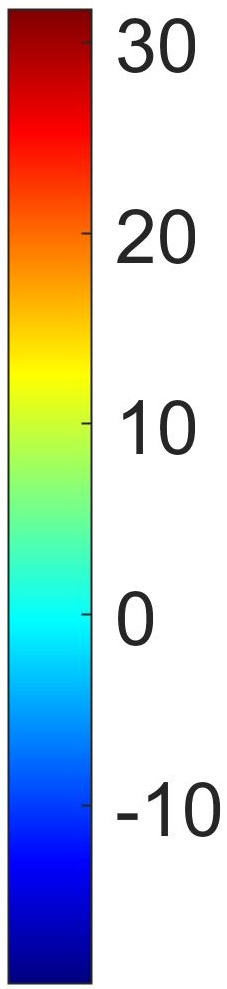}}

\end{picture}
\vspace{-5mm}
\caption{CNC twisting: Comparison of tr($\bsig_{\text{KL}}$) [$\text{N/m}$] at the twisting angles (side view)
(a) $\theta=4^{\circ}$,
(b) $\theta=7.5^{\circ}$,
(c) $\theta=9^{\circ}$ and
(d) $\theta=15^{\circ}$,
(e) $\theta=20^{\circ}$,
(f) $\theta=28.35^{\circ}$. CNC($19.2^{\circ}$) with the length 12.04 nm is used.}
\label{f:CNC_torsion_energy_contours_ISO}
\end{center}
\end{figure}

\begin{figure}
\begin{center} \unitlength1cm
\begin{picture}(18,9)
\put(1.25,5){\includegraphics[width=50mm,trim=10cm 3cm 10cm 0cm,clip]{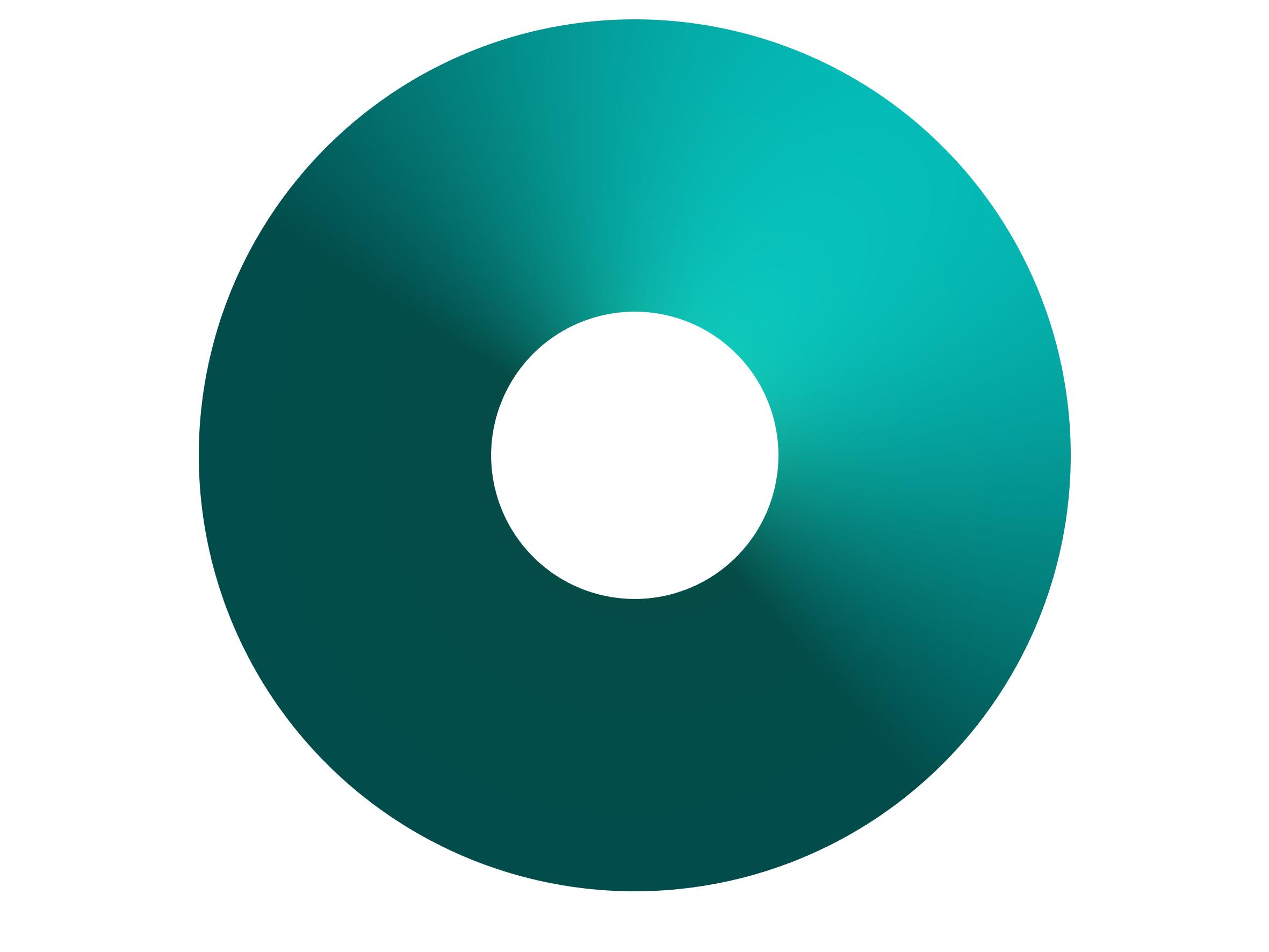}}
\put(6.15,5){\includegraphics[width=50mm,trim=10cm 3cm 10cm 0cm,clip]{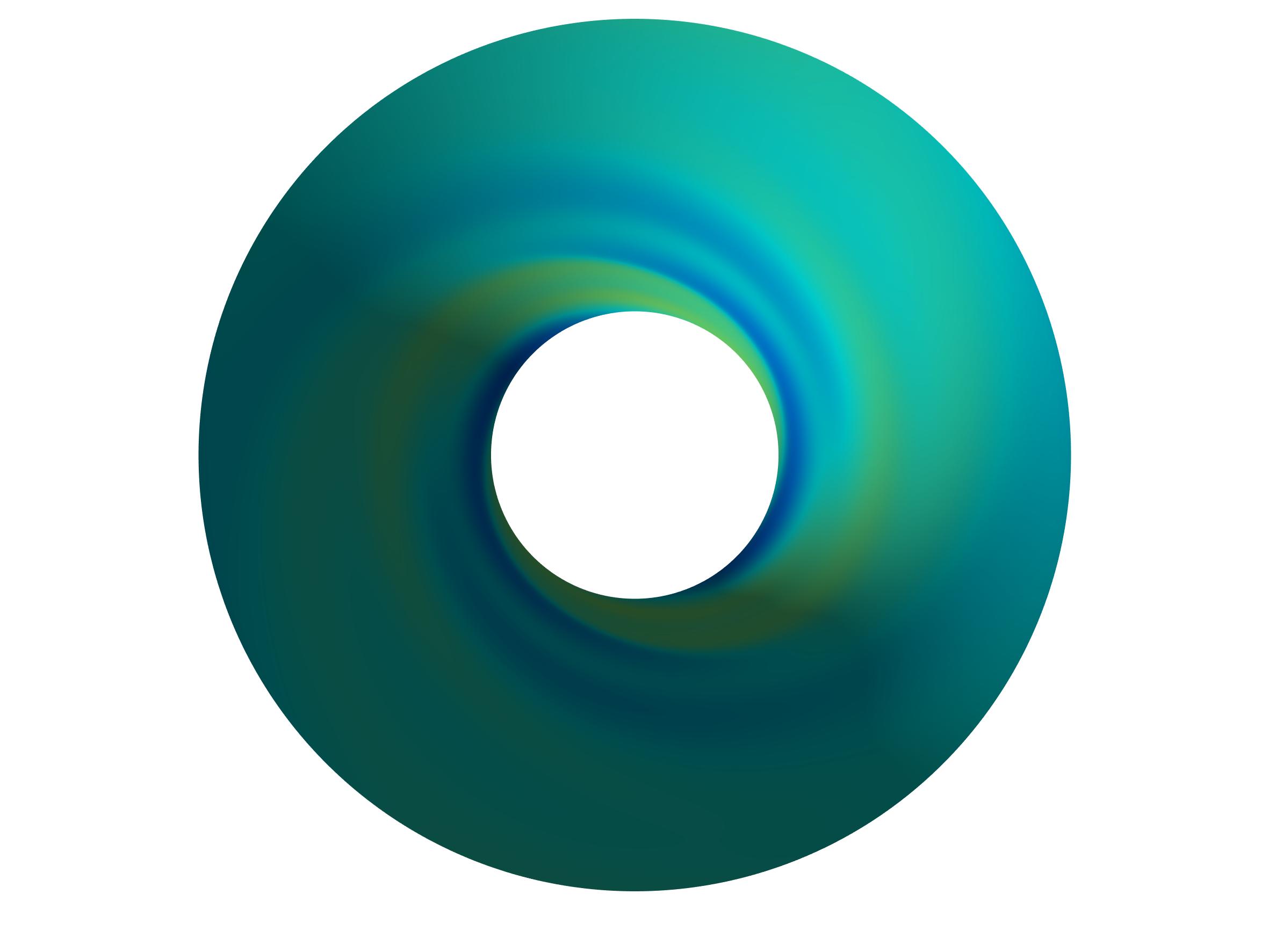}}
\put(11,5){\includegraphics[width=50mm,trim=10cm 3cm 10cm 0cm,clip]{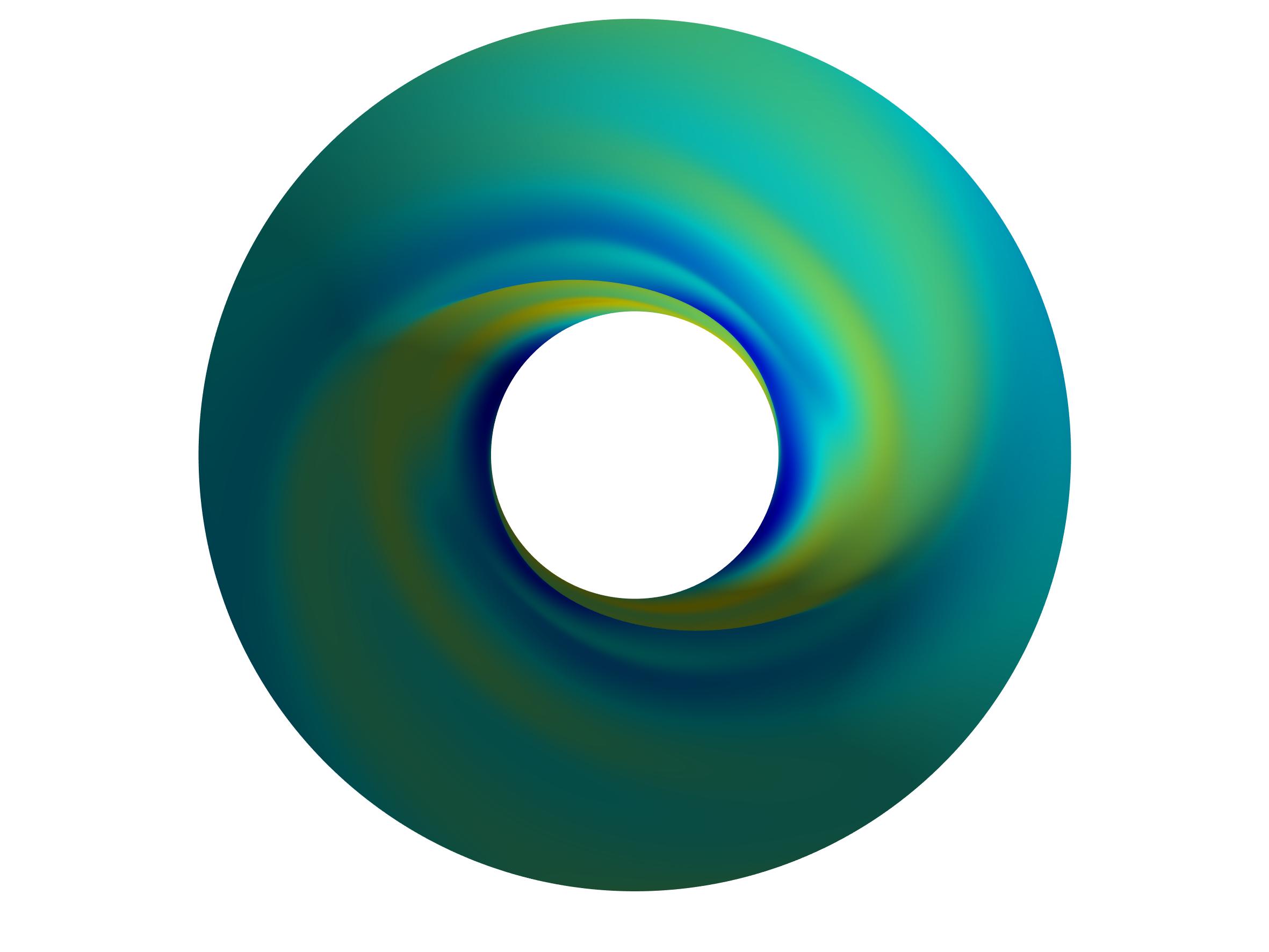}}

\put(1.25,0){\includegraphics[width=50mm,trim=10cm 3cm 10cm 0cm,clip]{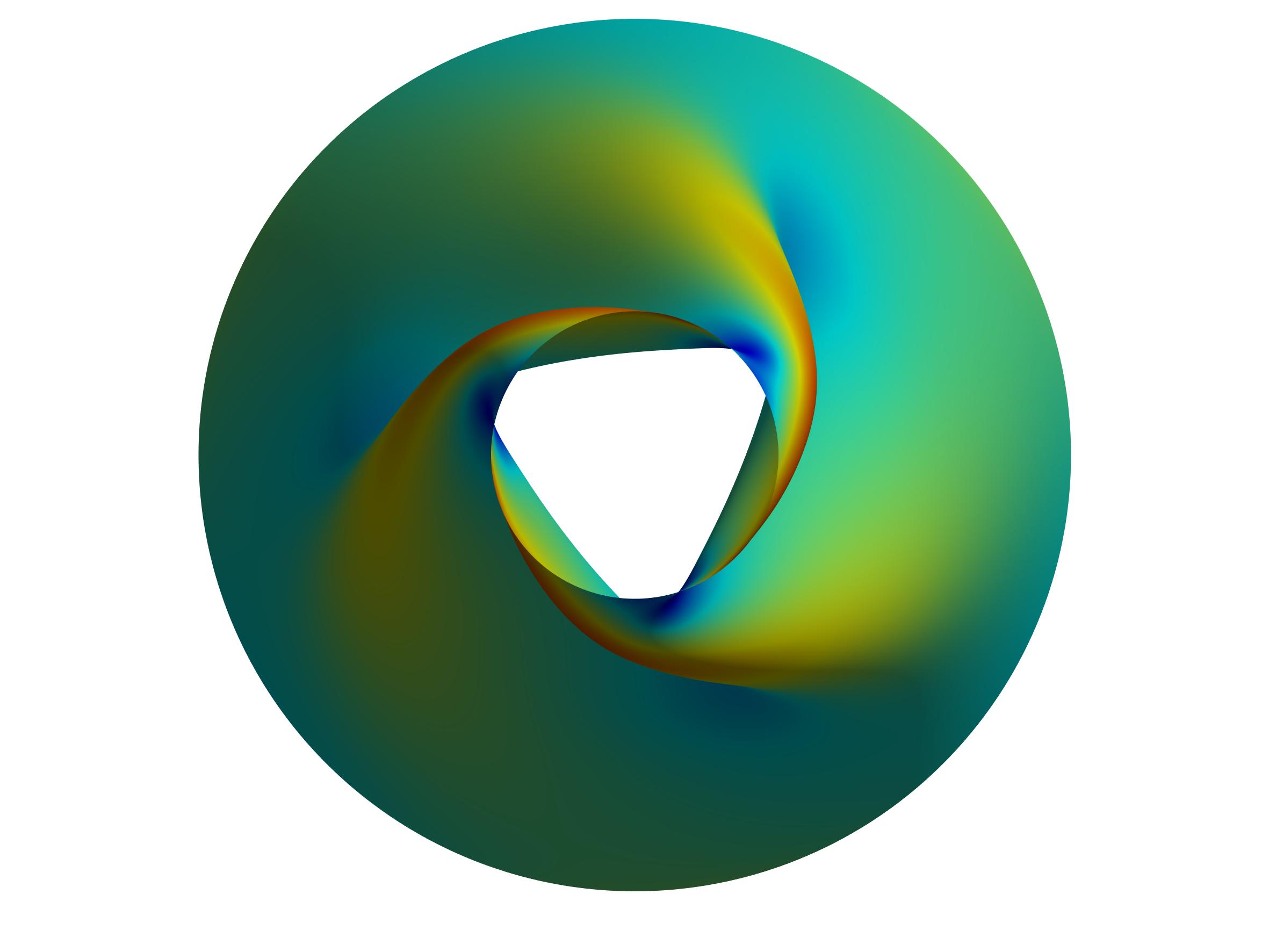}}
\put(6.15,0){\includegraphics[width=50mm,trim=10cm 3cm 10cm 0cm,clip]{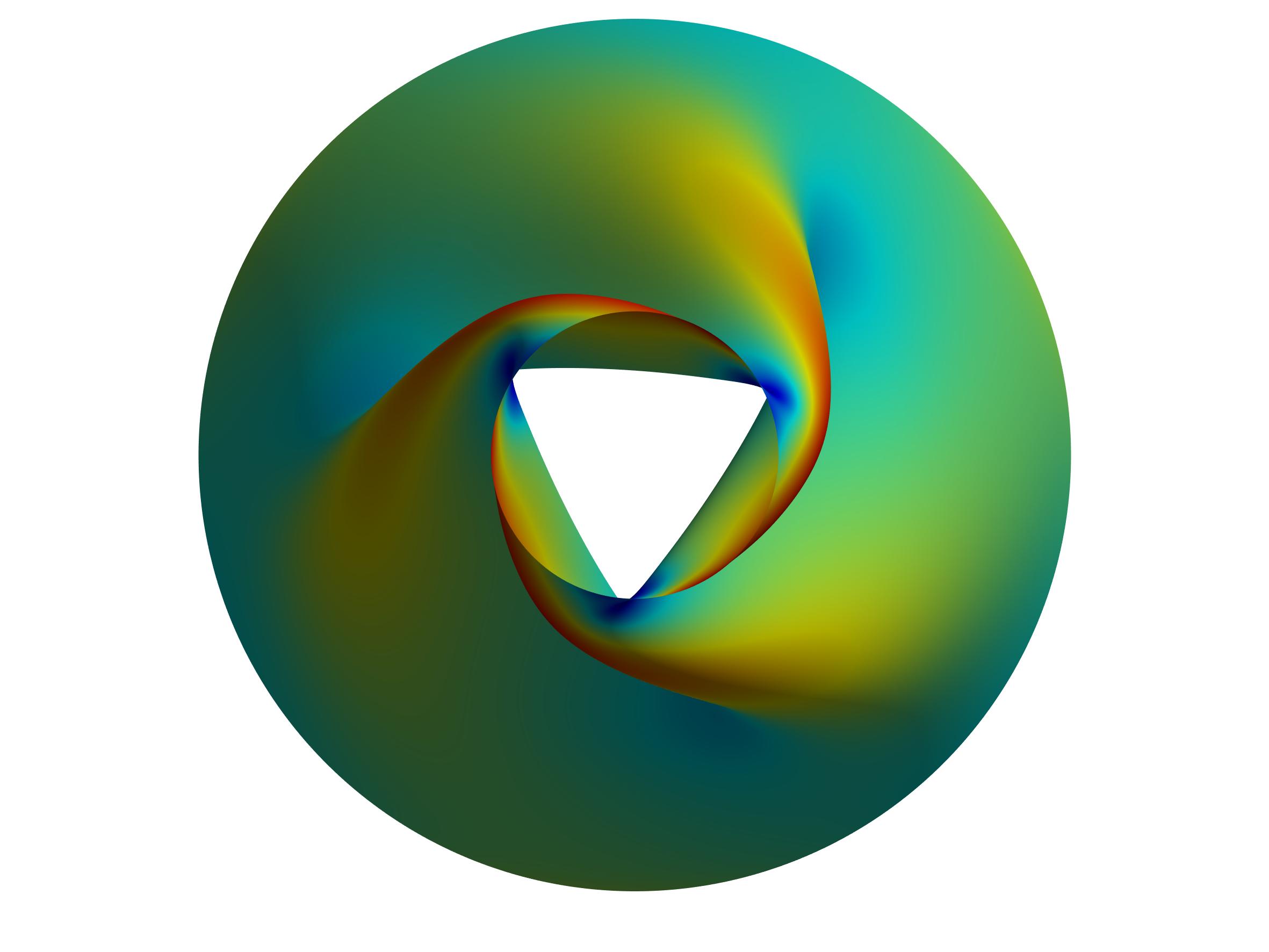}}
\put(11,0){\includegraphics[width=50mm,trim=10cm 3cm 10cm 0cm,clip]{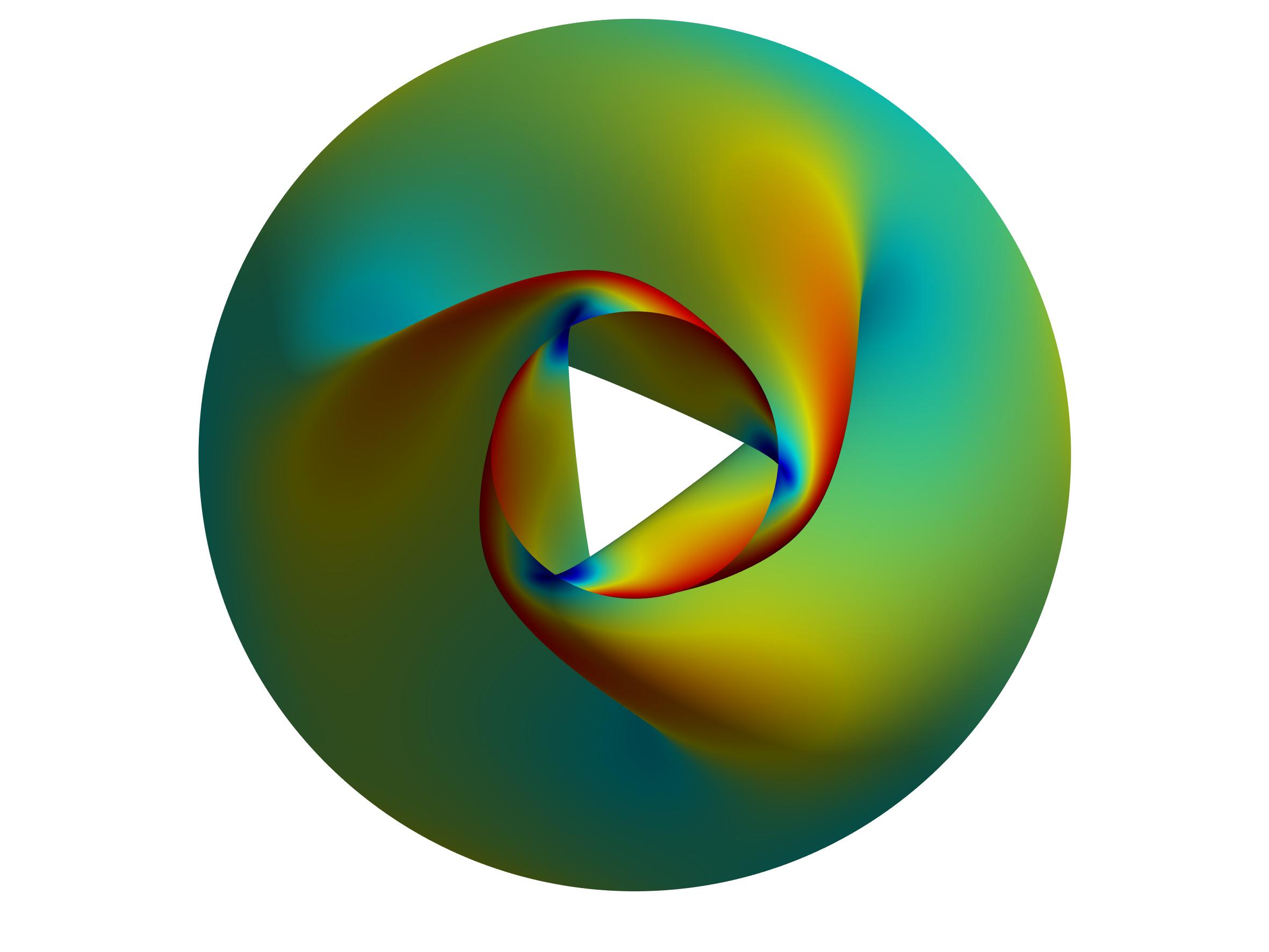}}

\put(1.55,5.3){(a)}
\put(6.6,5.3){(b)}
\put(11.6,5.3){(c)}
\put(1.55,0.25){(d)}
\put(6.6,0.25){(e)}
\put(11.6,0.25){(f)}
\put(0,0){\includegraphics[height=95mm]{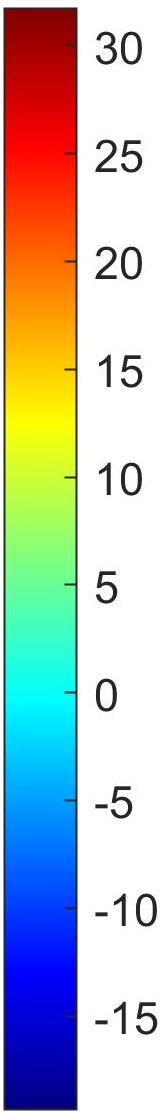}}

\end{picture}
\vspace{-5mm}
\caption{CNC twisting: Comparison of tr($\bsig_{\text{KL}}$) [$\text{N/m}$] at the twisting angles  (front view)
(a) $\theta=4^{\circ}$,
(b) $\theta=7.5^{\circ}$,
(c) $\theta=9^{\circ}$ and
(d) $\theta=15^{\circ}$,
(e) $\theta=20^{\circ}$,
(f) $\theta=28.35^{\circ}$.  CNC($19.2^{\circ}$) with the length and tip radius 12.04 nm and 1 nm is used.}
\label{f:CNC_torsion_energy_contours_front}
\end{center}
\end{figure}

The second example considers CNC twisting. The boundary conditions of twisting are shown in Fig.~\ref{f:CNC_torsion_BC}. The end faces of the CNC is kept rigid, the torsion angle is applied to them equally and its length is kept fix. Fig.~\ref{f:Torsion_CNC_L12_A60_strainEnergy_conv} demonstrates the FE convergence under mesh refinement examining the error measure of Eq.~\eqref{e:error_dif}. \textcolor{cgn}{Quadratic NURBS meshes with $m\times m$ elements, for m = 10, 20, 40, 80, 100, 120, 140, 160, 200, are used for the convergence study. The energy difference between the finest and second finest mesh is about $0.01\%$.}
\noindent
The strain energy per atom and the ratio of the membrane energy to the total energy as a function of the twisting angle are given in Figs.~\ref{f:Torsion_CNC_L12_A60_Energy} and \ref{f:Torsion_CNC_L12_A60_Energy_ratio}. The structure buckles around $\theta=7^{\circ}$ and $\theta=9.85^{\circ}$, and these points can be precisely obtained from the ratio of the membrane energy to the total energy (see Fig.~\ref{f:Torsion_CNC_L12_A60_Energy_ratio}). Two stable paths \textcolor{cgm}{appear} after the second buckling point and the path with the lower level of energy is more favorable (see \citet{wriggers2008} for a discussion of bifurcation \textcolor{cgm}{in FE analysis}). The deformation follows one of the two paths depending on the load step and arclength parameter. Thermal fluctuations at finite temperatures should be sufficient to provide the model with enough energy to overcome the energy barrier between the two paths and go from the higher energy level to the lower one. $\tr(\bsig_{\text{K}})$ is shown at different twisting angles in side and front views in Figs.~\ref{f:CNC_torsion_energy_contours_ISO} and \ref{f:CNC_torsion_energy_contours_front}, respectively. The buckling geometry has rotational symmetry of 180$^{\circ}$ at low twisting angle (below $\theta=10^{\circ}$) and 120$^{\circ}$ at high twisting angle (above $\theta=10^{\circ}$) (see Fig.~\ref{f:CNC_torsion_energy_contours_front}c-d). \\
\noindent
\noindent
\noindent
\subsubsection{CNC contact}\label{s:CNC_contact_wall}
\begin{figure}
    \begin{subfigure}{0.49\textwidth}
        \centering
    \includegraphics[width=80mm,trim=5cm 12cm 0cm 5cm,clip]{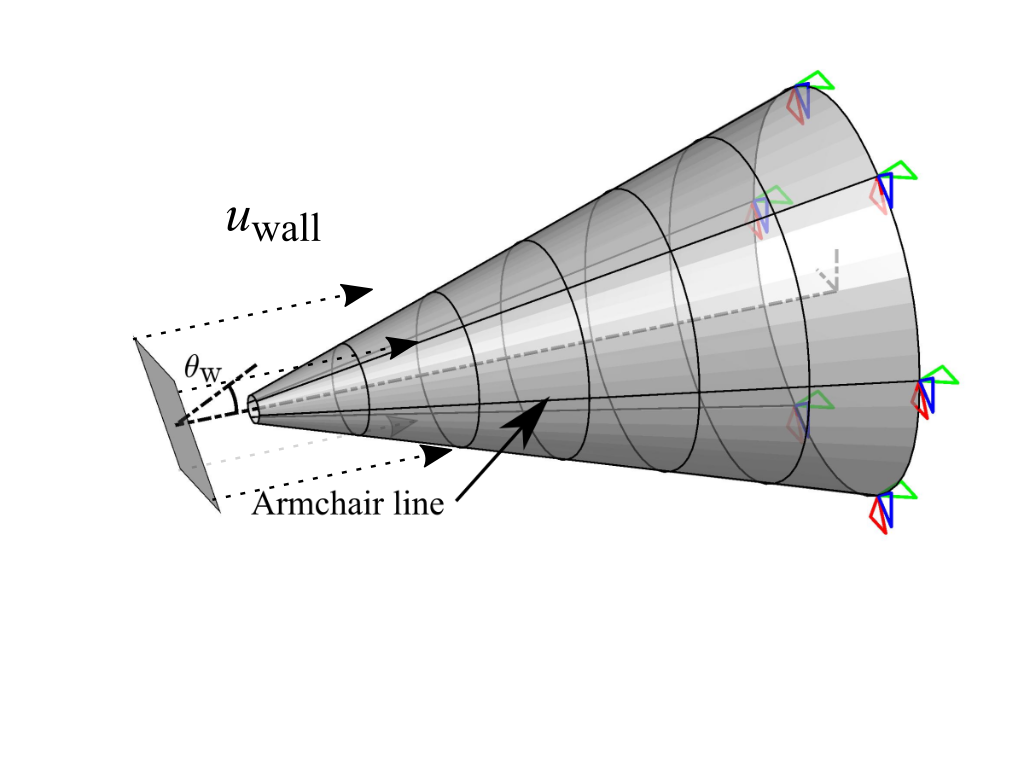}
        \subcaption{}
        \label{f:BC_CNC_LJ_wall}
        \vspace{-13mm}
    \end{subfigure}
    \begin{subfigure}{0.49\textwidth}
        \centering
    \includegraphics[width=80mm,trim=25cm 12cm 15cm 0cm,clip]{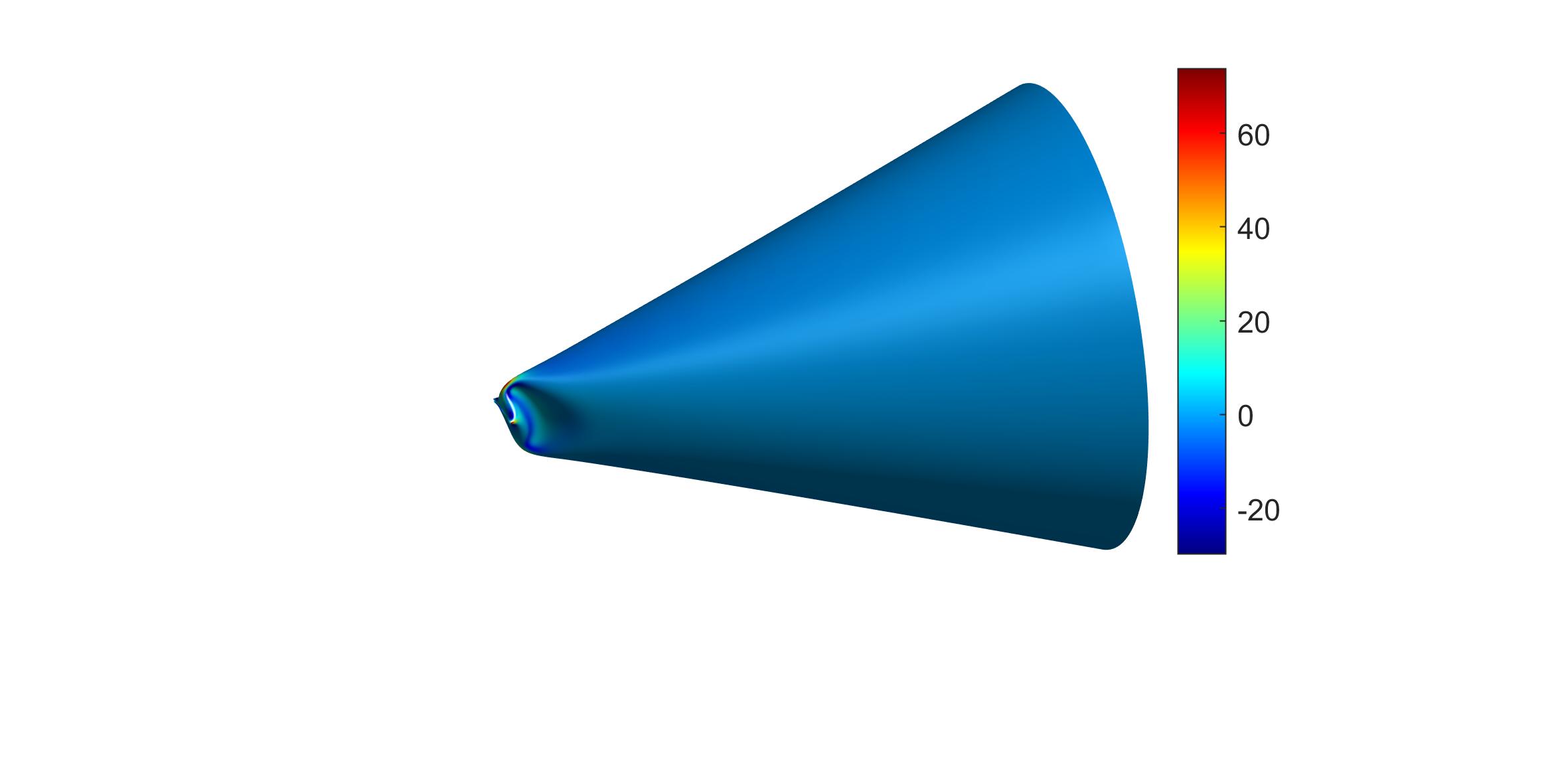}
        \subcaption{}
        \label{f:LJ_wall_CNC_L38_19}
    \end{subfigure}
    \vspace{-3.5mm}
    \caption{Contact of a CNC with a Lennard-Jones wall: (\subref{f:BC_CNC_LJ_wall}) boundary conditions; (\subref{f:LJ_wall_CNC_L38_19}) deformed geometry colored with tr($\bsig_{\text{KL}}$) [$\text{N/m}$]. CNC($38.94^{\circ}$) with the length 38.19 nm and the contact angle $\theta_{\text{w}}=17.45^{\circ}$ are used.}
\end{figure}

\begin{figure}
    \begin{subfigure}{1\textwidth}
        \centering
    \includegraphics[height=58mm]{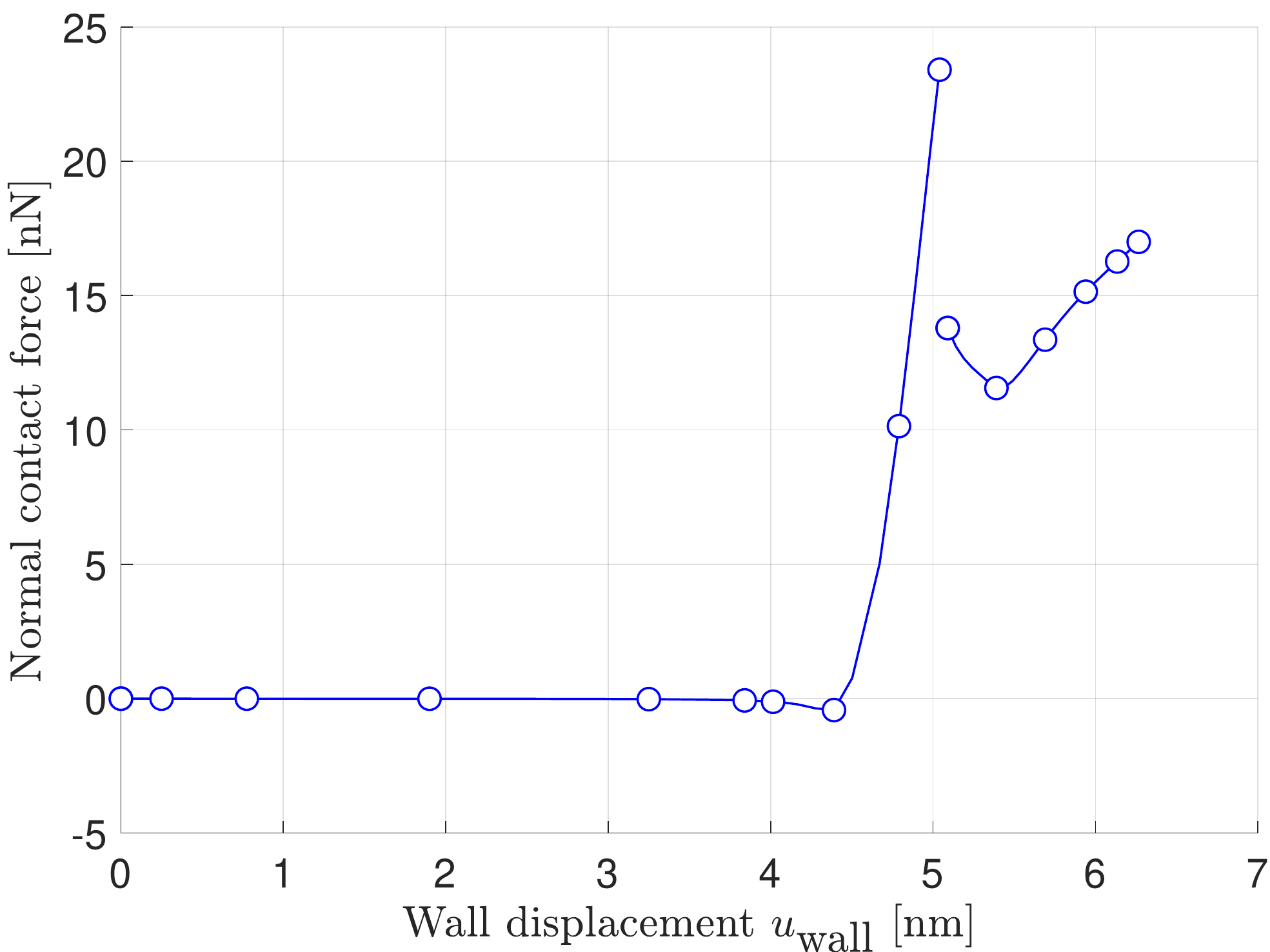}
    \end{subfigure}
    \vspace{-2.5mm}
    \caption{Contact of a CNC with a Lennard-Jones wall: Normal contact force of the wall. CNC($38.94^{\circ}$) with the length 38.19 nm and $\theta_{\text{w}}=17.45^{\circ}$ is used. The data between markers is continuous.}\label{f:CNC_LJ_wall_reaction}
\end{figure}

CNTs are good candidates for AFM tips, but they buckle fast due to their high aspect ratio. Therefore, CNCs are better candidates for AFM tips \citep{Chen2006_01,Huang2016_01}. To illustrate this point, contact between a CNC and a rigid wall is considered here. In order to compare CNC and CNT, the length and tip radius $r_{\text{tip}}$ of the CNC are selected such that they are equal to the length and radius of the CNT used before. In this case they would be able to measure with the same resolution. The boundary conditions and the deformed geometry of the CNC are shown in Figs.~\ref{f:BC_CNC_LJ_wall} and \ref{f:LJ_wall_CNC_L38_19}, respectively. The CNC is simply supported and the wall is rigid and moves in the axial direction of the CNC. The normal contact force is given in Fig.~\ref{f:CNC_LJ_wall_reaction}. It reaches about 23.41 nN at the buckling point. The buckling force of the CNC is several times larger than the buckling force of the CNT (see Tab.~\ref{t:CNC_CNT_LJ_wall}). The contact force decreases sharply after the buckling point. Then, it decreases smoothly to a local minimum and begins to increase after that minimum. The buckling force of the CNC is 18.44 times larger than for the CNT. \textcolor{cgn}{$250\times150$ quadratic NURBS elements are used in this study.}
\noindent
\begin{table}[h]
  \centering
    \begin{tabular}{l c c }
      \hline
        & CNT & CNCs    \\
      \hline
      Normal contact force [nN] &  1.27 & 23.40\\
      \hline
    \end{tabular}
      \caption{Comparison of the buckling force for the CNT and CNC considered in Secs.~\ref{s:CNT_contact_wall} and \ref{s:CNC_contact_wall} (see Fig.~\ref{f:LJ_wall_reaction} and Fig.~\ref{f:CNC_LJ_wall_reaction}), respectively.}\label{t:CNC_CNT_LJ_wall}
\end{table}
\section{Conclusion}\label{s:conclusion}
A new hyperelastic material model is proposed for graphene-based structures. The symmetry group and the structural tensor of graphene are used to obtain a set of invariants. These invariants are directly based on the right surface Cauchy-Green deformation tensor $\bC$, so its derivatives with respect to $\bC$ can be taken easily. The first and second invariants capture pure dilatation and shear, while the third one captures anisotropic behavior. This model is based on existing quantum data \citep{Kudin2001_01,Kumar2014_01}. The speedup of the model is 1.5 compared to the earlier model of \citet{Ghaffari2017_01}. Further, it is simpler to implement than the model of \citet{Ghaffari2017_01}. The material model is formulated such that it can be easily implemented within the rotation-free isogeometric shell formulation of \citet{Duong2016_01}. The elementary behavior of the new model is validated by uniaxial tension and pure shear tests. The strain energy of CNTs and CNCs under bending and twisting are computed. The buckling points are calculated by examining the ratio of membrane energy to total energy. The postbuckling behavior of CNTs and CNCs are simulated. The modified arclength method of \citet{Ghaffari2015} and a line search method are used to obtain convergence, and the finite element formulation fails to converge for some of the examples without these methods. CNCs buckle even without applying an imperfection due to inherent anisotropy along the different tangential coordinates. CNTs and CNCs can be used for an atomic force microscope (AFM) tip. Contact of a CNT and CNC with a Lennard-Jones wall is simulated. The reaction forces are computed and it is shown that loading and unloading paths are different for contact of the CNT with the Lennard-Jones wall. A CNT and CNC with the same tip radius are selected so they would have the same measurement precision. It is shown that the buckling force of the CNC is 18.44 times larger than the buckling force of the CNT. Hence, CNCs are much better candidates for AFM tips than CNTs.\\
\textcolor{cgn}{The proposed model is obtained from recent ab-intio results and is thus very accurate. MD and multiscale methods based on the first and second Brenner potential on the other hand usually underestimate the elastic modulus by one-third (see \citet{Cao2014_01} and \citet{Ghaffari2017_01} for a comparison of elastic moduli obtained from different potentials). More accurate potentials such as MM3 and REBO+LJ should be used to resolve this inaccuracy. This will be considered in future work.}
\section*{Acknowledgment}{Financial support from the German Research Foundation (DFG) through grant GSC 111 is
gratefully acknowledged.}
\appendix
\section{Kinematics of deforming surfaces}\label{s:curvilinear_desc}
Here, the curvilinear description of deforming surfaces is summarized following \cite{Sauer2017_01}. The surface in the reference and the current configuration can be written as
\eqb{l}
\bX = \ds \bX(\xi^{\alpha})~,
\eqe
\eqb{l}
\bx = \ds \bx(\xi^{\alpha})~,
\eqe
where $\xi^{\alpha}$ ($\alpha=1,2$) are the parametric coordinates. The tangent vectors of the reference and current configuration are
\eqb{l}
\bA_{\alpha} = \ds \pa{\bX}{\xi^{\alpha}}~,
\eqe
\eqb{l}
\ba_{\alpha} = \ds \pa{\bx}{\xi^{\alpha}}~.
\eqe
The co-variant components of the metric tensors are defined by using the inner product as
\eqb{l}
\Auab = \bA_{\alpha}\cdot\bA_{\beta}~,
\eqe
\eqb{l}
\auab = \ba_{\alpha}\cdot\ba_{\beta}~.
\eqe
The contra-variant components of the metric tensors are defined as
\eqb{l}
[\Aab] = [\Auab]^{-1}~,
\eqe
\eqb{l}
[\aab] = [\auab]^{-1}~.
\eqe
The dual tangent vectors can then be defined as
\eqb{lll}
\bA^{\alpha} \dis \Aab\,\bA_{\beta}~,
\eqe
\eqb{lll}
\ba^{\alpha} \dis \aab\,\ba_{\beta}~.
\eqe
The normal unit vector of the surface in the reference and current configuration can be obtained by using the cross product of the tangent vectors as
\eqb{l}
\bN =\ds \frac{\bA_{1}\times\bA_{2}}{\|\bA_{1}\times\bA_{2}\|}~,
\eqe
\eqb{l}
\bn =\ds \frac{\ba_{1}\times\ba_{2}}{\|\ba_{1}\times\ba_{2}\|}~.
\eqe
The 3D identity tensor $\bolds{1}$ can be then written as
\eqb{l}
\bolds{1} = \bI+\bN\otimes\bN = \bi+\bn\otimes\bn~,
\eqe
where $\bI$ and $\bi$ are the surface identity tensor in the reference and current configuration. They are
\eqb{l}
\bI = \bA_{\alpha}\otimes\bA^{\alpha} = \Auab\,\bA^{\alpha}\otimes\bA^{\beta} = \Aab\,\bA_{\alpha}\otimes\bA_{\beta}~,
\eqe
\eqb{l}
\bi =\ba_{\alpha}\otimes\ba^{\alpha}  = \auab\,\ba^{\alpha}\otimes\ba^{\beta} = \aab\,\ba_{\alpha}\otimes\ba_{\beta}~.
\eqe
The co-variant components of the curvature tensor are defined as
\eqb{l}
\buab := \bn\cdot\ba_{\alpha,\beta}=\bn\cdot\ba_{\alpha;\beta}~,
\eqe
where $\ba_{\alpha,\beta}$ and $\ba_{\alpha;\beta}$ are the parametric and co-variant derivatives of the tangent vectors. They are connected by
\eqb{l}
\ba_{\alpha;\beta} = \ba_{\alpha,\beta} -\Gamma^{\gamma}_{\alpha\beta}\,\ba_{\gamma}~,
\eqe
where $\Gamma^{\gamma}_{\alpha\beta}$ is the Christoffel symbol of the second kind. It is defined as
\eqb{lll}
\Gamma^{\gamma}_{\alpha\beta}\ \dis \ba_{\alpha,\beta}\cdot\ba^{\gamma}~.
\eqe
The mixed and contra-variant components of the curvature tensor, $b_{\alpha}^{\beta}$ and $\bab$, are defined as
\eqb{lll}
b_{\alpha}^{\beta} \is b_{\alpha\gamma}\,a^{\gamma\beta}~,\\[3mm]
\bab \dis \aag\,b_{\gamma\delta}\,\adb~.
\eqe
The mean and Gaussian curvatures are defined as
\eqb{l}
H := \ds \frac{1}{2}b^{\alpha}_{\alpha}=\frac{1}{2}(\kappa_1+\kappa_2)~,
\eqe
\eqb{l}
\kappa := \ds \frac{\det[\buab]}{\det[\augd]}=\kappa_1\,\kappa_2~,
\eqe
where $\kappa_{\alpha}$ are the principal curvatures that follow as the eigenvalues of matrix $\left[b_{\alpha}^{\beta}\right]$.\\
\section{Symmetry group and material invariants}\label{s:symmetry_group_matraial_invatriant}
In this section, the structural tensor for structures with $n$-fold rotational symmetry are reviewed. Two set of invariants are introduced by using the introduced structural tensor. They are based on the logarithmic strain tensor, \textcolor{cgm}{following} \citet{Kumar2014_01}, and a new set based on the right Cauchy-Green deformation tensor.
\subsection{Structural tensor of $C_{nv}$}
A 2D structure can be modeled based on its lattice structure and symmetry group. A symmetry group is a certain type of operations that leave the lattice indistinguishable from its initial configuration \citep{tadmor_2011_01}. The usual operations are identity mapping, inversion, rotation and reflection. They can be used to formulate structural tensors for the modeling of anisotropic materials \citep{Zheng_1993_01}. The structural tensor\textcolor{cgm}{s} of a lattice with symmetry group of $n$-fold rotational symmetry and additional reflection plane $C_{nv}$ are given as \citep{Zheng_1993_01}
\eqb{lll}
\ds \bbH_{n}\dis \Re\left[(\hat{\bx}+i\hat{\by})^{(n)}\right]=
\left\{ {\begin{array}{*{20}{l}}
{\Re\left[\left(\hat{\bM}+i\hat{\bN}\right)^{(m)}\right];~~n=2m}~,\\[5mm]
{\Re\left[\left(\hat{\bx}+i\hat{\by}\right)\otimes\left(\hat{\bM}+i\hat{\bN}\right)^{(m)}\right];~~n=2m+1~,}
\end{array}} \right.
\eqe
where $\hat{\bx}$ and $\hat{\by}$ are two orthonormal vectors (see Fig.~\ref{f:lattice}) and at least one of them is in the crystal symmetry plane, $i$ is the imaginary unit number, $(\bullet)^{(n)}:=(\bullet)\otimes(\bullet)...(\bullet)$ is the tensor product of $n$ times, $m$ is a integer number, $\Re$ indicates real part of its argument, and $\hat{\bM}$ and $\hat{\bN}$ are defined as
\eqb{lll}
  \hat{\bM} \dis \hat{\bx}\,\otimes\,\hat{\bx}-\hat{\by}\,\otimes\,\hat{\by}~,\\[2mm]
  \hat{\bN} \dis \hat{\bx}\,\otimes\,\hat{\by}+\hat{\by}\,\otimes\,\hat{\bx}~.
  \label{e:tensors_M_N}
\eqe
Graphene has a hexagonal lattice and its structure repeats after $60^{\circ}$ of rotation. This lattice relates to the symmetry group of $C_{nv}$ and has 14 symmetry operations\textcolor{cgm}{,} which are an identity mapping, an inversion, six mirror planes and six rotations. $\hat{\bx}$ taken in the armchair direction (see Fig.~\ref{f:lattice}) and $n=6$ for graphene. Directions of $\hat{\bx}$ and $\hat{\by}$ can be transformed to $\hat{\bx}^{\diamond}$ and $\hat{\by}^{\diamond}$ by a rotation of $\theta$ in the counter-clock-wise direction such that
\eqb{lll}
\ds  \hat{\bx}^{\diamond}+i\,\hat{\by}^{\diamond}\is e^{-i\,\theta}\left(\hat{\bf{x}}+i\,\hat{\bf{y}}\right)~,\\[2mm]
 \hat{\bM}^{\diamond}+i\,\hat{\bN}^{\diamond}\is \ds e^{-2\,i\,\theta}\left(\hat{\bM}+i\,\hat{\bN}\right)~,\\[2mm]
\ds  \bbH_{n} \is \ds \left[\hat{\bM}\,\otimes\,\hat{\bM}\,\otimes\,\hat{\bM} -\left(\hat{\bM}\,\otimes\,\hat{\bN}\,\otimes\,\hat{\bN} + \hat{\bN}\,\otimes\,\hat{\bM}\,\otimes\,\hat{\bN}+\hat{\bN}\,\otimes\,\hat{\bN}\,\otimes\,\hat{\bM}\right)\right]\\[2mm] \is \ds \Re\left[{e^{i\,n\,\theta}\left(\hat{\bx}^{\diamond}+i\hat{\by}^{\diamond}\right)^{(n)}}\right]~.
\label{e:graphene_structural_tensor}
\eqe
The invariants \textcolor{cgm}{of} a symmetric surface tensor of rank 2 $\bA$ $\in \bbR^3\times\bbR^3$ can be written as  \citep{Zheng_1993_01,Zheng_1994_01}
\eqb{lll}
 \ds \sJ_{1\bA}^{\vdash}\is \ds \tr(\bA)~,\\[2mm]
 \ds \bar{\sJ}_{2\bA}^{\vdash} \is \ds \frac{1}{2}\,\tr\left(\bA^2\right)~,\\[2mm]
  \sJ_{3\bA}^{\vdash} \is \ds \frac{1}{8}\tr\left(\Pi^{\bA}_{n}\,\bA\right)=\frac{1}{8}\Pi^{\bA}_{n}:\bA~,
  \label{e:general_invariant_form}
\eqe
where $\Pi^{\bA}_{n}$ is defined as
\eqb{lll}
\Pi^{\bA}_{n}\dis A^{m-1}\,\Re\left[e^{i\,n\,\theta+i\,(m-1)\,\theta_{\bA}} \left(\hat{\bM}^{\diamond}+i\,\hat{\bN}^{\diamond}\right)\right]~\text{with}~n=2m~,~m=1, 2, 3,...~,
\eqe
and $\theta_{\bA}$ and $A$ can be obtained from
\eqb{lll}
2\bA \is \tr(\bA)\,\bI+A\, \left[\cos(\theta_{\bA})\,\hat{\bM}^{\diamond}+\sin(\theta_{\bA})\hat{\bN}^{\diamond}\right]~.
\eqe
$\theta_{\bA}$ and $A$ can be easily obtained by using the spectral decomposition of $\bA$, i.e
\eqb{lll}
\bA \is \ds \sum_{\alpha=1}^{2}{\lambda_{\alpha\bA}\bY_{\!\alpha\bA}\otimes\bY_{\!\alpha\bA}}~,
\eqe
where $\lambda_{\alpha\bA}$ and $\bY_{\!\alpha\bA}$ are the eigenvalues and eigenvectors of $\bA$.
If $\theta$ selected such that $\hat{\bx}^{\diamond}$ and $\bY_{\!1\bA}$ are in the same direction, then $A=\lambda_{1\bA}-\lambda_{2\bA}; (\lambda_{1\bA}>\lambda_{2\bA})$ and $\theta_{\bA}=0$.\\
\subsection{Invariants based on the logarithmic strain}
The logarithmic strain is a good candidate for the development of material models. It can be used to additively decompose finite strains into volumetric/deviatoric and elastic/plastic strains. This additive decomposition simplifies the formulation of constitutive laws. In addition, it can capture micro-mechanical behavior of materials very \textcolor{cgm}{well} \citep{Neff_2013_01,Neff_2015_01,Neff_2015_02,Neff_2016_01}. Using Eq.~(\ref{e:general_invariant_form}), the invariant of the logarithmic strain $\bE^{(0)}=\ln(\bU)$ can be obtained by taking $\bA=\bE^{(0)}$ as
\eqb{lll}
 \ds  \sJ_{1\bE^{(0)}}\is \tr\left(\bE^{(0)}\right)=\ln(\lambda_1\,\lambda_2)=\ln(J)~,\\[2mm]
 \ds \sJ_{2\bE^{(0)}}^{\vdash} \is \ds \frac{1}{2}\,\tr\left(\left(\bE^{(0)}\right)^2\right) = \frac{1}{2}\,\left(\frac{1}{2}\ln(J)\bI+\bE^{(0)}_{\text{dev}}\right):\left(\frac{1}{2}\ln(J)\bI+\bE^{(0)}_{\text{dev}}\right) \\[4mm] \is \ds \frac{1}{4}\ln(J)+\frac{1}{2}\bE^{(0)}_{\text{dev}}:\bE^{(0)}_{\text{dev}}~,\\[2mm]
  \ds \sJ_{3\bE^{(0)}} \is \ds \frac{1}{8}\bbH\left(\bE^{(0)},\bE^{(0)},\bE^{(0)}\right)=\left(E^{(0)}\right)^3\,\cos(6\theta) = \left(\ln\left(\frac{\lambda_1}{\lambda_2}\right)\right)^3\,\cos(6\theta)~,
\eqe
where $\lambda_{\alpha}$ are the eigenvalues of the surface stretch tensor and $\bE^{(0)}_{\text{dev}}=\bE^{(0)}-1/2\ln(J)\bI$ is deviatoric part of the strain. These set of invariants can be simplified by eliminating the first invariant from the second invariant as
\eqb{lll}
 \ds \sJ_{1\bE^{(0)}} =\ds\ln(J)~,\\[2mm]
 \ds \sJ_{2\bE^{(0)}} =\ds \frac{1}{2}\bE^{(0)}_{\text{dev}}:\bE^{(0)}_{\text{dev}}= (\ln(\lambda))^2~,\\[2mm]
 \ds \sJ_{3\bE^{(0)}} =\ds \frac{1}{8}\bbH\left(\bE^{(0)},\bE^{(0)},\bE^{(0)}\right)=(\ln(\lambda))^3\,\cos(6\theta)~,
\eqe
with $\ds\lambda=\sqrt{\frac{\lambda_1}{\lambda_2}};\lambda_1>\lambda_2$.\\

\subsection{Invariant\textcolor{cgm}{s} based on the right Cauchy-Green tensor}
The logarithmic strain facilitates the development of material models and strain energy densities. But for the classical numerical description and FE implementation, derivatives of the strain energy density with respect to $\bC$ are needed. So, the chain rule should be utilized for material models based on $\bE^{(0)}$. The first and second derivative are the stress and elasticity tensor. These tensors can be directly obtained for isotropic materials \textcolor{cgm}{that} are developed based on $\bE^{(0)}$ without using the chain rule \citep{bonet_2008_01}, but it is not the case for anisotropic materials. Using Eq.~(\ref{e:general_invariant_form}), a set of invariants based on $\bC$ can be written as
\eqb{lll}
 \ds \sJ_{1\bC}^{\vdash} \is \ds \tr(\bC)~,\\[2mm]
 \ds \sJ_{2\bC}^{\vdash} \is \ds \frac{1}{2}\bC:\bC = \frac{1}{2}\left(\Lambda_1^2+\Lambda_2^2\right)~,\\[2mm]
 \ds \sJ_{3\bC}^{\vdash} \is \ds \frac{1}{8}\bbH(\bC,\bC,\bC)
 =\frac{1}{8}\left(\Lambda_1-\Lambda_2\right)^3\,\cos(6\theta)~.
\eqe
The material model will be simplified by using a set of invariants which correspond to \textcolor{cgm}{area-changing} and \textcolor{cgn}{area-invariant} deformations. So, $J$ is assumed to be an additional invariant and the set of invariants will be
\eqb{lll}
 \ds \bar{\sJ}_{1\bC}^{\vdash} \is \ds \tr(\bar{\bC})~,\\[2mm]
 \ds \bar{\sJ}_{2\bC}^{\vdash} \is \ds \frac{1}{2}\bar{\bC}:\bar{\bC} = \frac{1}{2}\left(\frac{\Lambda_1}{\Lambda_2}+\frac{\Lambda_2}{\Lambda_1}\right)~,\\[2mm]
 \ds \bar{\sJ}_{3\bC}^{\vdash} \is \ds \frac{1}{8}\bbH(\bar{\bC},\bar{\bC},\bar{\bC})
 =\frac{1}{8}\left(\frac{\lambda_1}{\lambda_2}-\frac{\lambda_2}{\lambda_1}\right)^3\,\cos(6\theta)~,\\
\bar{\sJ}_{4\bC}^{\vdash} \is J~,
\eqe
where $\bar{\bC}$ is \textcolor{cgn}{the area-invariant} part of $\bC$ which is defined based on the \textcolor{cgn}{area-invariant} deformation gradient $\bar{\bF}$ (see Eq.~\eqref{e:F_C_tlC}).
The material model will be more simple, if the set of the invariants are irreducible. $\bar{\sJ}_{1\bC}^{\vdash}$ can be written based on $\bar{\sJ}_{2\bC}^{\vdash}$ as
\eqb{lll}
(\bar{\sJ}_{1\bC}^{\vdash})^{2} \is \ds \frac{1}{2}\bar{\sJ}_{2\bC}^{\vdash}+2~.
\eqe
So, $\bar{\sJ}_{1\bC}^{\vdash}$ is dependent on $\bar{\sJ}_{2\bC}^{\vdash}$ and can be excluded from the list of invariants. In addition, $\bar{\sJ}_{2\bC}$ can be written as
\eqb{lll}
\bar{\sJ}_{2\bC}^{\vdash} \is 2\sJ_{2\bC}+1~,
\eqe
where $\sJ_{2\bC}$ and the final form of the other invariants are given in Eq.~\eqref{e:graphene_invar_C}.
\subsection{Invariant\textcolor{cgm}{s} based on the right Cauchy-Green and curvature tensors}
A surface can be described by the first and second fundamental forms of the surface\textcolor{cgm}{,} which are the metric and curvature tensors. Hence, the curvature tensor $\bkappa$ can be considered as the second tensorial object next to $\bC$. A set of nine invariants based on $\bC$ and $\bkappa$ can be obtained as
\eqb{lll}
 \ds \sJ_{1\bC} \is \ds \sqrt{\det(\bC)}=J~,\\[2mm]
 \ds \sJ_{2\bC} \is \ds \frac{1}{2}\bar{\bC}^{\bot}:\bar{\bC}^{\bot}~,\\[2mm]
 \ds \sJ_{3\bC} \is \ds \frac{1}{8}\bbH\left(\bar{\bC},\bar{\bC},\bar{\bC}\right) =\frac{1}{8}\left(\frac{\lambda_1}{\lambda_2}-\frac{\lambda_2}{\lambda_1}\right)^3\cos(6\varphi_{\bC})~,\\[2mm]
 \ds \sJ_{4\bkappa} \is \ds\frac{1}{2}\tr(\bkappa)=H~,\\[2mm]
 \sJ_{5\bkappa} \is \ds \kappa~,\\[2mm]
 \sJ_{6\bkappa} \is \ds \frac{1}{8}\bbH(\bkappa,\bkappa,\bkappa) =\frac{1}{8}(\kappa_1-\kappa_2)^3\cos(6\varphi_{\bkappa})~,\\[2mm]
 \ds \sJ_{7\bC\,\bkappa} \is \ds \frac{1}{2}\bar{\bC}:\bkappa~,\\[2mm]
 \ds \sJ_{8\bC\,\bkappa} \is \ds \frac{1}{8}\bbH\left(\bar{\bC},\bar{\bC},\bkappa\right) =\frac{1}{8}\left(\frac{\lambda_1}{\lambda_2}-\frac{\lambda_2}{\lambda_1}\right)^2(\kappa_1-\kappa_2)\cos(4\varphi_{\bC}+2\varphi_{\bkappa})~,\\[2mm]
 \ds \sJ_{9\bC\,\bkappa} \is \ds \frac{1}{8}\bbH\left(\bkappa,\bkappa,\bar{\bC}\right) =\frac{1}{8}\left(\frac{\lambda_1}{\lambda_2}-\frac{\lambda_2}{\lambda_1}\right)(\kappa_1-\kappa_2)^2\cos(2\varphi_{\bC}+4\varphi_{\bkappa})~,
\eqe
where $H$ and $\kappa$ are the mean and Gaussian curvatures and $\kappa_{\alpha}$ are the eigenvalues of the curvature tensor such that $\kappa_1>\kappa_2$. $\varphi_{\bC}$ and $\varphi_{\bkappa}$ are the angles between the eigenvectors of $\bC$ and $\bkappa$ relative to $\hat{\bx}$ and are defined as
\eqb{lll}
  \cos(\varphi_{\bkappa}) \dis \hat{\bx}\cdot\bY_{\!1\bkappa}~, \\
  \cos(\varphi_{\bC}) \dis \hat{\bx}\cdot\bY_{\!1\bC}~,
\eqe
where $\bY_{\!1\bkappa}$ and $\bY_{\!1\bC}$ are the eigenvectors of $\bC$ and $\bkappa$ corresponding to the largest eigenvalue. In the current model, the bending energy is assumed to be isotropic.\\
\section{Various derivatives}\label{s:some_coef}
The derivatives of the invariants of $\bC=\bF^{\text{T}}\,\bF$ can be written as
\eqb{lll}
  \ds \pa{\sJ_{1\bC}}{\bC} \is \ds \frac{1}{2}J\,\bC^{-1}~,\\[3mm]
  \ds \pa{\sJ_{2\bC}}{\bC} \is \ds \frac{1}{J}\bar{\bC}^{\bot}-\sJ_{2\bC}\,\bC^{-1}~,\\[3mm]
  \ds \pa{\sJ_{3\bC}}{\bC} \is \ds -\frac{3}{2}\sJ_{3\bC}\,\bC^{-1}+\frac{1}{8J}\left[a_{\hat{\bM}}\,\hat{\bM}+ a_{\hat{\bN}}\,\hat{\bN}\right]~,
  \label{e:derivative_metric_invariants_graphene}
\eqe
and derivative of strain energy density as
\eqb{lll}
\ds \pa{W}{\sJ_{1\bC}} \is \ds \frac{\varepsilon}{J}\,\hat{\alpha}^2\ln(J)e^{-\hat{\alpha}\ln(J)}+2\mu^{\prime}f_1 + \eta^{\prime}f_2~,\\[4mm]
\ds \pa{W}{\sJ_{2\bC}} \is \ds 2\mu(e_1-2e_2\,\sJ_{2\bC})-g_2\,\eta\,\sJ_{3\bC}~,\\[4mm]
\ds \pa{W}{\sJ_{3\bC}} \is \ds \eta(g_1-g_2\,\sJ_{2\bC})~,
\eqe
where $\mu^{\prime}$ and $\eta^{\prime}$ are
\eqb{lll}
\mu^{\prime} \dis \ds -\mu_1\,\hat{\beta}\,J^{\hat{\beta}-1}~,\\[2mm]
\eta^{\prime} \dis \ds -2\frac{\eta_1}{J}\ln(J)~.
\eqe
$H_i$ are needed in the computation of the 2.PK stress and its corresponding elasticity tensor, see Eqs.~\eqref{e:metric_2PK_tensorial} and \eqref{e:mem_2PK_elastiicty_tensor_metric}. They are defined as
\eqb{lll}
H_1 \dis \ds \varepsilon\,\hat{\alpha}^2\ln(J)\,e^{-\hat{\alpha}\ln(J)}-2\mu_1\,\hat{\beta}\,J^{\hat{\beta}}\,f_1-2\eta_1\ln(J)\,f_2 -H_2\,\sJ_{2\bC}- 3H_3\,\sJ_{3\bC}~,\\[2mm]
H_2 \dis \ds 2\left(2\mu(e_1-2e_2\,\sJ_{2\bC})- g_2\,\eta\,\sJ_{3\bC}\right)~,\\[2mm]
H_3 \dis \ds \eta(g_1-g_2\,\sJ_{2\bC})~,
\eqe
and their derivatives w.r.t. the invariants of $\bC$ are
\eqb{lll}
\ds \pa{H_1}{\sJ_{1\bC}} \is \ds \frac{\varepsilon\,\hat{\alpha}^2}{J}\,(1-\hat{\alpha}\,\ln(J))\,e^{-\hat{\alpha}\,\ln(J)}
-2\mu_1\,\hat{\beta}^2\,J^{\hat{\beta}-1}\,f_1-2\frac{\eta_1}{J}\,f_2-\pa{H_2}{\sJ_{1\bC}}\sJ_{2\bC}-3\pa{H_3}{\sJ_{1\bC}}\,\sJ_{3\bC}~,\\[4mm]
\ds \pa{H_2}{\sJ_{1\bC}} \is \ds 2\left[2\left(-\mu_1\,\hat{\beta}\,J^{\hat{\beta}-1}(e_1-2\,e_2\,\sJ_{2\bC})\right)+2\frac{g_2\,\eta_1}{J}\, \ln(J)\,\sJ_{3\bC} \right]~,\\[4mm]
\ds \pa{H_3}{\sJ_{1\bC}} \is \ds -2\frac{\eta_1}{J}\ln(J)\,(g_1-g_2\,\sJ_{2\bC})~,
\eqe

\eqb{lll}
\ds \pa{H_1}{\sJ_2} \is \ds -2\mu_1\,\hat{\beta}\,J^{\hat{\beta}}\,(e_1-2\,e_2\,\sJ_{2\bC})+2g_2\,\eta_1\,\ln(J)\,\sJ_{3\bC} -H_2-\pa{H_2}{\sJ_{2\bC}}\sJ_{2\bC}-3\sJ_{3\bC}\,\pa{H_3}{\sJ_{2\bC}}~,\\[4mm]
\ds \pa{H_2}{\sJ_{2\bC}} \is \ds -8\mu\,e_2~,\\[4mm]
\ds \pa{H_3}{\sJ_{2\bC}} \is \ds -\eta\,g_2~,
\eqe
\textcolor{cgm}{and}
\eqb{lll}
\ds \pa{H_1}{\sJ_{3\bC}} \is \ds -2\eta_1\,(g_1-g_2\,\sJ_{2\bC})\,\ln(J)-\pa{H_2}{\sJ_{3\bC}}\,\sJ_{2\bC}-3\,H_3~,\\[4mm]
\ds \pa{H_2}{\sJ_{3\bC}} \is \ds -2g_2\,\eta~.
\eqe
Furthermore, $a_{\hat{\bM}}$ and $a_{\hat{\bN}}$ and their derivatives, and derivative of  $\bC^{-1}$ and $\bar{\bC}^{\bot}$ w.r.t $\bC$ are needed in the computation of the 2.PK stress and its elasticity tensor. They are
\eqb{lll}
a_{\hat{\bM}} \is 3\left[\left(\hat{\bM} : \bar{\bC}\right)^{2}- \left(\hat{\bN} : \bar{\bC}\right)^{2}\right]~,\\[4mm]
a_{\hat{\bN}} \is -6\left[\left(\hat{\bM} : \bar{\bC}\right)\left(\hat{\bN} : \bar{\bC}\right)\right]~,
\eqe
\eqb{lll}
\ds \pa{a_{\hat{\bM}}}{\bC} \is \ds \frac{6}{J}\,\left[\left(\hat{\bM}:\bar{\bC}\right)\,\hat{\bM}-\left(\hat{\bN}:\bar{\bC}\right)\,\hat{\bN}\right]
-a_{\hat{\bM}}\,\bC^{-1}~,\\[4mm]
\ds \pa{a_{\hat{\bN}}}{\bC} \is \ds -\frac{6}{J}\,\left[\left(\hat{\bM}:\bar{\bC}\right)\,\hat{\bN}+\left(\hat{\bN}:\bar{\bC}\right)\,\hat{\bM}\right]
-a_{\hat{\bN}}\,\bC^{-1}~.
\eqe
and
\eqb{lll}
\ds \frac{\partial\bC^{-1}}{\oplus\partial\bC} \is \ds -\frac{1}{2}(\bI \otimes \bI + \bI \boxtimes \bI)~,\\[3mm]
\ds \frac{\partial\bar{\bC}^{\bot}}{\oplus\partial\bC} \is \ds -\frac{1}{2}\,\bar{\bC}^{\bot}\oplus \bC^{-1}+\frac{1}{2J}(\bI \otimes \bI + \bI \boxtimes \bI-\bI \oplus \bI)~,
\eqe
where $\partial \hat{\bullet}/\oplus\partial \hat{\bullet}$ is defined as
\eqb{lll}
\ds \frac{\partial\bA}{\oplus\partial\bB} \dis \ds \frac{\partial A^{\alpha\beta}}{\partial B_{\gamma\delta}}\,\bA_{\alpha}\otimes\bA_{\gamma}\otimes\bA_{\delta}\otimes\bA_{\beta}~.
\eqe
The chain rule of \citet{Kintzel_2006_01} and \citet{Kintzel_2006_02} simplifies and expedites the derivation of elasticity tensors. They define
the elasticity tensor (related to the membrane strain energy density) as
\eqb{lll}
\ds \mathbb{C}^{\mathrm{L}}_{\text{m}} \dis \ds \frac{\partial^2{W_{\text{m}}}}{\partial{\bC}\oplus\partial{\bC}}= \frac{\partial^2{W_{\text{m}}}}{\partial{C_{\alpha\beta}}\partial{C_{\gamma\delta}}} \,\bA_{\alpha}\otimes\bA_{\gamma}\otimes\bA_{\delta}\otimes\bA_{\beta}~,
\eqe
which, for the proposed membrane strain energy density, is
\eqb{lll}
\ds \mathbb{C}^{\mathrm{L}}_{\text{m}} \is 2\biggl\{ \ds \left(\frac{J}{2}\pa{H_1}{\sJ_{1\bC}}-\sJ_{2\bC}\pa{H_1}{\sJ_{2\bC}} -\frac{3}{2}\sJ_{3\bC}\,\pa{H_1}{\sJ_{3\bC}}\right)\,\bC^{-1} \oplus \bC^{-1}+\frac{1}{J^2}\pa{H_2}{\sJ_2}\,\bar{\bC}^{\bot}\oplus\,\bar{\bC}^{\bot}\\[4mm]
\plus \ds \frac{2}{J}\pa{H_1}{\sJ_2}\,\left[\bC^{-1} \oplus \bar{\bC}^{\bot}\right]^{\text{S}}
+  \frac{1}{4J}\,\pa{H_1}{\sJ_3}\,\left[\bC^{-1} \oplus \bZ \right]^{\text{S}}
+ \ds \frac{1}{2J^2}\pa{H_3}{\sJ_2}\left[\bZ\oplus \bar{\bC}^{\bot} \right]^{\text{S}}\\[4mm]
\mi \ds \frac{1}{2}H_1\,(\bC^{-1} \otimes \bC^{-1} + \bC^{-1} \boxtimes \bC^{-1})
+ \ds \frac{H_2}{2J^2}\,(\bI \otimes \bI + \bI \boxtimes \bI-\bI \oplus \bI)\\[4mm]
\plus \ds \frac{3H_3}{2J^2}\,\left[(\hat{\bM}:\bar{\bC})\,(\hat{\bM} \oplus \hat{\bM}-\hat{\bN}\oplus\hat{\bN})-(\hat{\bN}:\bar{\bC})(\hat{\bM}\oplus\,\hat{\bN}+\hat{\bN}\oplus\,\hat{\bM})\right] \biggr\}~,
\eqe
with
\eqb{lll}
\bZ \dis \ds a_{\hat{\bM}}\,\hat{\bM}+a_{\hat{\bN}}\,\hat{\bN}~,
\eqe
\eqb{lll}
\ds (\bA\oplus\bB)^{\text{S}} \dis \ds \frac{1}{2}(\bA\oplus\bB+\bB\oplus\bA)~.
\eqe
The components of $\mathbb{C}^{\mathrm{L}}_{\text{m}}$ should be rearranged for a FE implementation (see \citet{Kintzel_2006_01} and \citet{Kintzel_2006_02}). This rearrangement can be written as
\eqb{lll}
\mathbb{C}_{\text{m}} \dis \ds \bigl(\mathbb{C}^{\mathrm{L}}_{\text{m}}\bigr)^{\text{R}}= \bigl(C^{\mathrm{L}~\alpha\beta\gamma\delta}_{\text{m}}\,\bA_{\alpha}\otimes\bA_{\beta}\otimes\bA_{\gamma}\otimes\bA_{\delta}\bigr)^{\text{R}} =C^{\mathrm{L}~\alpha\gamma\delta\beta}_{\text{m}}\,\bA_{\alpha}\otimes\bA_{\beta}\otimes\bA_{\gamma}\otimes\bA_{\delta}~,
\eqe
where $C^{\mathrm{L}~\alpha\gamma\delta\beta}_{\text{m}}$ can be written as
\eqb{lll}
C^{\mathrm{L}~\alpha\gamma\delta\beta}_{\text{m}} \is \textcolor{cgm}{\bA^{\alpha}\otimes\bA^{\beta}:\mathbb{C}_{\text{m}}: \bA^{\gamma}\otimes\bA^{\delta}}~.
\eqe
This rearrangement can be written for two second order tensors of $\bA$ and $\bB$ as
\eqb{lll}
(\bA\oplus\bB)^{\text{R}} \is \bA\otimes\bB~,\\[2mm]
(\bA\otimes\bB)^{\text{R}} \is \bA\boxtimes\bB^{\text{T}}~,\\[2mm]
(\bA\boxtimes\bB)^{\text{R}} \is \bA\oplus\bB^{\text{T}}~,\\[2mm]
\left[(\bullet)^{\text{R}}\right]^{\text{L}}=\left[(\bullet)^{\text{L}}\right]^{\text{R}}\is (\bullet)~.
\eqe
\section{Analytical solution of for bending a cylindrical thin beam}\label{s:circ_beam_analy}
An analytical solution based on linear Euler--Bernoulli beam theory of thin structures is given in this section. The considered beam is a hollow cylinder with inner and outer radii, thickness and length of $r_{\text{i}}$ and $r_{\text{o}}$, $t$ and $L$, respectively. The polar moment of inertia of a thin ring is
\eqb{lll}
\tilde{I}_{\text{p}} \is \ds 2\pi\,r_{\text{m}}^3\,t~,
\eqe
where $r_{\text{m}}$ is
\eqb{lll}
r_{\text{m}} \is \ds \frac{r_{\text{i}}+r_{\text{o}}}{2}~.
\eqe
Exploiting the symmetry, the moment of inertia is
\eqb{lll}
\tilde{I}_{w} \is \ds \frac{\tilde{I}_{\text{p}}}{2}=\pi\,r_{\text{m}}^3\,t~.
\eqe
The analytical solution for a cantilever beam with a concentrate force $F_{w}$ at its tip can be written as \citep{beer2014}
\eqb{lll}
F_{w} \is \ds \frac{3\tilde{E}\,\tilde{I}_{w} }{L^3}\,w~,
\eqe
where $\tilde{E}$ is 3D elastic modulus with unit of $\text{N/m}^2$. The 2D elastic modulus $E$ \citep{Kumar2014_01} and moment of inertia $I_{w}$ can be written as
\eqb{lll}
E \is \tilde{E}\,t~,\\[3mm]
I_{\text{y}} \is \ds \pi\,r_{\text{m}}^3~.
\eqe
$w$ and the axial displacement of the wall can be connected as
\eqb{lll}
w \is \ds \tan\left(\frac{\pi}{2}-\theta_{\text{w}}\right)\,\Delta L_{\text{axial}}~,
\eqe
and the final perpendicular force-displacement relation is
\eqb{lll}
F_{w} \is \ds \frac{3E\,I_{w} }{L^3}\,\tan\left(\frac{\pi}{2}-\theta_{\text{w}}\right)\,\Delta L_{\text{axial}}~.
\eqe
\textcolor{cgm}{The axial force $F_{\text{A}}$ can be related to $F_{w}$ as}
\eqb{lll}
\ds \textcolor{cgm}{F_{\text{A}}} \is \ds \textcolor{cgm}{F_{w}\,\tan\left(\frac{\pi}{2}-\theta_{\text{w}}\right)~.}
\eqe
\bibliographystyle{model1-num-names}
\bibliography{bibliography}

\end{document}